\documentclass{aa}
\usepackage[varg]{txfonts}
\usepackage{amsmath}
\bibpunct{(}{)}{;}{a}{}{,}
\usepackage{epsfig}
\usepackage{graphicx}
\usepackage{epstopdf}
\usepackage{amssymb}
\usepackage{ulem}

\begin{document}
\title{WSRT observations and surface photometry \\ of two unusual spiral galaxies.\thanks{FITS files of  \ion{H}{I} data cubes  are available at the CDS via anonymous ftp to cdsarc.u-strasbg.fr (130.79.128.5) } }

\author{A.~S.~Saburova\inst{\ref{inst1}}
\and  G.~I.~G.~J\'ozsa \inst{\ref{inst2},\ref{inst3}}\and
A.~V.~Zasov \inst{\ref{inst1}}
\and D.~V.~ Bizyaev \inst{\ref{inst4},\ref{inst1}}}

\institute{ Moscow M.V. Lomonosov State University, Sternberg Astronomical Institute, Universitetskii pr. 13, 119992, Moscow, Russia
\email{saburovaann@gmail.com}\label{inst1}
\and
Netherlands Institute for Radio Astronomy, Postbus 2, 7990 AA Dwingeloo, The Netherlands \email{jozsa@astron.nl}\label{inst2}
\and
Argelander-Institut f\"ur Astronomie, Auf dem Hügel 71, D-53121 Bonn\label{inst3}
\and
New Mexico State University and Apache Point Observatory, Sunspot, NM 88349, USA \email{dmbiz@apo.nmsu.edu}\label{inst4}
}

\abstract{ We  discuss the results of a mass decomposition of two
  spiral galaxies, NGC~6824 and UGC~11919. In a previous analysis of
  the Hyperleda catalog, the galaxies were identified as having a
  peculiar dynamical $M/L$. The aim of 
this study is to confirm or disprove the preliminary
  findings, indicating a non-standard stellar initial mass function (IMF) for the galaxies. The surface photometry in B, V,  and R  bands was carried out with the Apache Point 0.5-m telescope and the \ion{H}{I}  data cubes were obtained with the Westerbork Synthesis Radio Telescope (WSRT). 
Photometric profiles were
  decomposed into bulge and exponential disk components. Using the
  obtained \ion{H}{I} data cubes, rotation curves of both
  galaxies were constructed. Employing the photometric
  profiles, the mass distribution of the galaxies was decomposed into
  mass components: bulge, stellar disk, gas, and pseudo-isothermal dark
  halo. We conclude that NGC~6824 possesses a stellar disk
  with mass-to-light ratio $(M/L_B)_{\rm disk} = 2.5$, in agreement with its color $(B-V)_0$. On the contrary, UGC~11919 appears to have a very lightweight disk. Its dynamically estimated mass corresponds to a low stellar disk mass-to-light ratio $(M/L_B)_{\rm disk} \approx 0.5$. Under standard assumptions, this ratio does not agree with the relatively red color of the disk, while a bottom light stellar initial mass function is needed to explain the observations. }
\titlerunning{Observations and surface photometry of two unusual spiral galaxies}
\maketitle

\sloppypar
\vspace{2mm}
\noindent
\date{Received <date> /
Accepted <date>}

\section{Introduction}
The total mass-to-light ratio of a galaxy ($M/L$) is a key observable
comprising information about the galaxy's evolutional status, its
stellar initial mass function (IMF), and the relative masses of its
stellar disk and  dark matter halo. Hence, for a given galaxy, a
peculiar $M/L$
for the observed color index may indicate both  unusual dynamical
properties and/or a non-standard stellar IMF.

The universality of the IMF is a matter of hot debate. There  is
evidence both in favor and against its uniformity
among galaxies and galaxy types.   
The current consensus
shared among a number of studies is that the IMF remains 
the same for different galaxies
(see \citealt {Kroupa2012} and references
therein). 
This is inferred from star counts in the Milky Way and in nearby galaxies such as the
LMC \citep { Kroupa2002, Gilmore,
  Bastian}. However, several authors argue against
  the universality of the IMF.

\citet{Meurer2009} present a correlation
between the ratio of the
$H_{\alpha}$ line and far UV fluxes $F_{H_{\alpha}}/F_{FUV}$  (corrected for
the extinction) and the surface brightness in R-band. This ratio also appears to correlate with the rotation
velocities, luminosities, and dynamical masses of galaxies. The observed
range of $F_{H_{\alpha}}/F_{FUV}$ could be explained in terms of
a variation of either the high mass end or the slope of the IMF. 

According to a study of $\sim 300$ galaxies with distances
lower than 11 Mpc and star formation
  rates (SFRs) spanning an interval
of five orders, the
ratio $F_{H_{\alpha}}/F_{FUV}$ is roughly constant only for normal
spirals with a star formation rate of $SFR\sim 1 M_{ \sun} /{\rm
year}$, but
for dwarf galaxies 
$F_{H_{\alpha}}/F_{FUV}$ decreases with decreasing luminosity
\citep{Lee_Gil}.

High dynamical $M/L$s in the V-band for ultra-compact dwarf galaxies may also indicate the presence of non-universal IMFs with a large fraction of stellar remnants or low-mass stars \citep{Dabringhausen2010}. 
\begin{table*}
\caption {Main properties of the observed galaxies based on Hyperleda. (1) Name;
(2) morphological type;
(3) accepted distance;
(4) inclination angle;
(5) rotational velocity;
(6) \ion{H}{I} total flux; (7) $(B-V)_0$ color index corrected for the Galactic extinction; (8) apparent B-band stellar magnitude corrected for the Galactic extinction.
\label{tab1}}
\centering
\begin{tabular}{c c c c c c c c c}

\hline\hline
{\small Galaxy} &type&{\small $D$, Mpc }&$i$($\degr$)&{\small$v_{\rm rot}$ }&{\small flux \ion{H}{I}}&$(B-V)_0$&$btc$\\
&&&&(${\rm km}\,{\rm s}^{-1}$)&(${\rm Jy}\,{\rm km}\,{\rm s}^{-1})$&&\\
(1)&(2)&(3)&(4)&(5)&(6)&(7)&(8)\\
\hline
{\small UGC~11919}&SABb&74.3&54.8&71.6&5.78&0.74&12.89\\
{\small NGC~6824}&Sab&50.5&46&359&3.24&0.65&12.19\\
\hline
\end{tabular}
\end{table*}

By comparison of
equivalent widths of the $H_{\alpha}$ line for a large sample of
galaxies, \citet{Hoversten2008} show 
that the IMF could change systematically
with the luminosity of galaxies. 
Using the ${\rm H}{\alpha}$ line width and
the color to  confine the IMF, they conclude that galaxies with lower luminosities
contain a lower fraction of massive stars. Near-infrared observations of disky galaxies
by  \citet{Zackrisson} show that their stellar halos possess
extremely red colors which could be explained by a non-standard
stellar IMF with an excess in low mass stars.
For a sample of $\sim 33000$ galaxies, \citet{Gunawardhana} show
that the IMF depends on the SFR: galaxies with higher SFR tend to contain more massive stars. Finally, \citet{Goudfrooij} find that the optical colors of
metal-rich globular clusters associated with seven massive elliptical
galaxies are bluer than those of parent galaxies at a given radius. They
explain the observed color difference by the presence of a bottom-heavy
IMF in the observed galaxies.

The question whether the IMF is universal is highly relevant. Stellar
population models, which connect star formation history and
stellar mass with photometric or spectrophotometric data, are
usually based on the assumption of the universality of the IMF
(see, e.g., \citealt{bdj}, \citealt{Bruzual2003}, \citealt{Into} and others). 

In the following, we identify the standard IMF (and hence the standard
IMF-based $M/L$ -- color relationship) with the scaled (bottom-light)
Salpeter-like IMF \citep{bdj}. For evolutionary models
with exponentially decaying star formation rates, this IMF leads to
a  $M/L$ -- color relationship which is practically identical to
the relationship based on the Kroupa' IMF (see Appendix B in
\citealt{Portinari}). It is worth mentioning that the
Salpeter-light IMF agrees with dynamical mass estimates of
stellar populations obtained both from rotation curves \citep{bdj}
as well as from measurements of the stellar velocity
dispersion of spiral disks 
with known color indices \citep{ Zasov2011}. 
It is not the aim of the present paper to study the universal
IMF applicable for most of spiral galaxies. Instead we try to find
 outliers, galaxies which may possess a
non-typical $M/L$ for
their color, and to confirm their existence.

Low surface
brightness galaxies (LSBs) may be a good example for a possibly peculiar IMF. Based on stellar population models with
the standard IMF they are believed to be dark halo dominated
systems. However, some independent estimates of their disk masses
give evidence that LSB disks may have extremely high disk $M/L$
(see, e.g., \citealt{Saburova2011}, \citealt{Fuchs2003}), which can
be explained if their IMF is bottom-heavy and hence deviates from the
standard IMF towards the excess of low-massive stars (see
\citealt{Lee2004}).  Alternatively, their disks may contain some fraction of
dark matter in non-baryonic \citep{Read} or baryonic form. We note
  that whereas the excess of massive stars may be revealed from  $H_{\alpha}$ or UV data, the
  excess or deficit of low massive stars can be found using mass decompositions only. 

In our previous studies we searched for high surface-brightness disk
galaxies that may  possess non-standard $M/L$,
either too high or too low for the observed color indices (see
\citealt{Saburova2009}), using catalog data, which might be
  contaminated by individual errors. 
Errors in  rotation velocity
or possibly in conjunction
with errors in orientation angle(s), which may be responsible for the
overestimation or underestimation of the galaxy's mass 
cannot be excluded in that study and should be
  confirmed independently. The other
aspect which should be taken into account is the presence of a dark
matter halo. Within the optical radial extent of the disk,
  its  contribution to the total mass of a galaxy can be comparable
to that of its stellar components \citep{Zasov2011}. This
means that $M/L$ of the stellar population 
should always be lower than the total $M/L$ of a galaxy
within any radial distance. Hence, if 
 the total
$M/L$ ratio is too low for
the color of a galaxy, including the dark matter (DM) halo
 makes the stellar
IMF even more peculiar.  In the opposite case, if the total
$M/L$ is unusually high for the color index of a
galaxy, the situation becomes ambiguous: 
either the galaxy possesses an extremely
massive dark halo within the optical radius, or it possesses a bottom-heavy stellar
IMF. A mass decomposition may solve the ambiguity, if some
additional information, e.g., the radial scalelength of
the disk, is known.

In \citet{Saburova2009}, making use of Hyperleda, we estimate  total $M/L$
ratios within the optical radius $R_{25}$ for a sample of $\sim 1300$
galaxies. The obtained
total $M/L$ ratios were correlated with the $(B-V)_0$ color
indices of the galaxies in our sample. Then we compared the positions of the galaxies on the
$(M/L_B)-(B-V)_0$ diagram with the relation based on stellar
population synthesis modeling by \citet{bdj} for 
the standard
IMF. We select the galaxies which significantly deviate from the
model relation, because peculiarly high $M/L$ ratios (if they
are not the result of observational errors) can either
be the result
of a very heavy disk with bottom heavy IMF (with a lack of massive
stars) or reflect an unusually large contribution of dark matter in
baryonic or non-baryonic form to the total mass of a
galaxy. Peculiarly low dynamical $M/L$ can indicate a
non-universal IMF with a deficit of low massive stars.

In this work we present the detailed follow-up observations of
the two  galaxies NGC~6824 and UGC~11919, which were
selected as extreme cases from the parent sample as described above. They were observed in the \ion{H}{I}
 emission line with the 
Westerbork Synthesis Radio Telescope
 (WSRT). In addition, multicolor surface photometry 
with the Apache-Point 0.5 m telescope was carried out.  The data were used to construct a mass model of  the galaxies and to estimate the masses of their stellar components. This allows us to check whether the peculiar values of $M/L$
of the galaxies can be confirmed. The main properties of the
 galaxies, taken from the Hyperleda database and the accepted
 distances from NED are given in Table \ref{tab1}.

\section{Observations and data reduction}

\subsection{Radio observations}
We conducted \ion{H}{I} observations of NGC~6824 and UGC~11919 with
the WSRT in August and September 2011. The observational parameters are
listed in Table \ref{tab2}. We used a bandwidth of 10 MHz, subdivided into 1024 channels observed in two parallel-handed
polarization products. The linewidth of the observed channels is
  2.06 ${\rm km}\,{\rm s}^{-1}$. We performed a standard data reduction with the
Miriad software package (\citealt{Sault}). 

After flagging bad data and a primary bandpass and gain calibration, a first continuum subtraction was
performed using a polynomial spectral fit to the visibilities. To correct the
frequency-independent gains, a self-calibration was applied to the
resulting continuum data (the fitted continuum model). The resulting gains were
then copied to the original data sets. For the final continuum
subtraction, we first subtracted the continuum model resulting from the
selfcal procedure directly from the visibilities, to then
perform a
second-order continuum subtraction, again on the visibilities. The
resulting continuum-subtracted visibilities were then inverted to
obtain the data cubes. We produced data cubes with different weighting
schemes. The cubes that finally adopted were gridded and inverted using a robust
weighting of 0.4 and averaged over four channels, representing the most suitable
scheme for our aims (to obtain the rotation curve). After the
inversion we applied a Hanning smoothing. Finally, the obtained
\ion{H}{I} data cubes were deconvolved, using the Clean algorithm with
clean masks centered on the emission, iteratively decreasing the
clean cutoff and increasing the size of the mask regions.

\begin{figure*}
\begin{center}
\begin{minipage}[h]{1.0\linewidth}

\includegraphics[ trim= 0 0 0 0,clip,height=5.5cm]{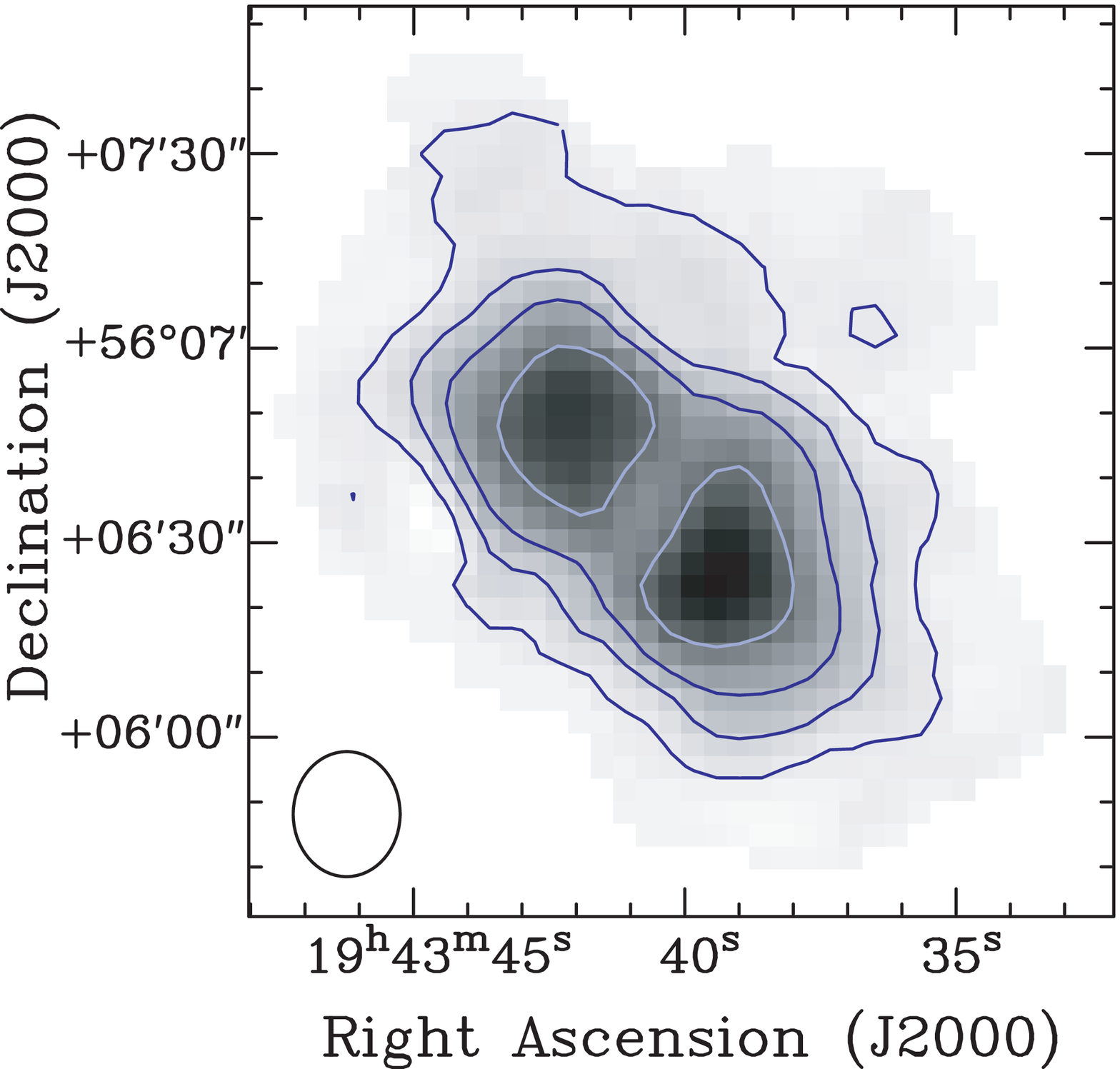}
\includegraphics[trim= 0 0 0 0,clip,height=5.5cm]{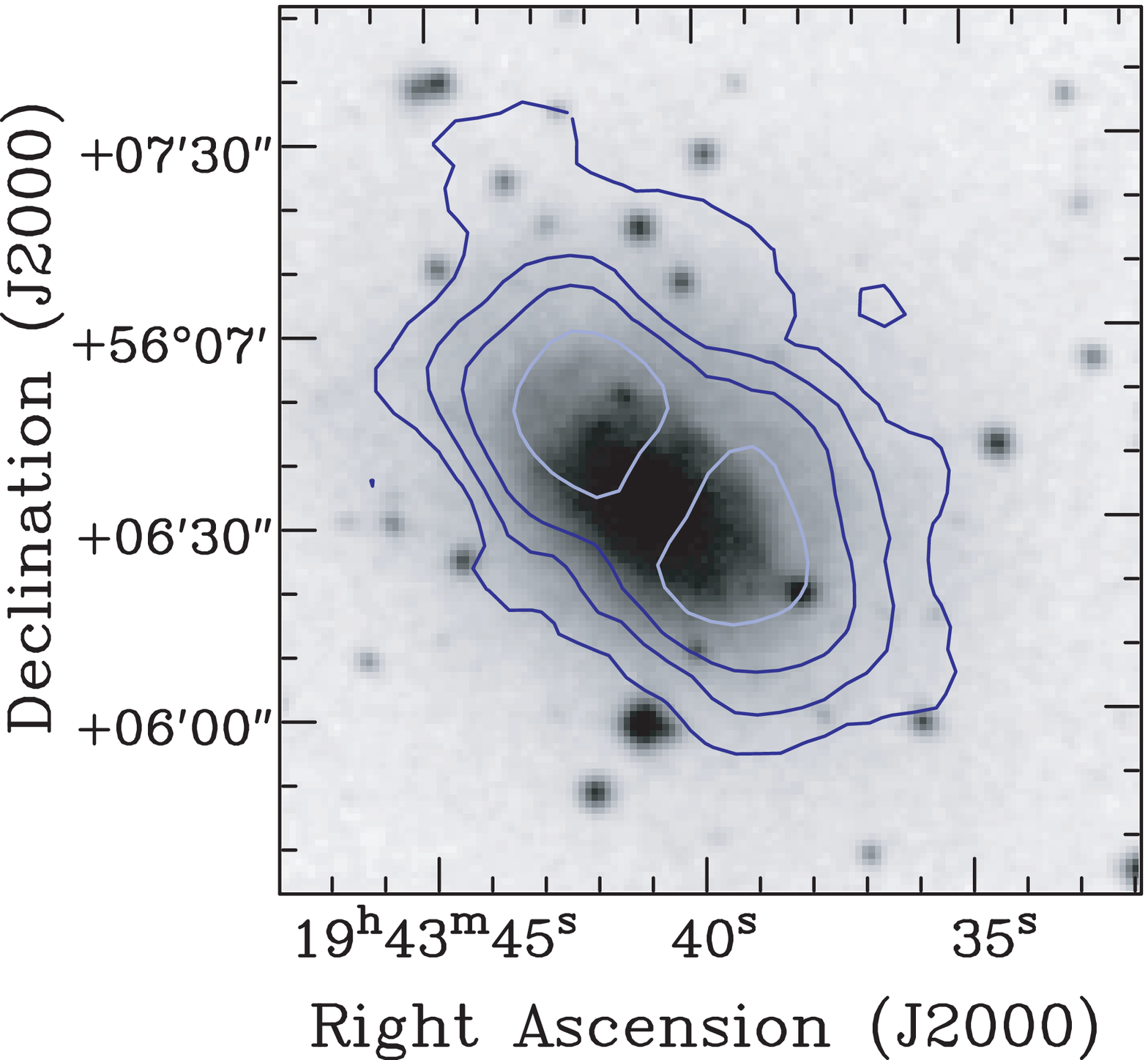}
\includegraphics[trim= 0 0 0 0,clip,height=5.5cm]{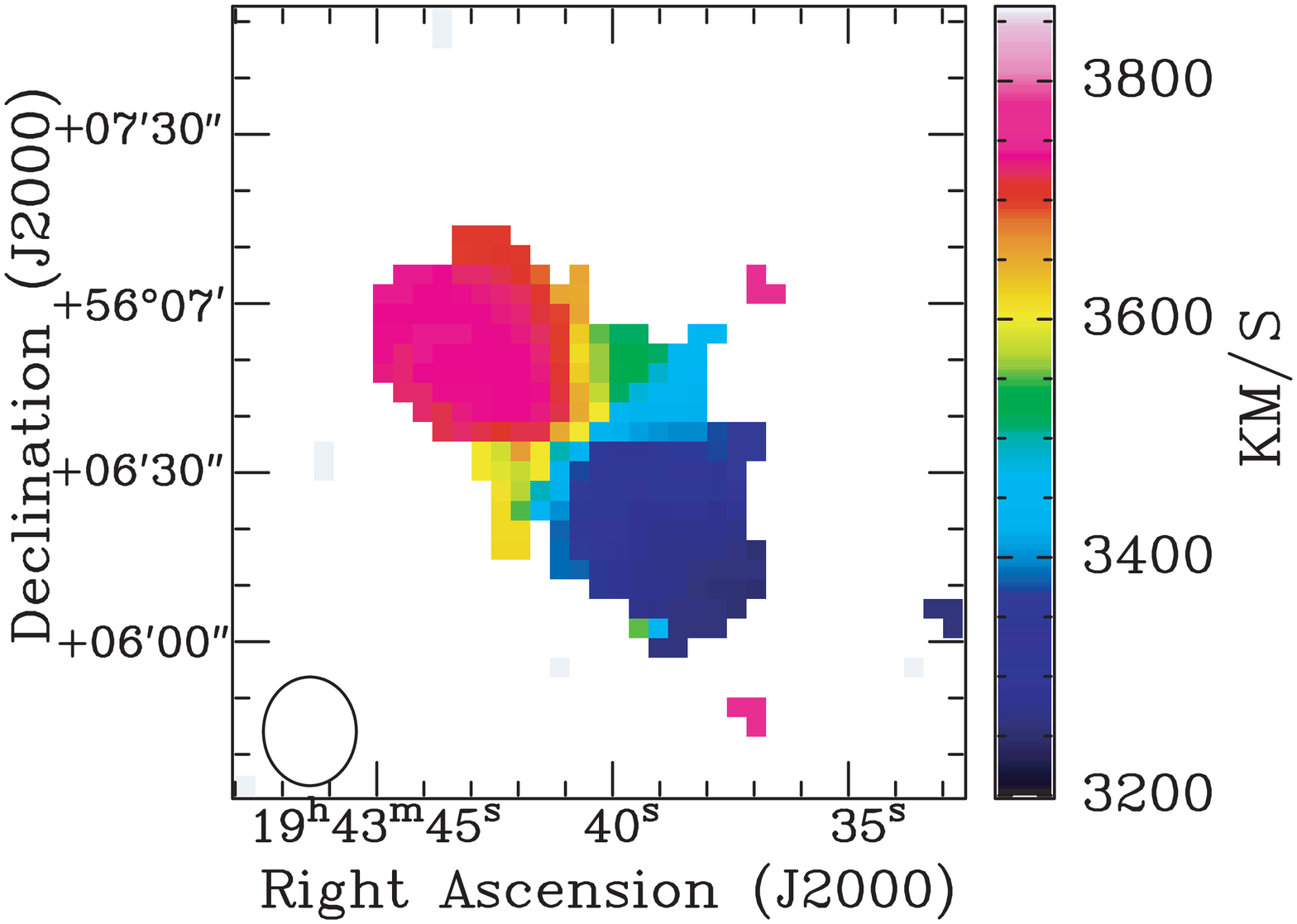}
\includegraphics[trim= 0 0 0 0,clip,height=5.5cm]{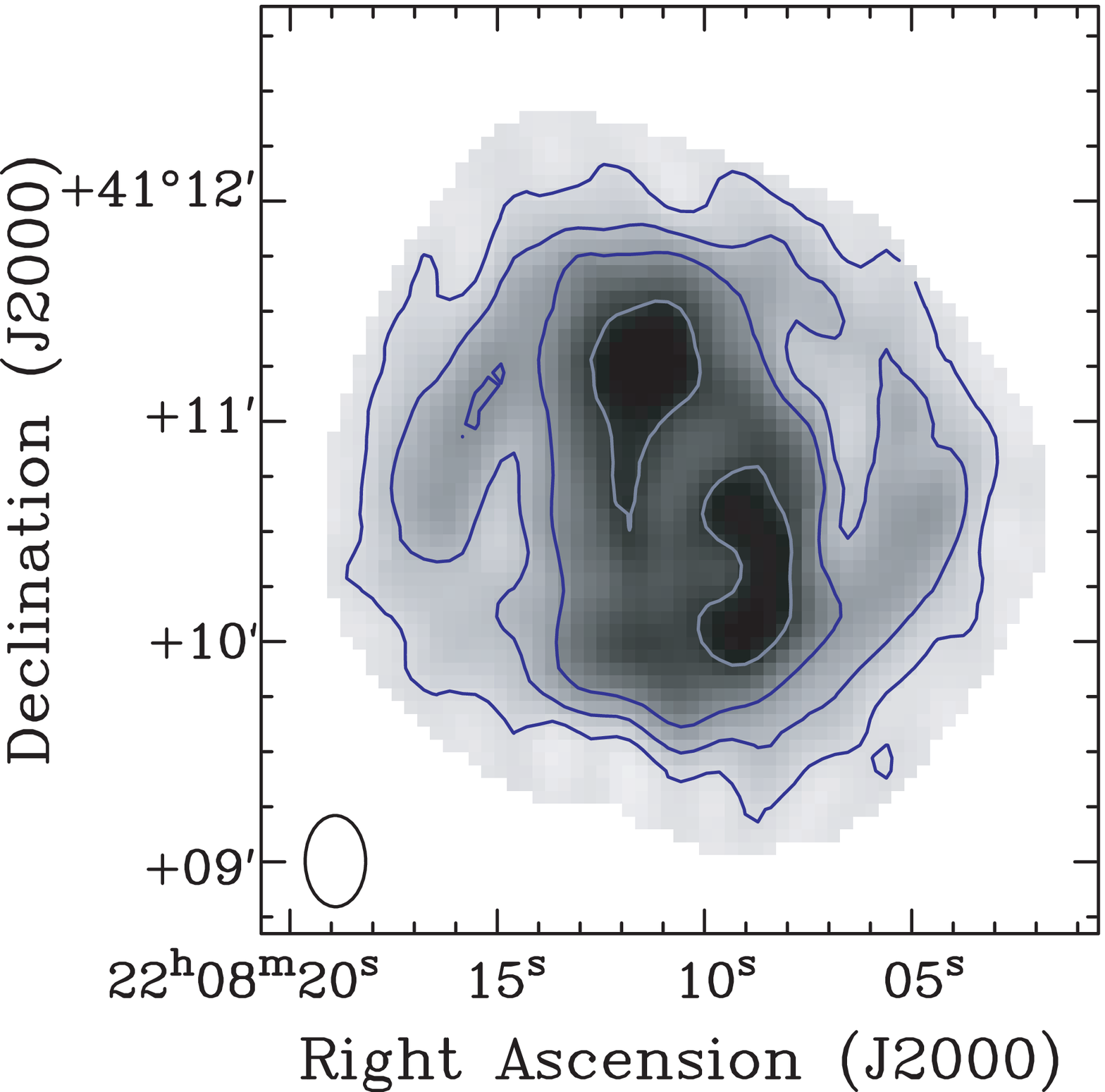}
\includegraphics[trim= 0 0 0 0,clip,height=5.5cm]{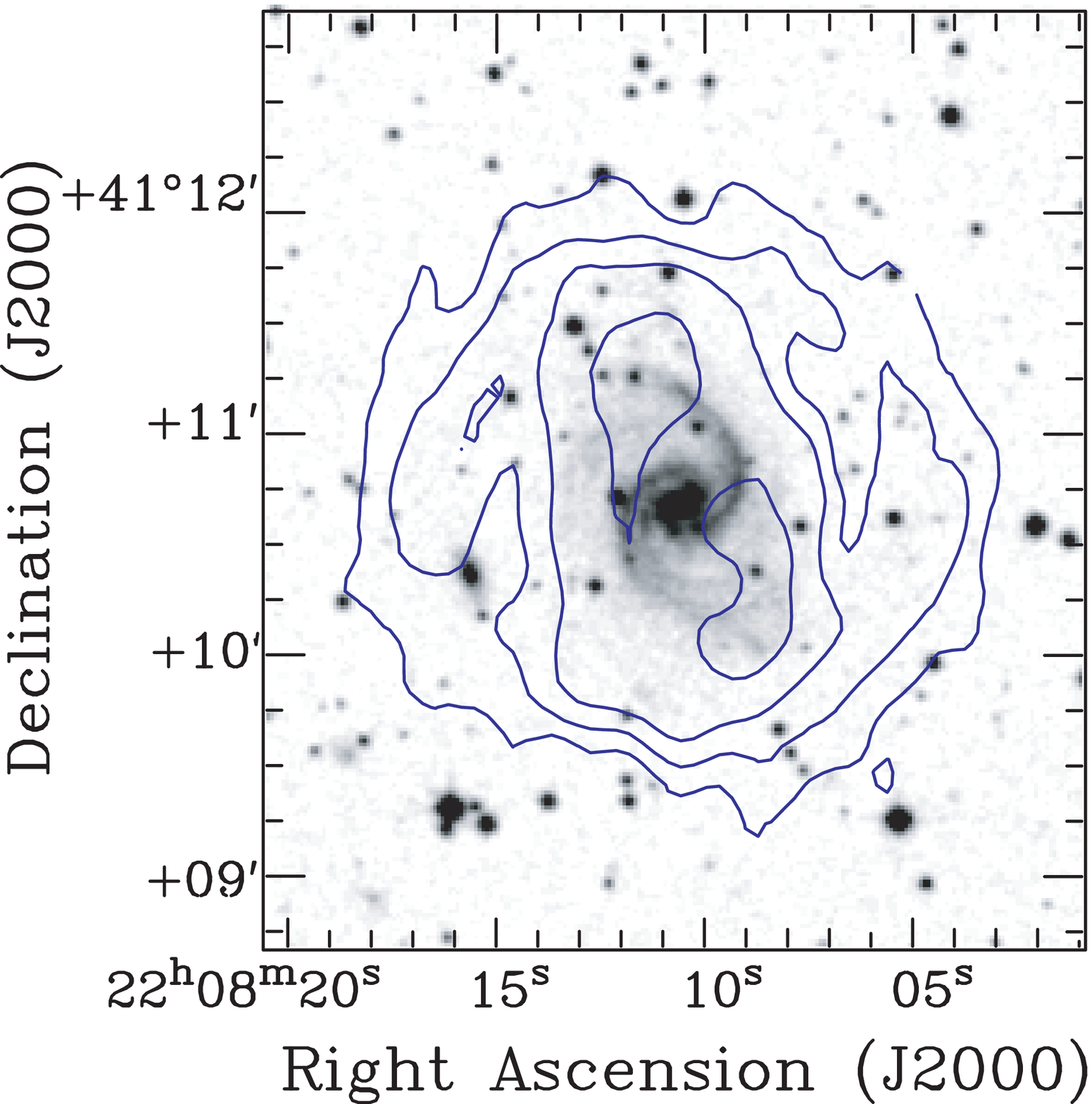}
\includegraphics[trim= 0 0 0 0,clip,height=5.5cm]{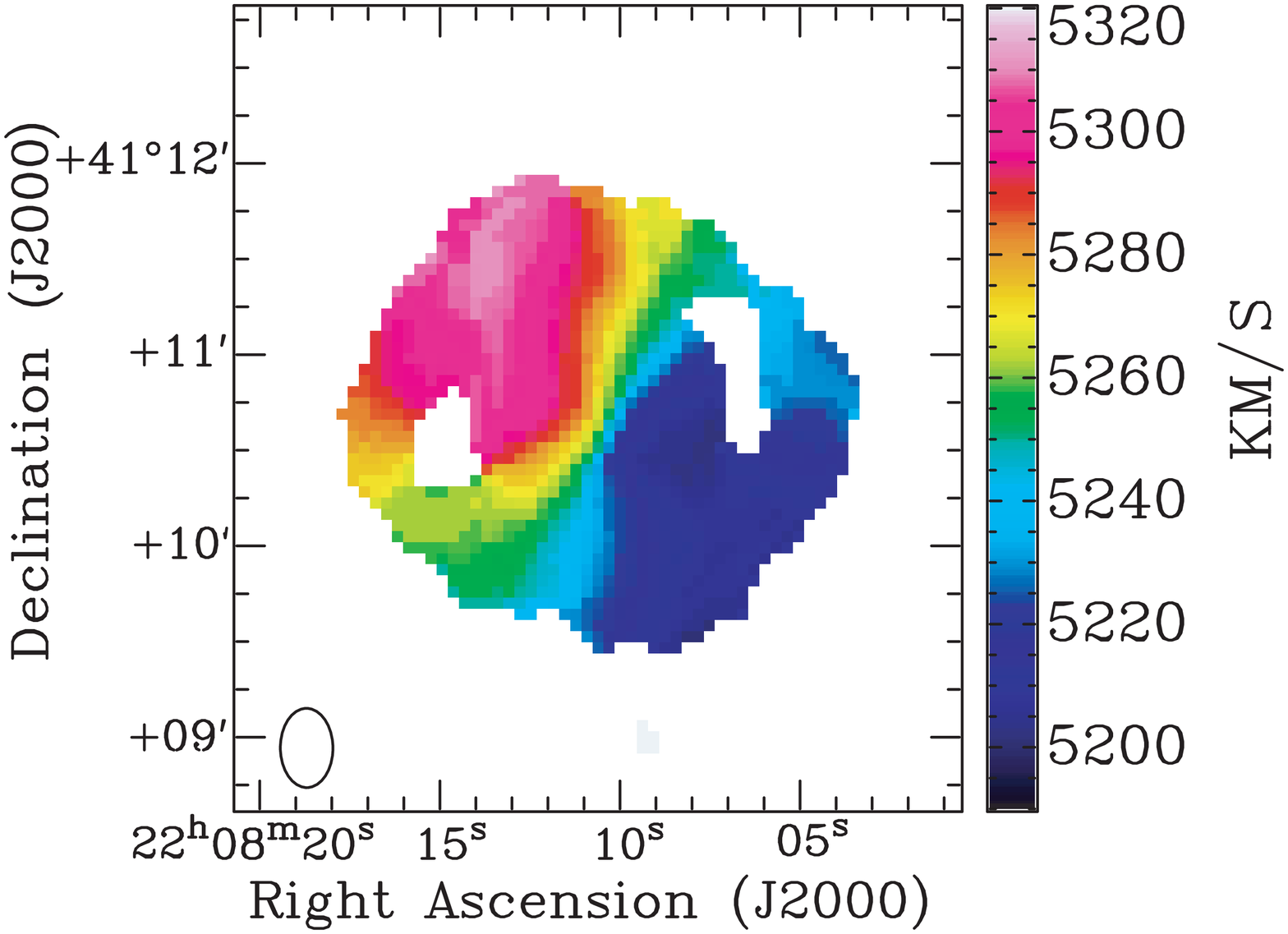}
\end{minipage}
\caption{Top: \ion{H}{I} total intensity map and R-band optical image of NGC~6824 with 
  overplotted \ion{H}{I} total intensity contours: 7.0, 26, 45, 104
 $\times 10^{19}$  ${\rm atoms \, cm}^{-2}$ (left and
 middle panels respectively) and  
first-moment map of NGC~6824 (right panel). 
Bottom: the same 
 for UGC~11919;  contours: 5.4, 20, 35, 80
 $\times 10^{19}$  ${\rm atoms \, cm}^{-2}$.}
 
\label{fig0}
\end{center}
\end{figure*}

Employing our clean masks, we generated total
intensity maps and first-moment maps as shown in
Fig. \ref{fig0}, which also shows the
optical images with overplotted \ion{H}{I} total-intensity maps. From
Fig. \ref{fig0} it is evident that UGC~11919 possesses
an extended \ion{H}{I} disk reaching beyond the optical
  body of the galaxy, while for NGC~6824 the size of the gaseous disk
  is comparable to that of the optical disk. 

\begin{table}[h!]
\small \caption{ \ion{H}{I} observational parameters and description of data cubes used in this study (see Sect. 2).  \label{tab2}}
  \begin{center}
\begin{tabular}{c  c c }
    \hline
\hline
&NGC~6824&UGC~11919\\
\hline
Observing date& 11.08.2011& 30.09.2011\\
Array configuration& Maxi-short&Maxi-short\\
On-source integration time (h)& 12&12\\
Scale (${\rm pix}/\arcsec$)& 3.5&3.5\\
Channel width $dv$ (${\rm km}\,{\rm s}^{-1}$)&8.24&8.24\\
Velocity resolution $FWHM_{\rm v}$ (${\rm km}\,{\rm s}^{-1}$)&16.5&16.5\\
Beam size ($\arcsec$)&$19.3$ $\times$ $16.5$&$24.86$ $\times$ $16.49$\\
Beam position angle ($\degr$) &-0.36&0.48\\
Velocity range (cz, ${\rm km}\,{\rm s}^{-1}$)&2400-4356 &4367-6348\\

 \hline
\end{tabular}
 \end{center}
\end{table}

\subsection{Photometric observations}
Photometric observations of NGC~6824 and UGC~11919 were made
with the 0.5-m Apache Point Observatory telescope  using
  the B, V, and R bands. The galaxies were observed
under photometric conditions in September
2011 and May 2012. The photometric data reduction 
was conducted following a standard procedure using the MIDAS software package. The images of
photometric standard stars from \citet{Landolt1992, Landolt2009}  were obtained
during the same nights and were used to
  calibrate the data. 
A bias and flat field correction
was applied to the data. The foreground stars were removed
from the galactic images and replaced by the mean fluxes of
surrounding regions. The integration
  times are specified in Table \ref{tab3}.

\begin{table}
\small \caption{Photometric observations: exposure times \label{tab3}}
  \begin{center}
\begin{tabular}{c  c c}
    \hline\hline
&NGC~6824&UGC~11919\\
\hline
Observing dates & 28.05.2012& 27.09.2011\\
&30.05.2012&\\
Exposure time (s) $B$&5400 &3000\\
Exposure time (s) $V$&5400 &1800\\
Exposure time (s) $R$&5400 &\\
\hline
\end{tabular}
 \end{center}
\end{table}
\section{Results}
We study two galaxies, NGC~6824 and UGC~11919, which, according to
preliminary studies, have unusually high (NGC~6824) and low (UGC~11919) $M/L$ ratios for their color indices. We use the photometrical data and \ion{H}{I} data cubes to prove or disprove the peculiarity of these objects.
\subsection{Photometric results}

\begin{figure*}
\centering
\includegraphics[width=7cm,keepaspectratio]{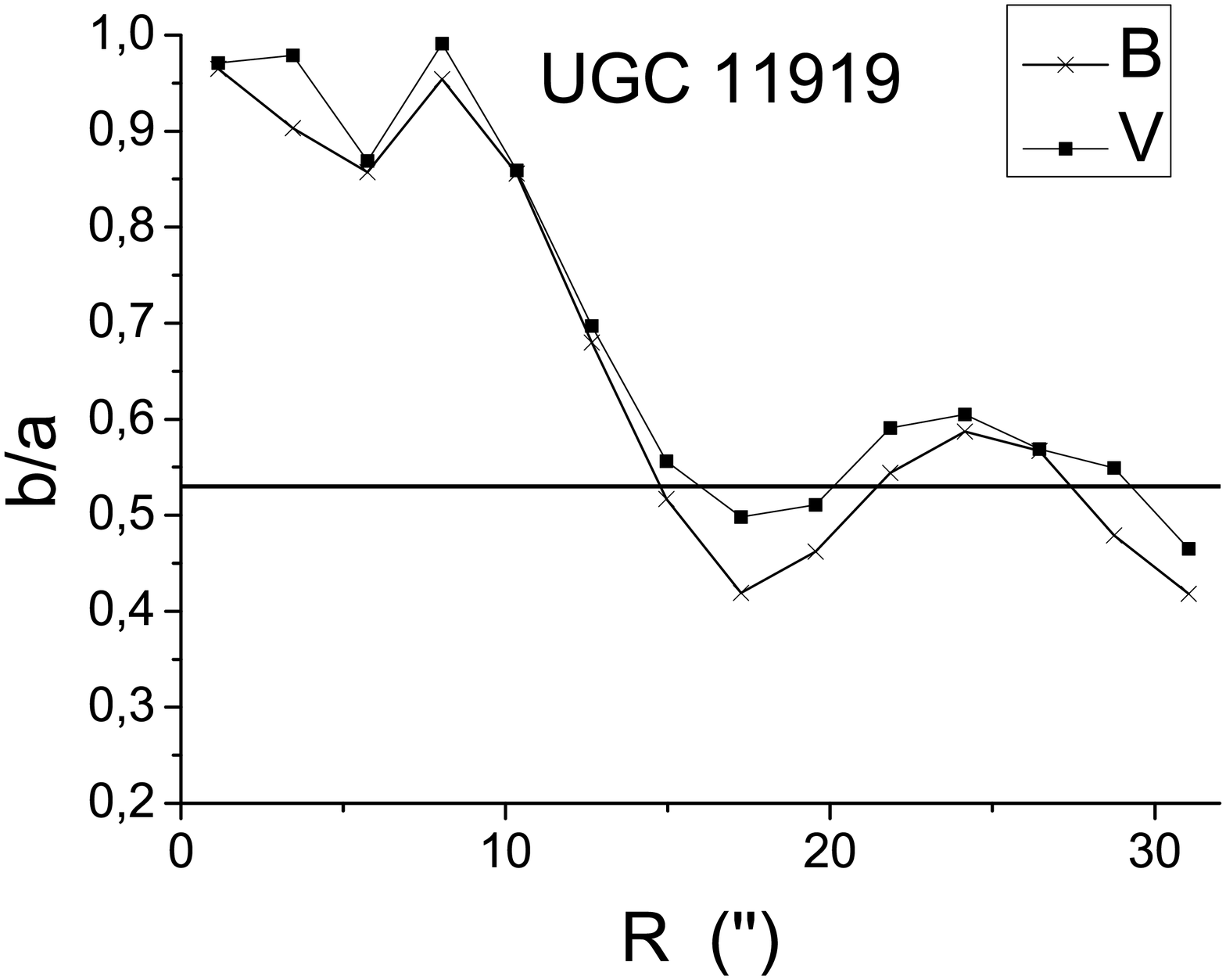}
\includegraphics[width=7cm,keepaspectratio]{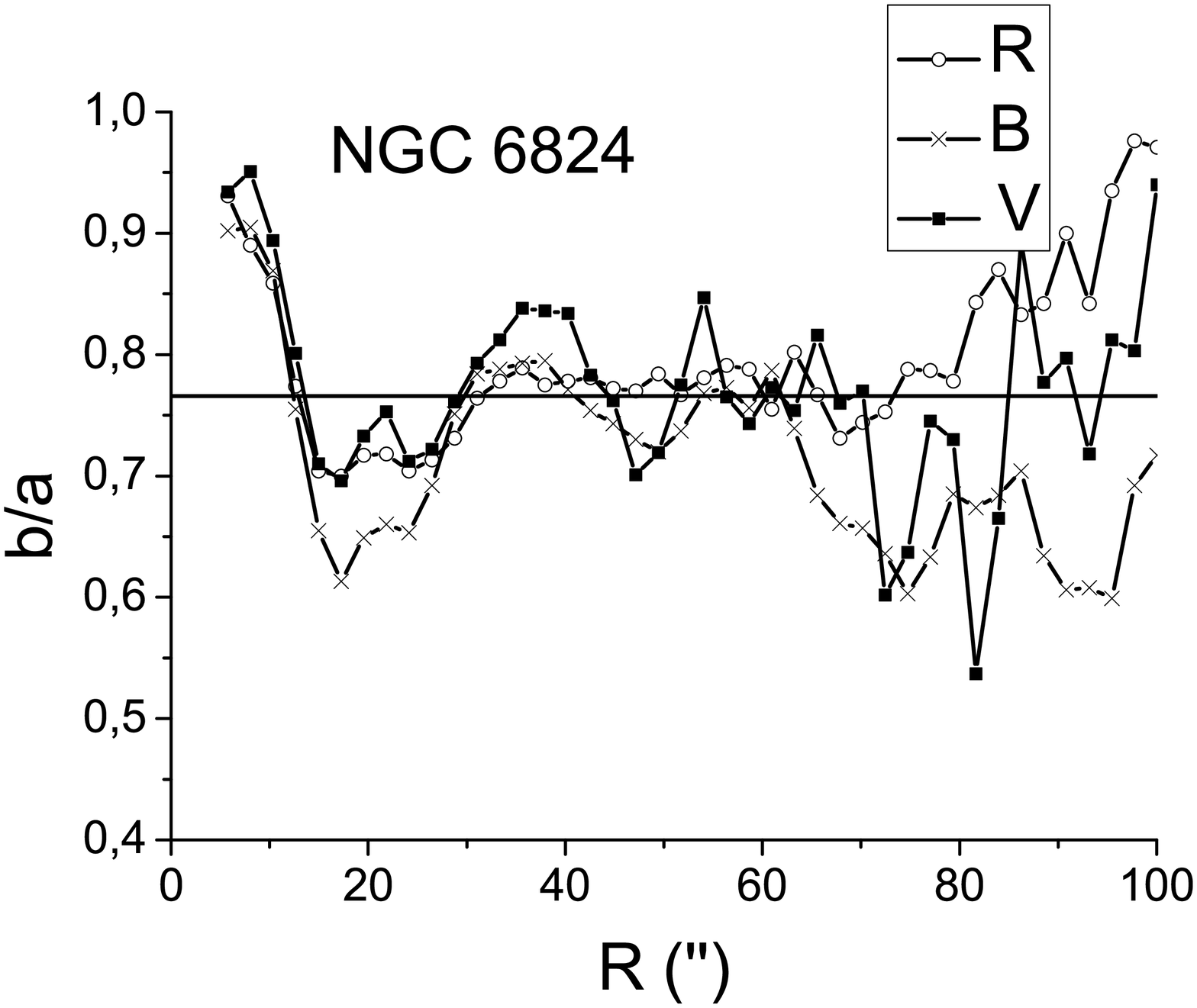}
\caption{ Radial profiles of isophotal flattening obtained in the B,V, and
  R bands. The horizontal lines correspond to the inclinations $i=58 \degr$   and $i=40 \degr$ for UGC~11919 and NGC~6824, respectively.} 
\label{fig1}
\end{figure*}

\begin{figure*}
\centering
\includegraphics[width=7cm,keepaspectratio]{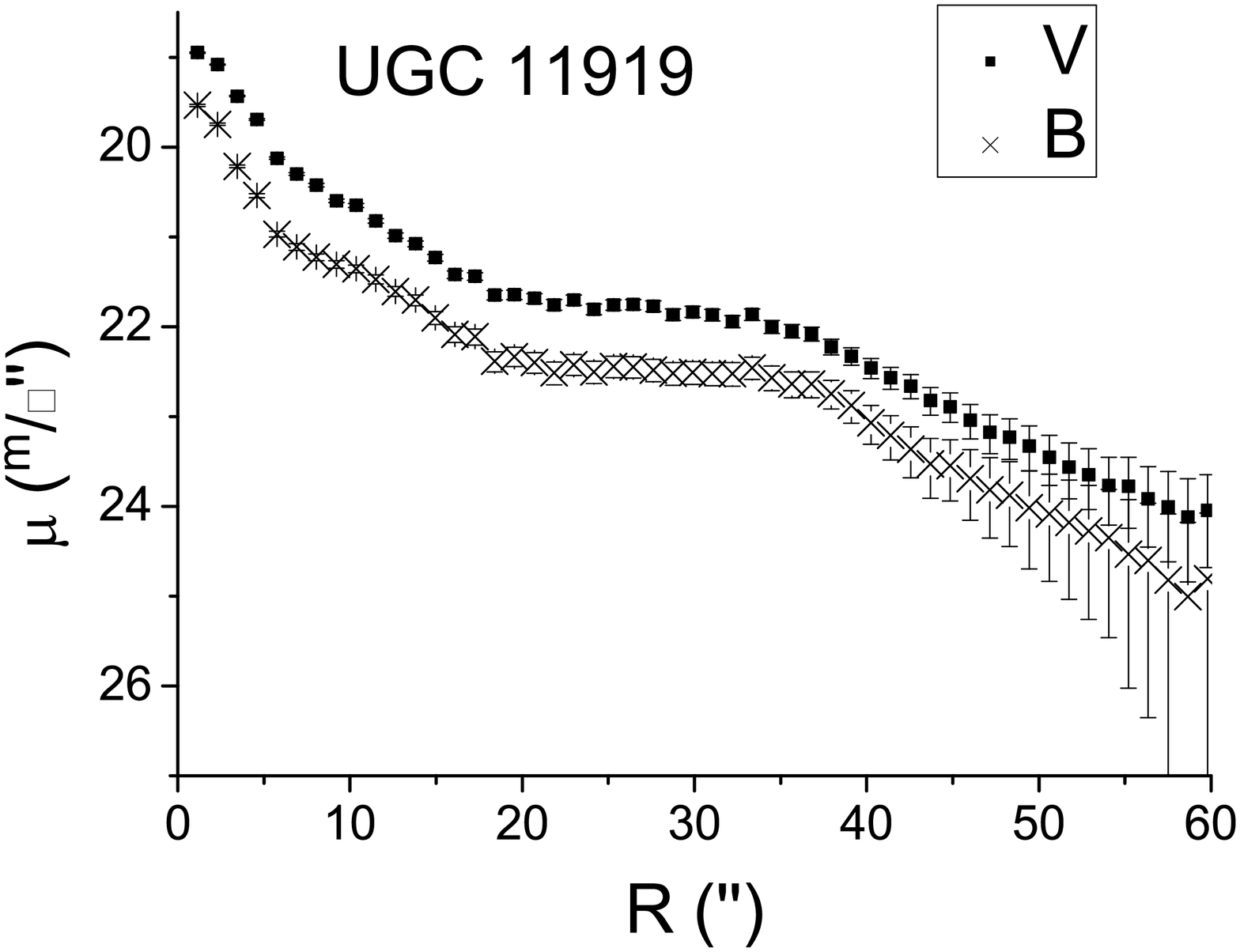}
\includegraphics[width=7cm,keepaspectratio]{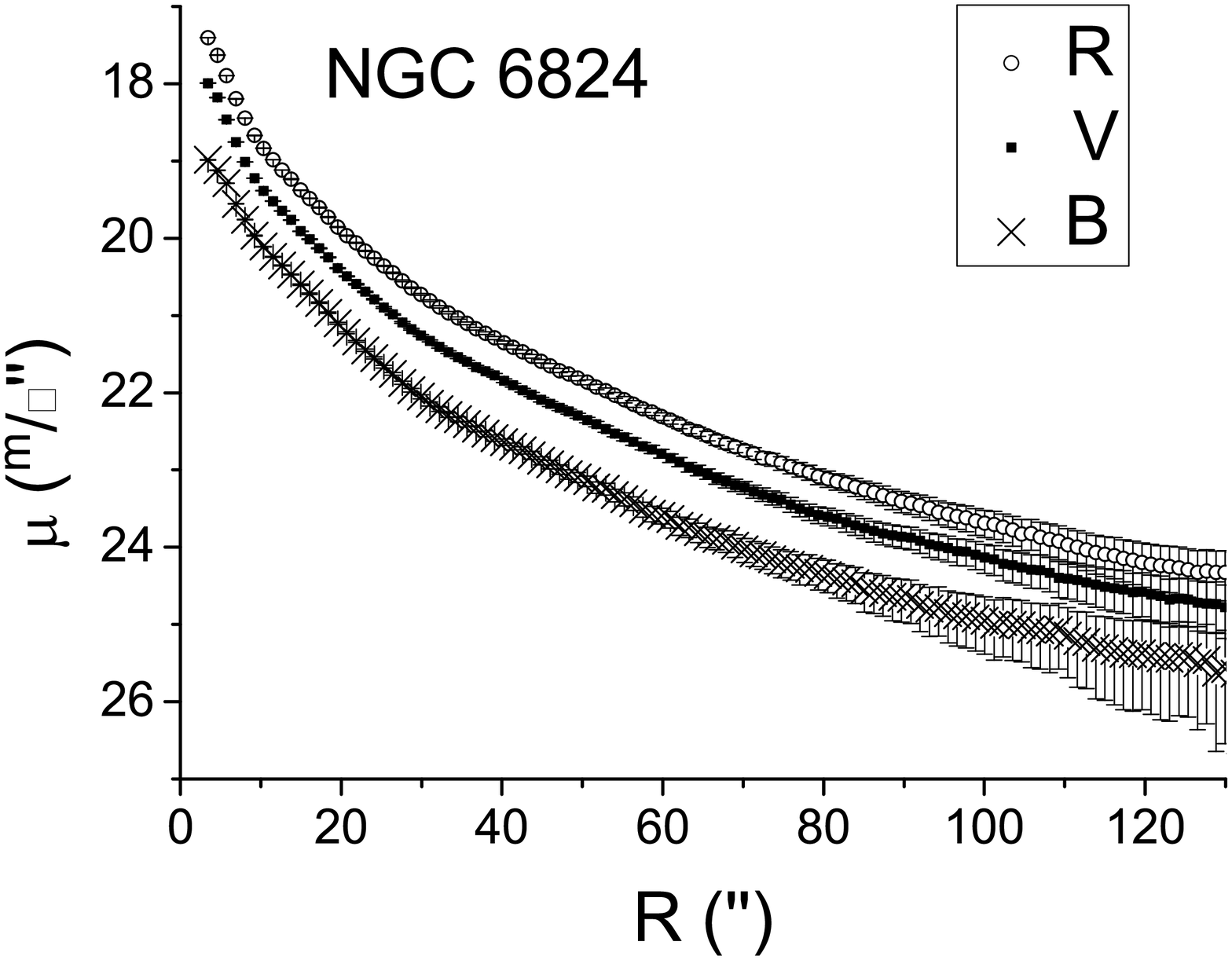}
\includegraphics[width=7cm,keepaspectratio]{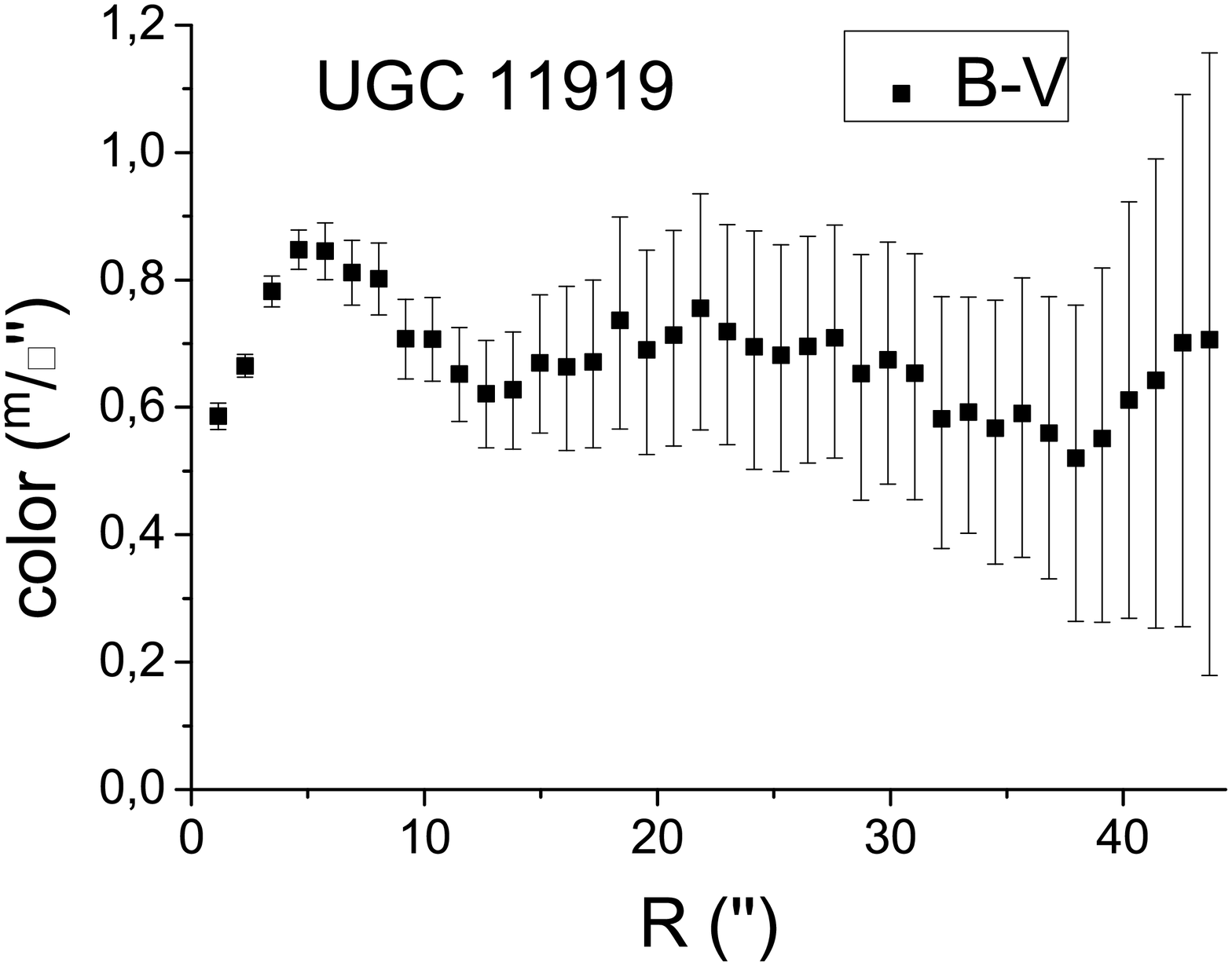}
\includegraphics[width=7cm,keepaspectratio]{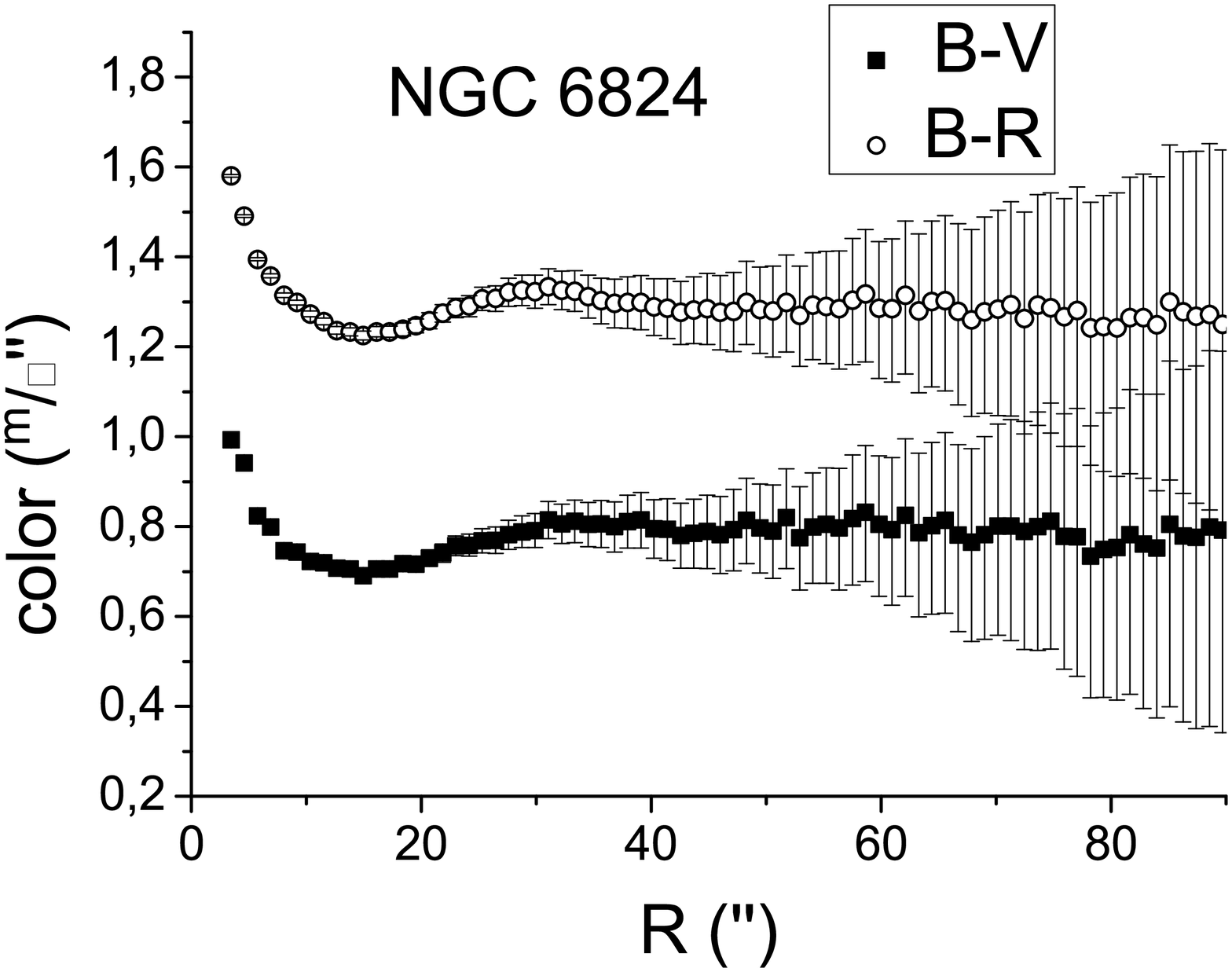}

\caption{Azimuthally averaged radial profiles of surface brightness
  (in B, V, and R bands) and color indices.}
\label{fig2}
\end{figure*}

Using the Midas software package we obtained the azimuthally averaged
profiles of isophotal flattening ($b/a$, Fig. \ref{fig1}) and radial
profiles of surface brightness and color indices for both
  galaxies (Fig. \ref{fig2}). The profiles 
are
 corrected for Galactic extinction according to \citet{Schlegel}, but are not corrected for internal extinction and disk orientation. 

From the isophotal flattening (Fig. \ref{fig1}) we derive an average
inclination $i=acos(b/a) = 40\degr \pm 13 \degr$  for NGC~6824 in the
radial range $30 \arcsec <R<70 \arcsec$, which is close to the value
obtained after masking the spiral arms. For UGC~11919 we derive $i= 58\degr  \pm 4 \degr$ for $R>15\arcsec$,
which changes substantially after masking the prominent spiral arms
in the innermost region. In that case we derive $i \geq 35 \degr$. The values  obtained from isophotal fitting without correcting for
  the appearance of spiral structure  are in good agreement with those given in Hyperleda (see Table \ref{tab1}).
\begin{figure*}
\centering
\includegraphics[width=5cm,keepaspectratio]{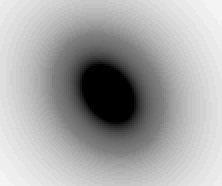}
\includegraphics[width=5cm,keepaspectratio]{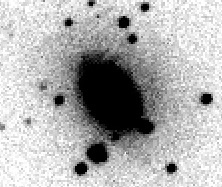}
\includegraphics[width=5cm,keepaspectratio]{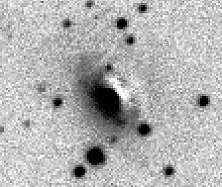}
\includegraphics[width=5cm,keepaspectratio]{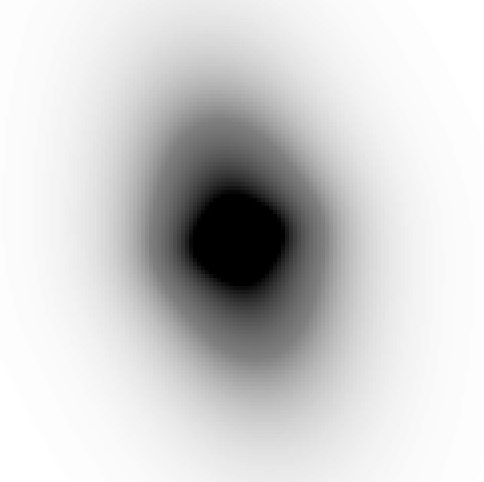}
\includegraphics[width=5cm,keepaspectratio]{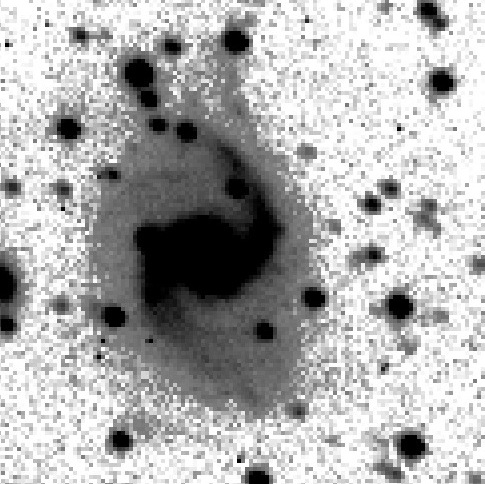}
\includegraphics[width=5cm,keepaspectratio]{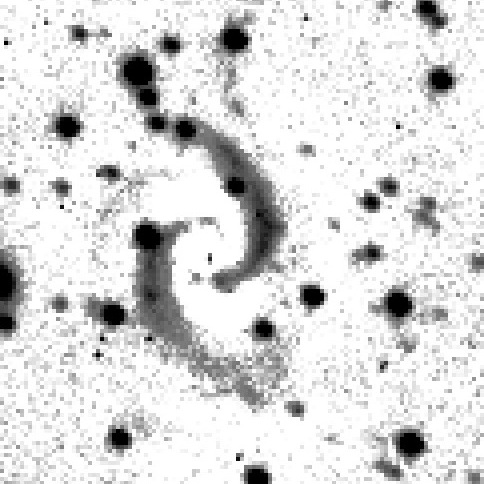}

\caption{Results of V-band bulge-disk decomposition for NGC~6824 (top) and UGC~11919 (bottom). From left to right: model, observed image, and  residual.}
\label{fig3}
\end{figure*}
Errors for the surface brightness and colors
(see Fig. \ref{fig2}) were calculated 
following \citet{Valder}
\begin{equation}\label{1}
\delta I = \sqrt{N_{\rm tot}+(\delta n_{\rm sky}A)^2},
\end{equation}
where $N_{tot}$ is the total flux in the aperture of size $A$ and
$\delta n_{sky}$ is the standard deviation of sky background flux
obtained in small apertures in the vicinity of a galaxy.

For UGC~11919 the radial profile of the $(B-V)_0$ color
index is in good
agreement with the total value of $(B-V)_0$ given in Hyperleda 
 (see Fig. \ref{fig2} and Table \ref{tab1}). This
means that the tentative peculiar value of $M/L$
of the galaxy is not a result of 
incorrect photometrical data given in Hyperleda.  For NGC~6824, we find 
the total color index  $(B-V)_0=0.74\pm 0.08$
  (corrected for extinction) which differs by $0^{\rm m}.09$ from the value given in Hyperleda (see Table \ref{tab1}). 

To construct dynamical models of the galaxies we  require
the structural parameters of their bulges and disks. For that purpose
we decomposed the images into two components: a disk with an exponential radial brightness
distribution  $\mu_d(R)=\mu_0+1.086R/R_d$ and
a Sersic bulge $\mu_b(R)=\mu_e+c_n((R/R_e)^{1/n}-1)$. We used
BUDDA (\citealt{Gadotti}, version 2.2), a
  dedicated standalone software, to perform this bulge-disk decomposition. The results are shown in Fig. \ref{fig3}. The model
and the observed V-band images are shown together with the
residuals. The images were generated using an
  identical scale and an identical contrast.

\begin{figure*}
\centering
\includegraphics[width=5cm,keepaspectratio]{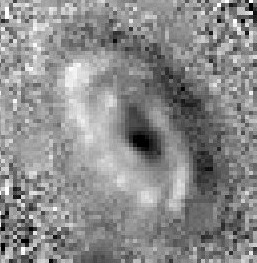}
\caption{$B-R$-color map of NGC~6824. The red dust lane (black) is the one most visible above the  galaxy center.}
\label{fig4}
\end{figure*}

The residual images (Fig. \ref{fig3}) reveal the trace of spiral arms
for UGC~11919 and a clear manifestation of an asymmetric
structure for NGC~6824. The contrast residuals for NGC~6824 are mainly
due to a bright spiral arm (dark) together with a
dust lane (light) which could not be taken into account in the
symmetrical model created by BUDDA. The blue spiral arm is also seen
in the $B-R$ color map of the galaxy (see
Fig. \ref{fig4}). A red dust lane
is also recognizable to the north of the center of NGC~6824  at the top
of the galaxy image. 

Neglecting the obvious asymmetric features, we derive a
deviation of the model from the observed surface brightness profile
which does not exceed $0^m.15 /\Box\arcsec$. 
With the aid
of BUDDA we are 
able to derive structural models with sufficient quality, despite neglecting local asymmetric features. In Table \ref{tab4} we list the obtained structural parameters of 
disk and bulge for the galaxies.

\begin{table}
  \begin{center}
\small \caption{Structural parameters of observed galaxies.
(1) Name; (2) exponential disk radial scalelength  $R_d$; (3) central surface brightness of the disk;
(4) effective radius of the bulge;
(5) effective surface brightness of the bulge;
(6) Sersic index of the bulge; (7) observing band. \label{tab4}}
\begin{tabular}{c c c c c c c}
    \hline  \hline
{\small Galaxy}&$R_d$ &$\mu_0$ &$R_{e}$ &$\mu_{e}$ &$n$&Filter\\
&({\small$\arcsec$})&({\small$^m/\Box\arcsec$})&({\small $\arcsec$})&({\small $^m/\Box\arcsec$})&&\\
(1) & (2) & (3) & (4)&(5)&(6)&(7)\\

\hline
{\small NGC~6824}&27.7&21.3&10.3&20.4&1.1&$B$\\
&27.1&20.6&10.5&19.8&1.1&$V$\\
&25.8&19.8&8.9&18.9&1.3&$R$\\
\hline
{\small UGC~11919}&20.6&21.6&9.6&22.3&2.3&$B$\\
&19.4&20.7&8.7&21.1&2.3&$V$\\
\hline

\end{tabular}
 \end{center}
\end{table}
\begin{figure*}
\includegraphics[ trim= 0 20 0 20,clip,height=5.8cm,keepaspectratio]{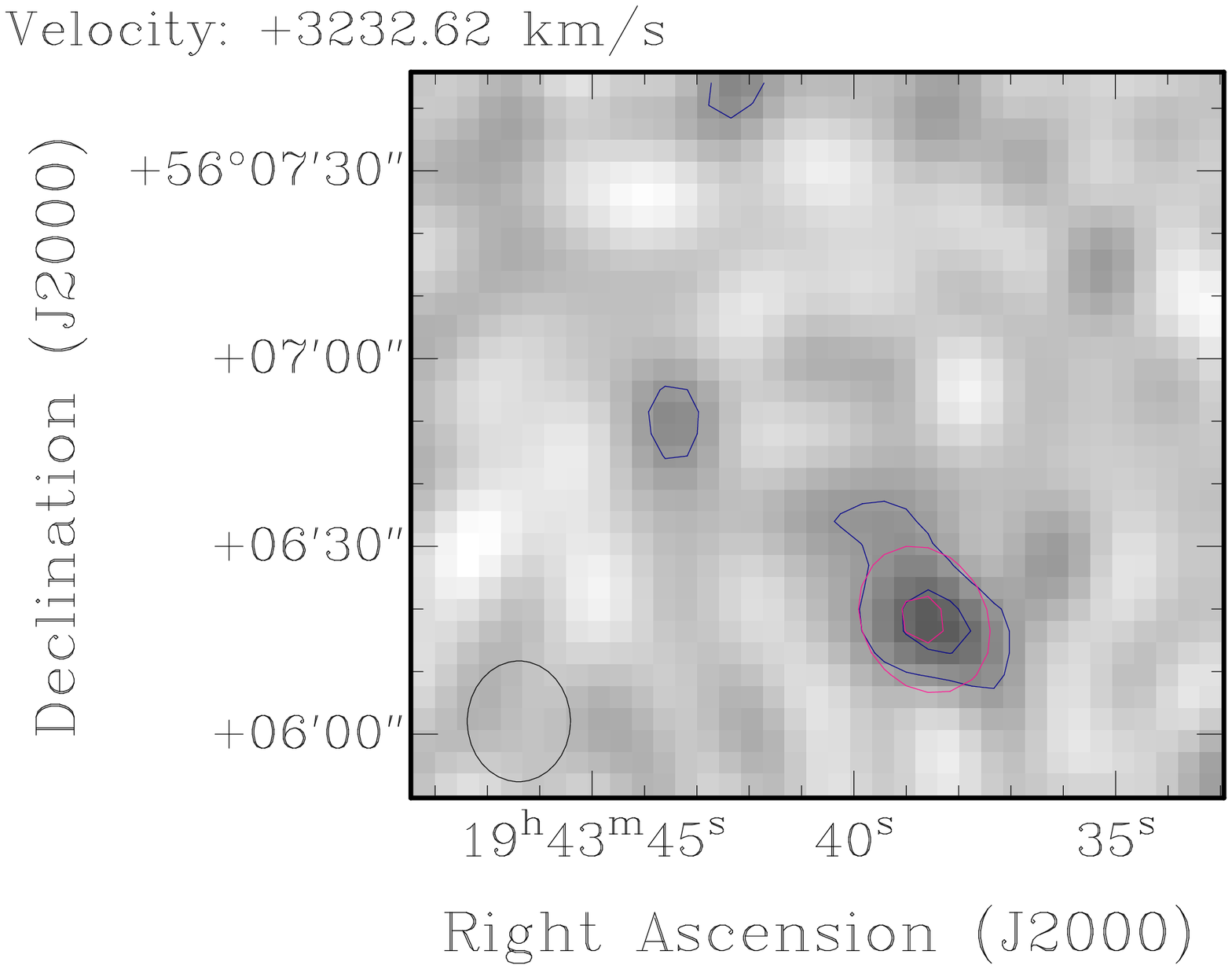}
\includegraphics[ trim= 0 20 0 20,clip,height=5.8cm,keepaspectratio]{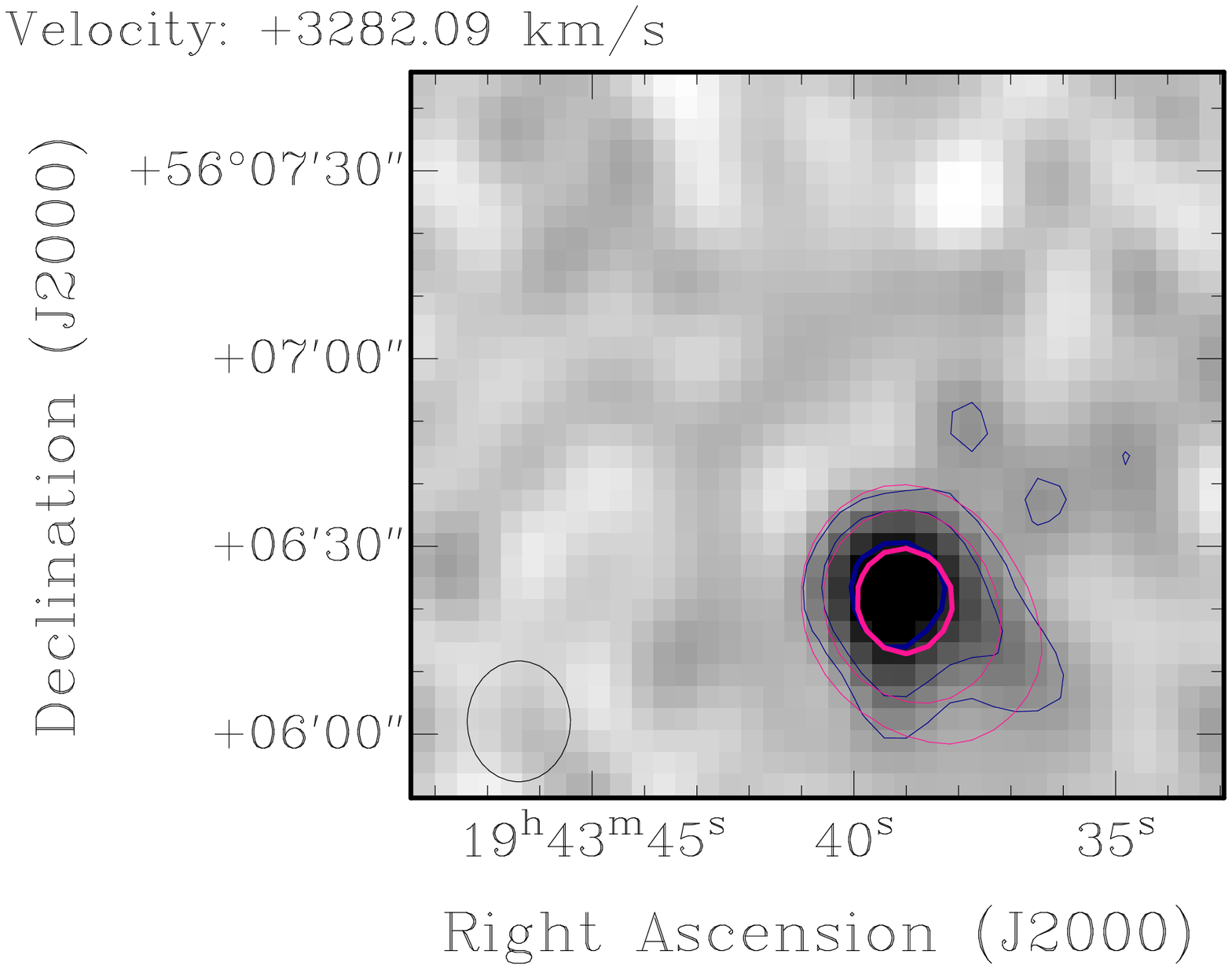}
\includegraphics[ trim= 0 20 0 20,clip,height=5.8cm,keepaspectratio]{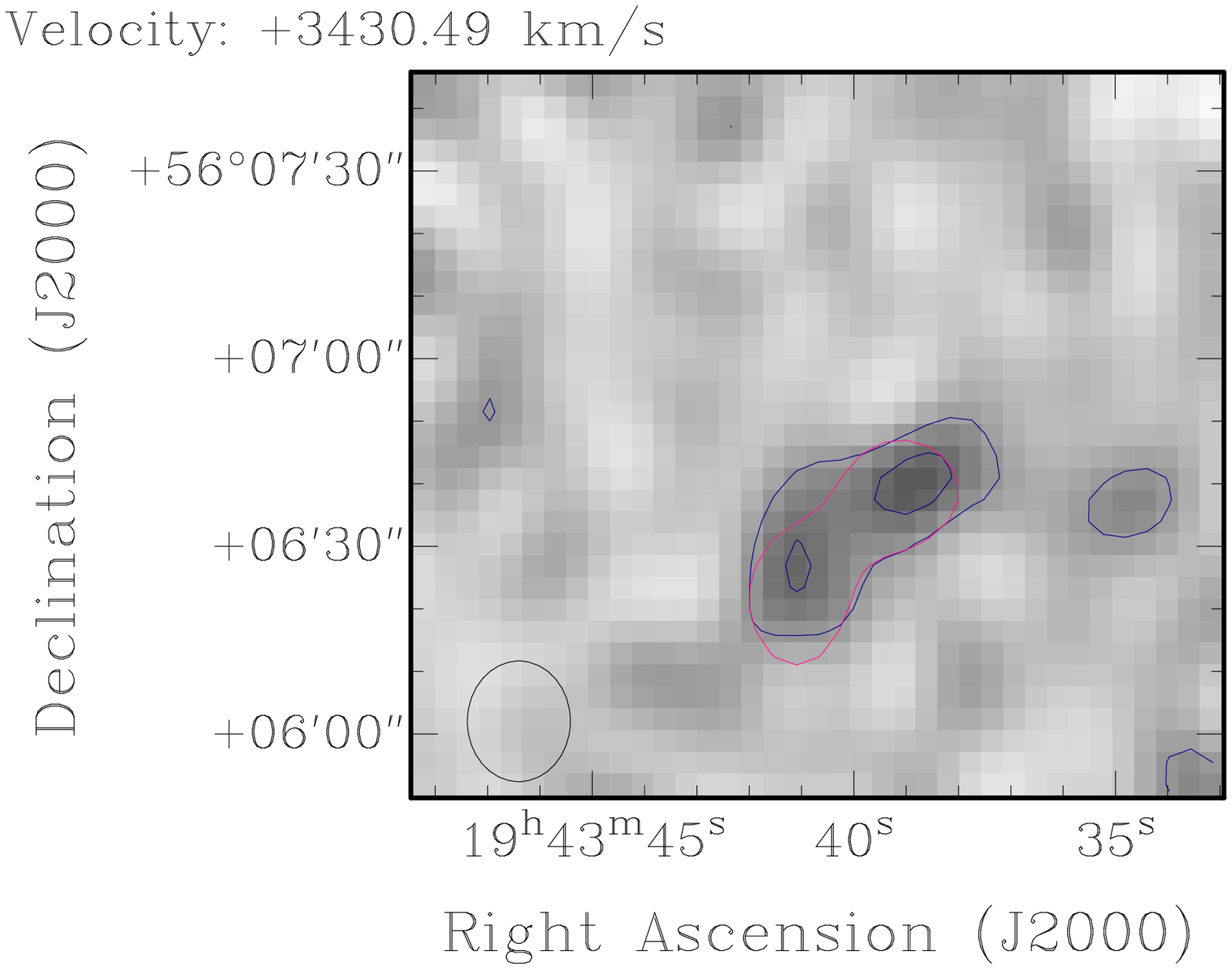}
\includegraphics[ trim= 0 20 0 20,clip,height=5.8cm,keepaspectratio]{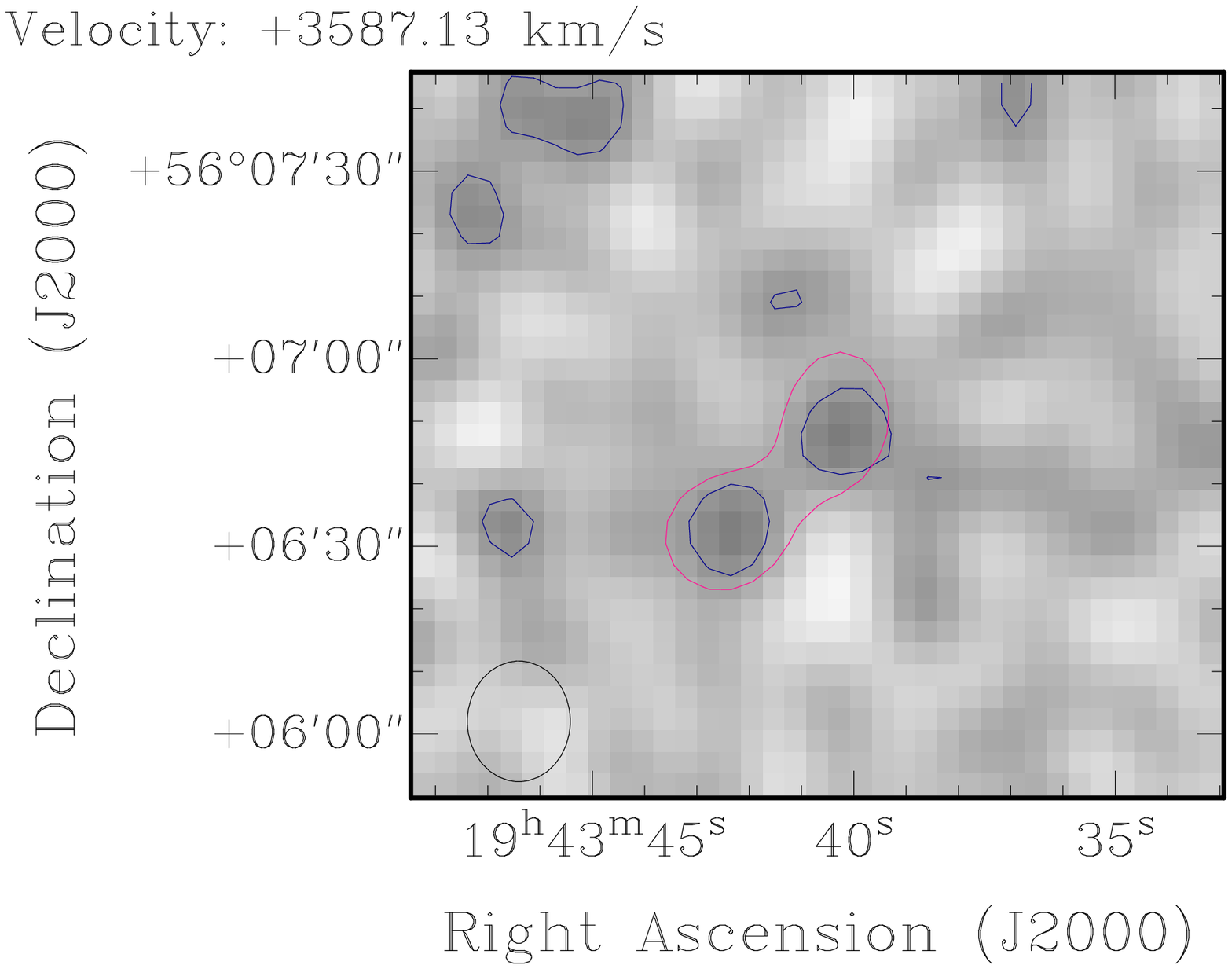}
\includegraphics[ trim= 0 20 0 20,clip,height=5.8cm,keepaspectratio]{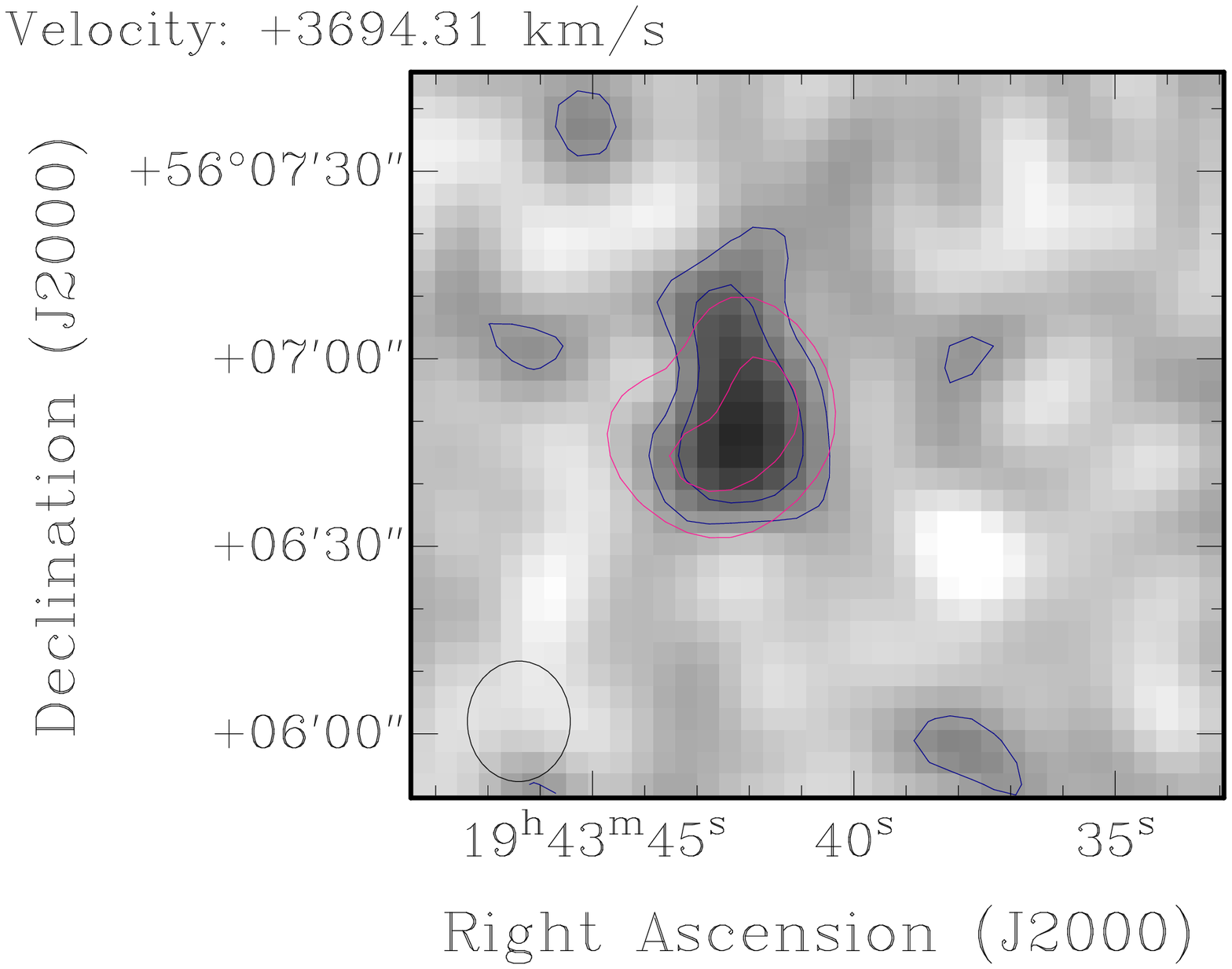}
\includegraphics[ trim= 0 20 0 20,clip,height=5.8cm,keepaspectratio]{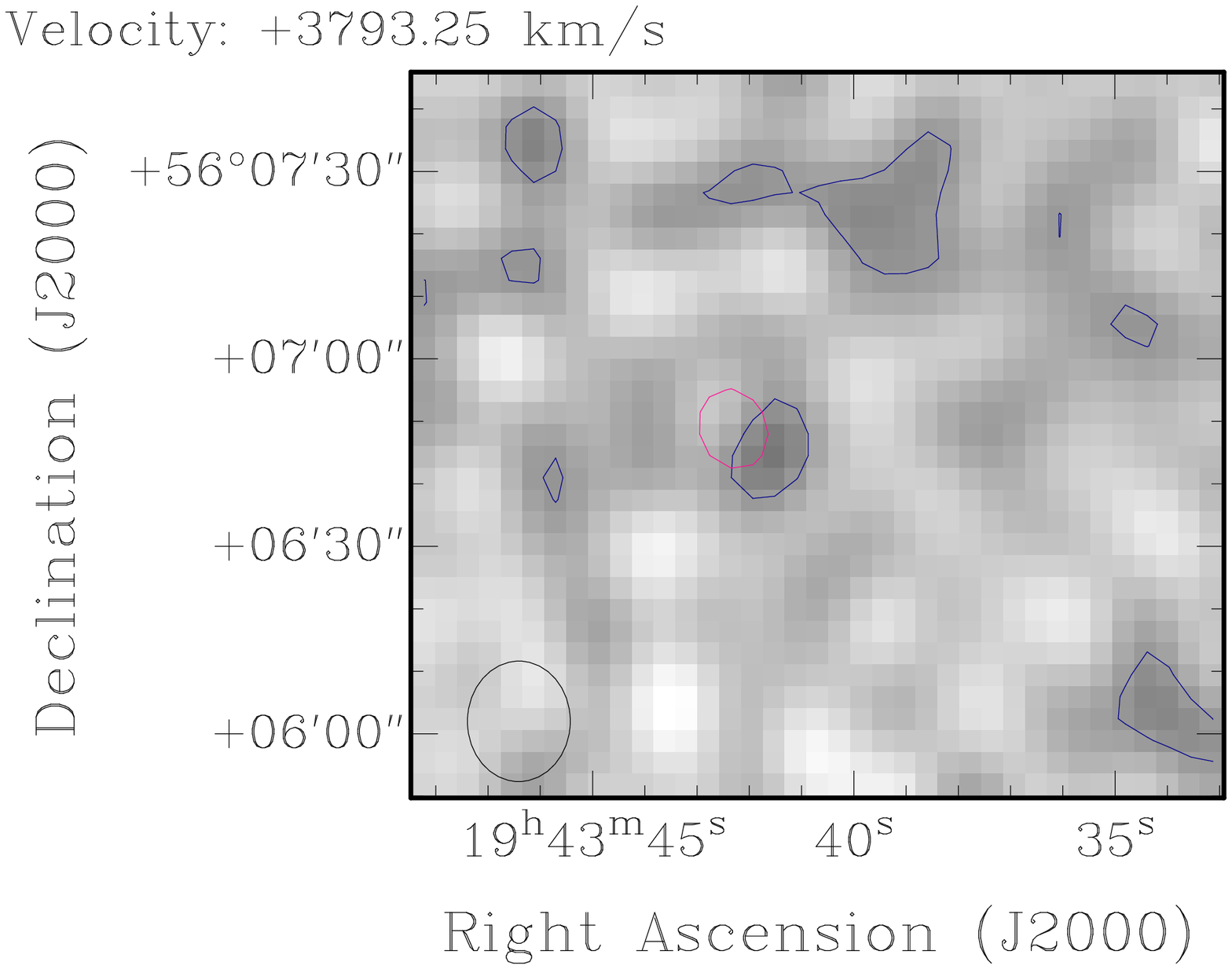}
\caption{\ion{H}{I} data cubes of NGC~6824. Blue and pink (black and grey in printed version)
  contours represent the 0.75 1.5, and 3 mJy/beam levels of the observed
  and the model data cube.}
\label{fig5}
\end{figure*}

\begin{figure*}
\centering
\includegraphics[height=6cm,keepaspectratio]{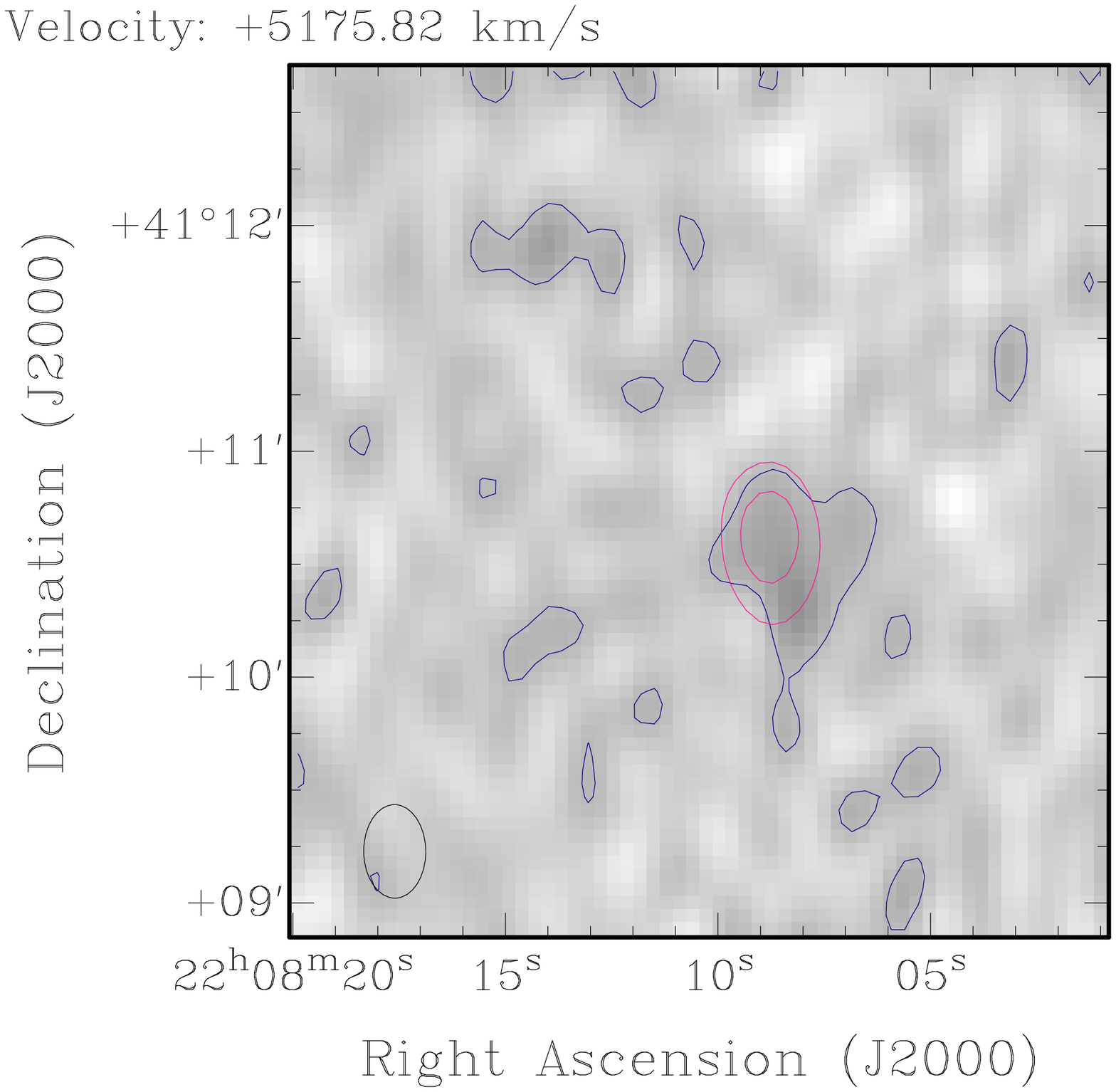}
\includegraphics[height=6cm,keepaspectratio]{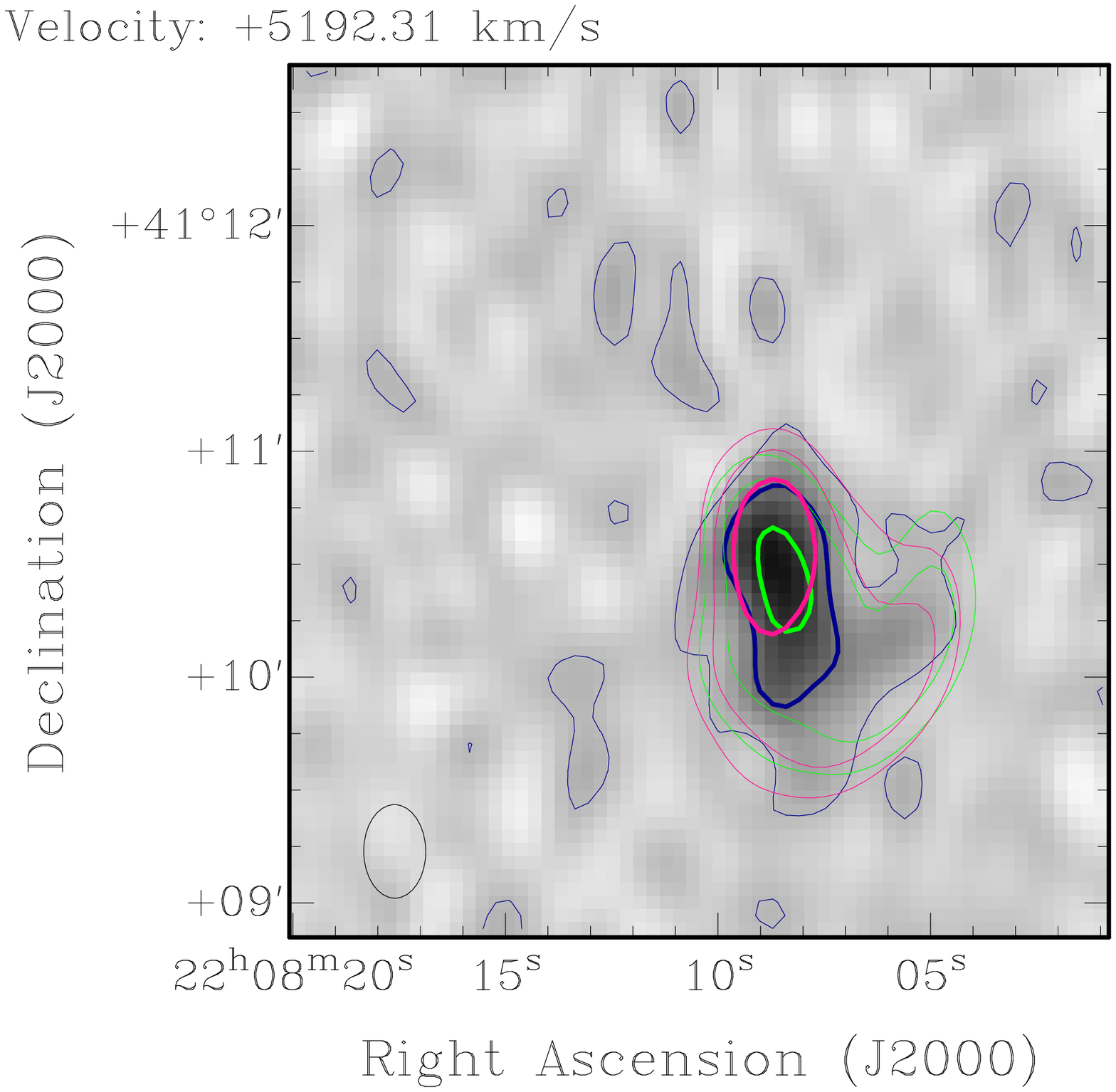}
\includegraphics[height=6cm,keepaspectratio]{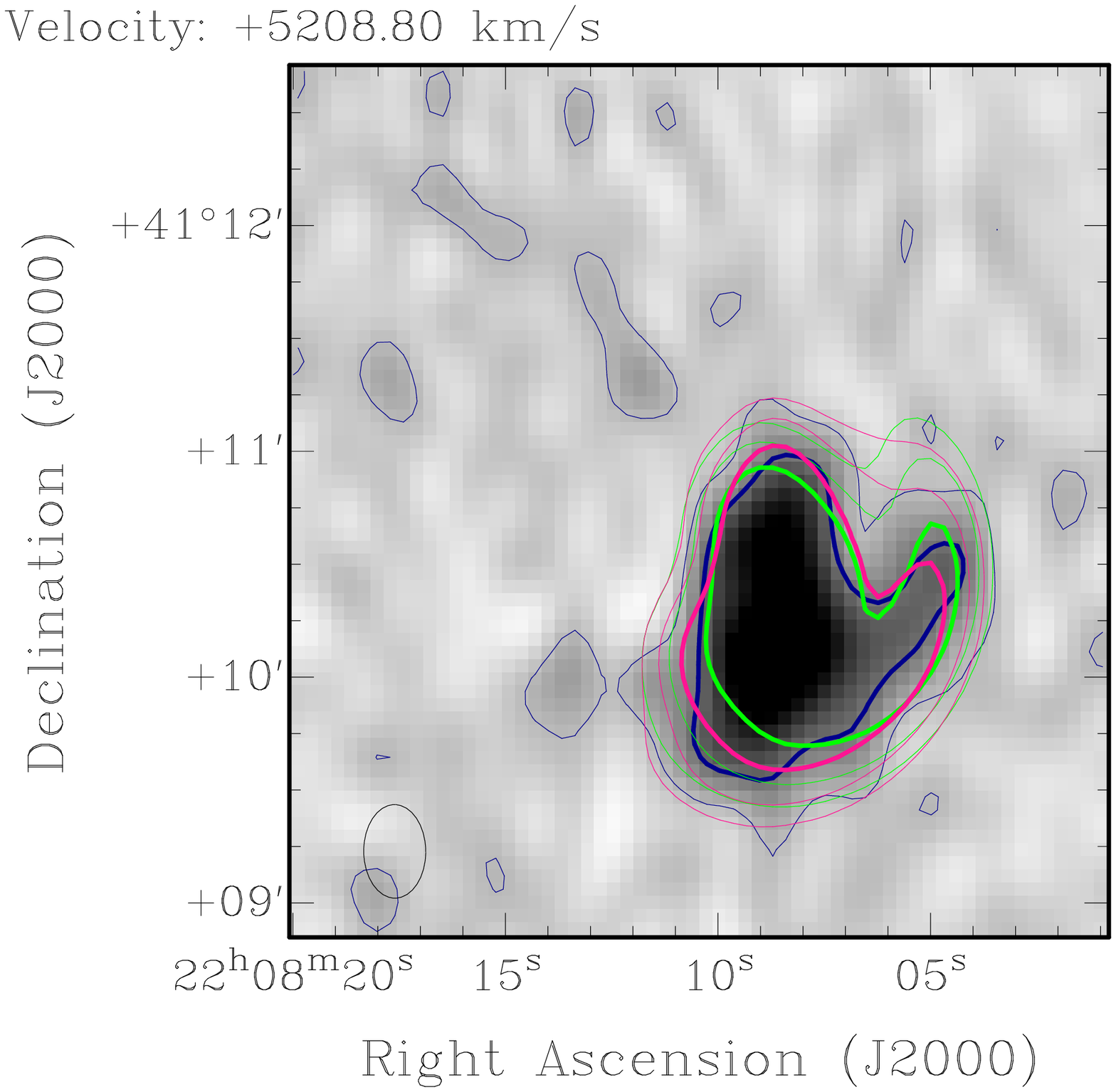}
\includegraphics[height=6cm,keepaspectratio]{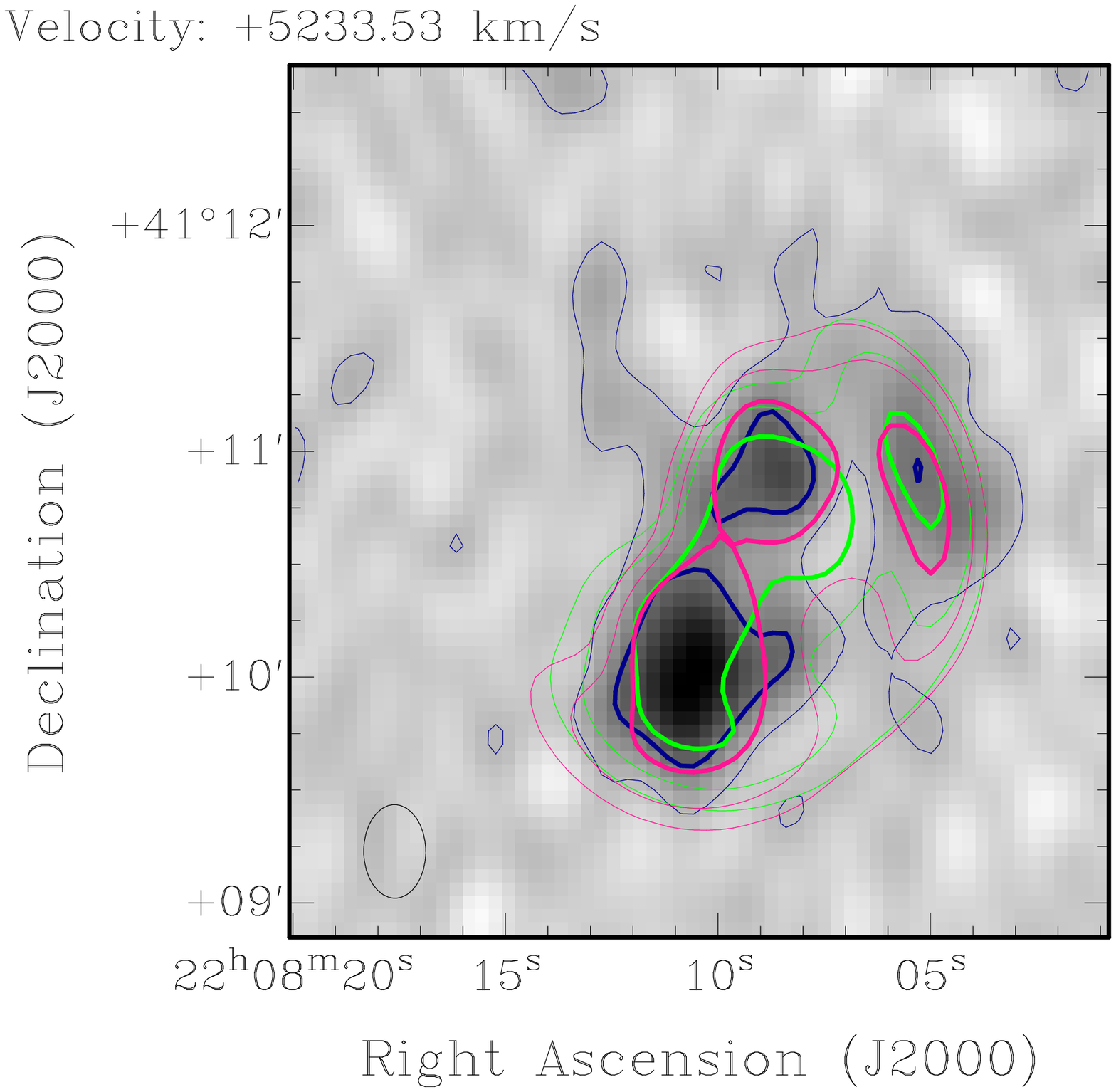}
\includegraphics[height=6cm,keepaspectratio]{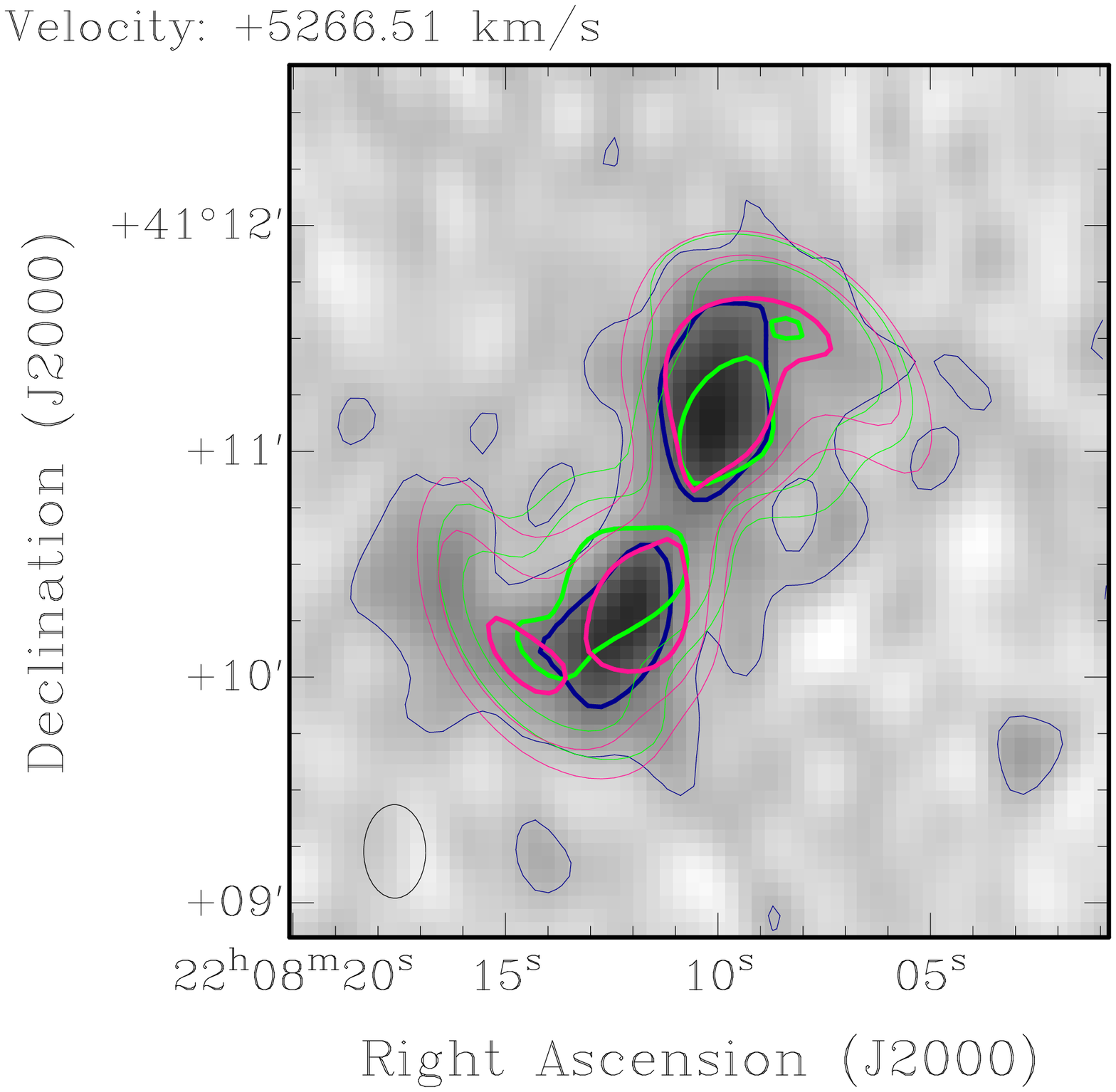}
\includegraphics[height=6cm,keepaspectratio]{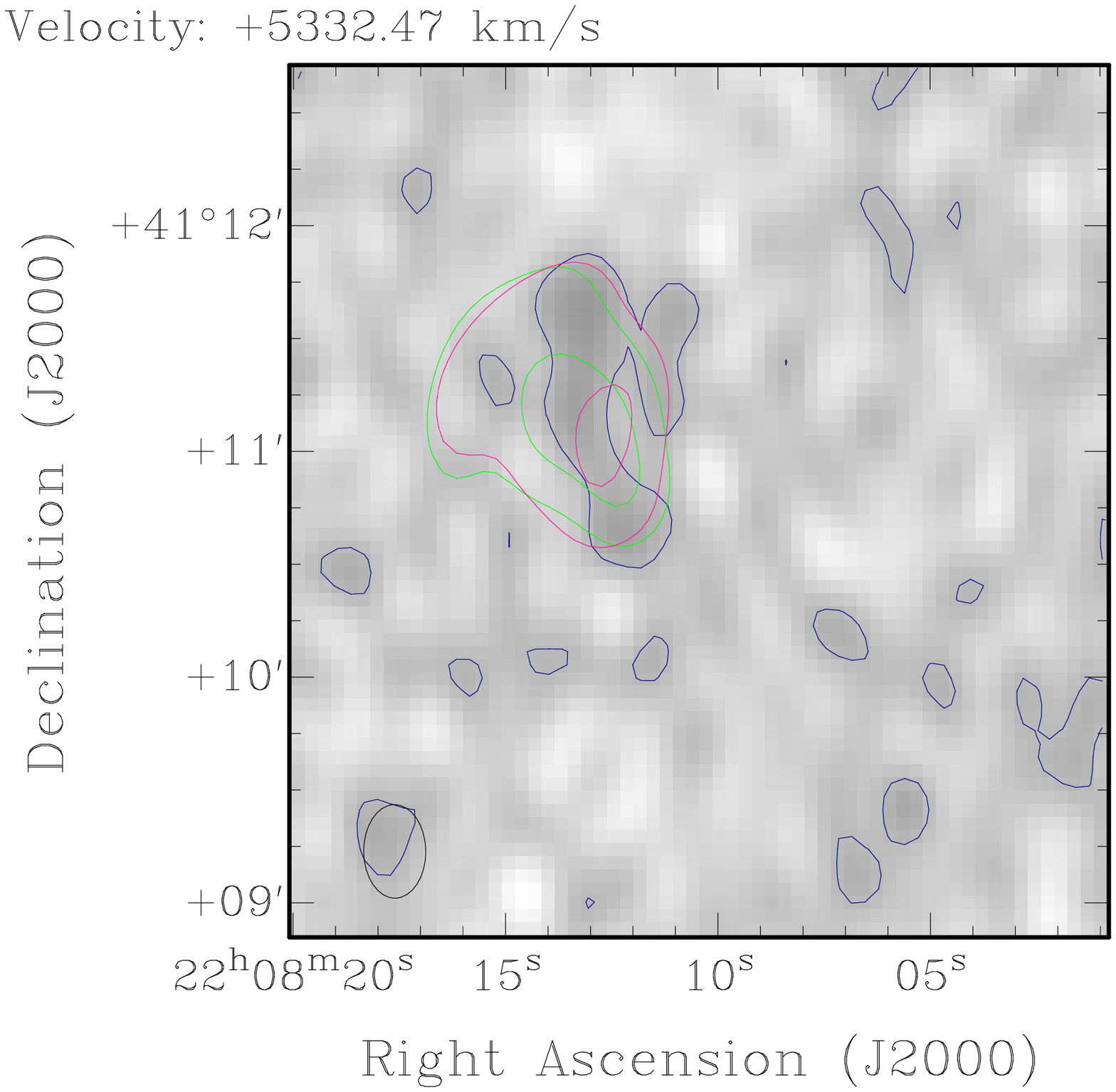}
\caption{ Selected images of \ion{H}{I} data cubes of UGC~11919. Blue, pink, and green (black, grey, and light grey in printed version) contours represent the 0.75, 1.5, and 3 mJy/beam levels of the observed, bisymmetric, and warped model data cubes, respectively.} 
\label{fig6}
\end{figure*}

\begin{figure*}
\centering
\includegraphics[width=7.cm,keepaspectratio]{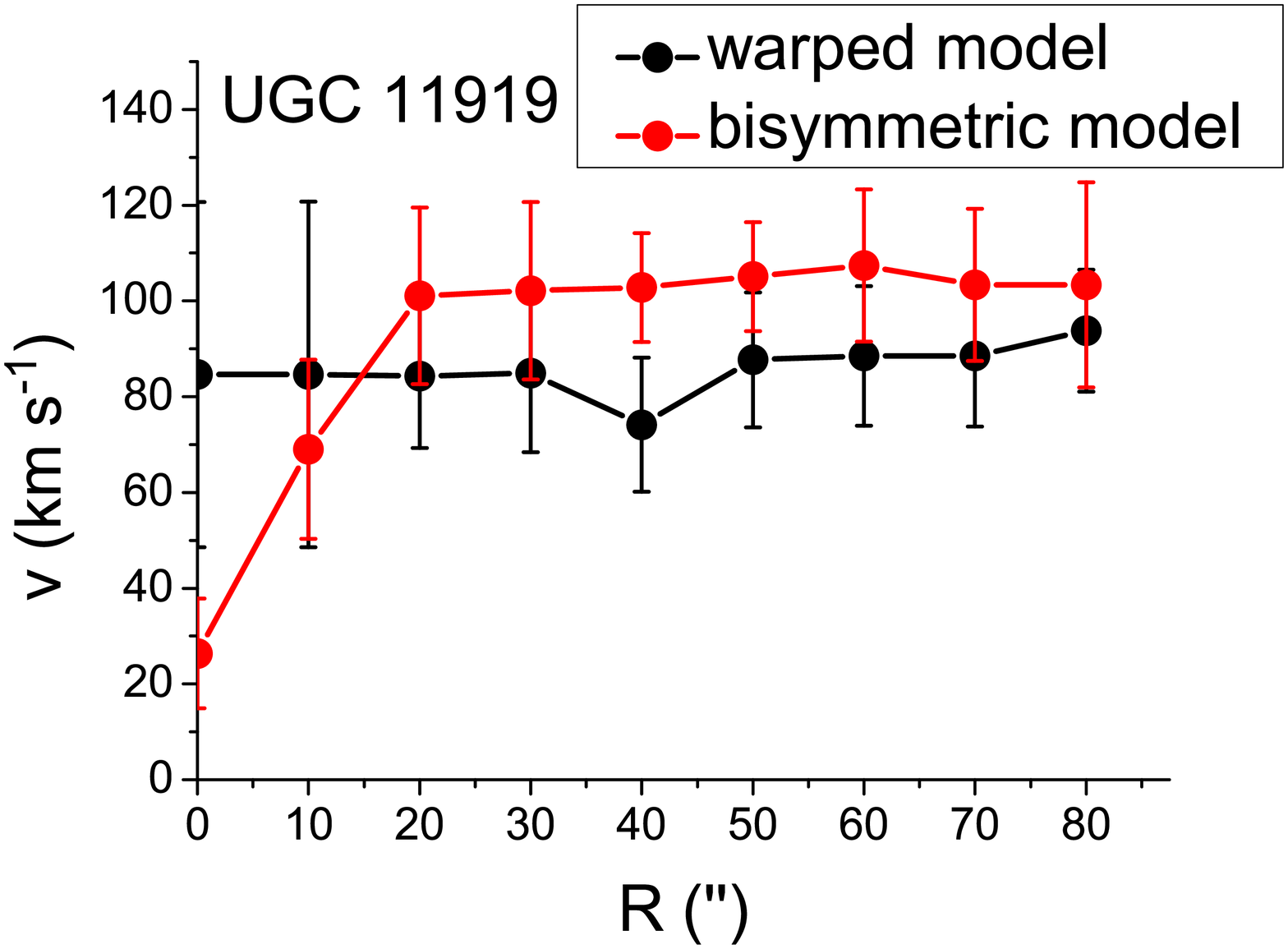}
\includegraphics[width=7.cm,keepaspectratio]{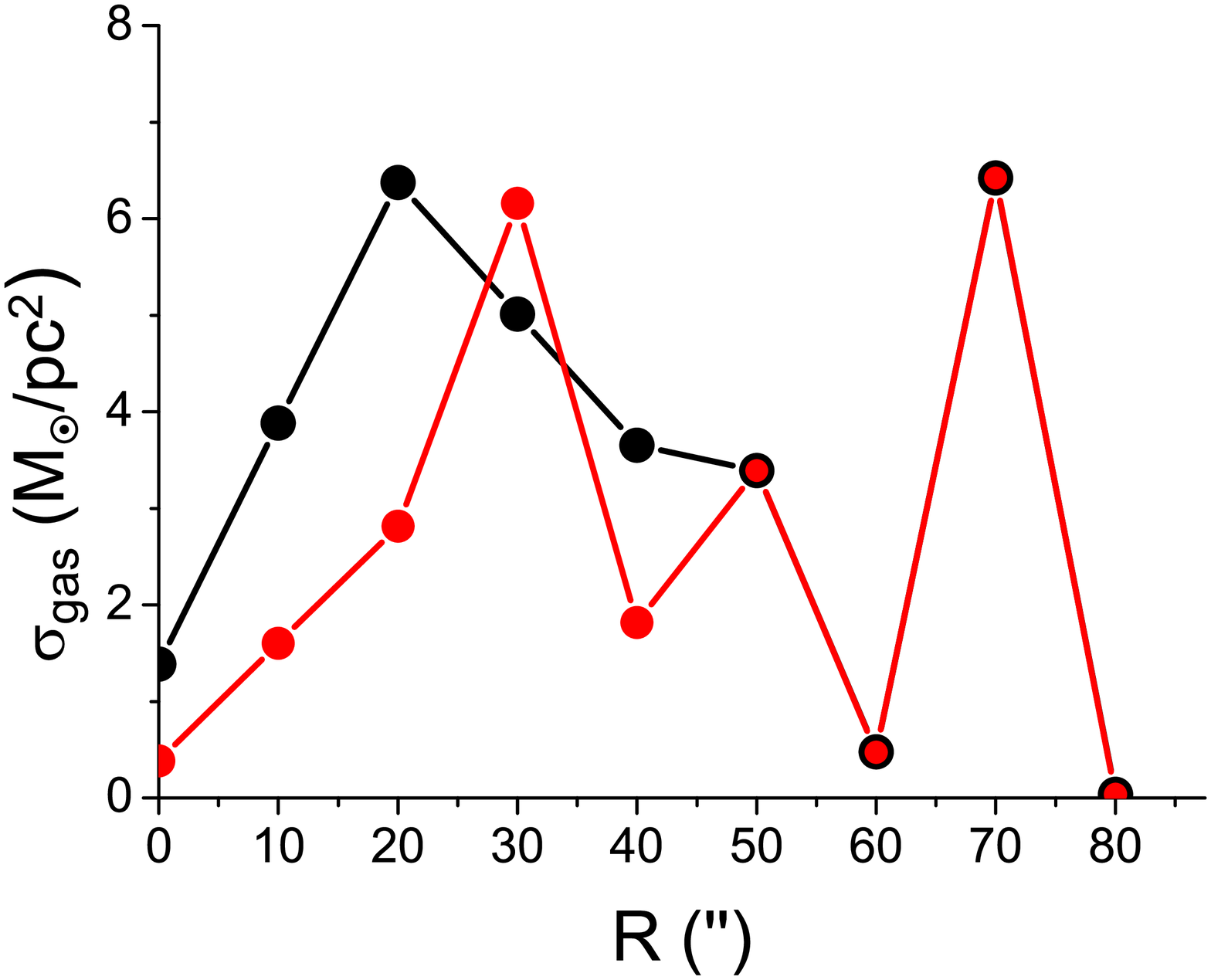}
\includegraphics[width=7cm,keepaspectratio]{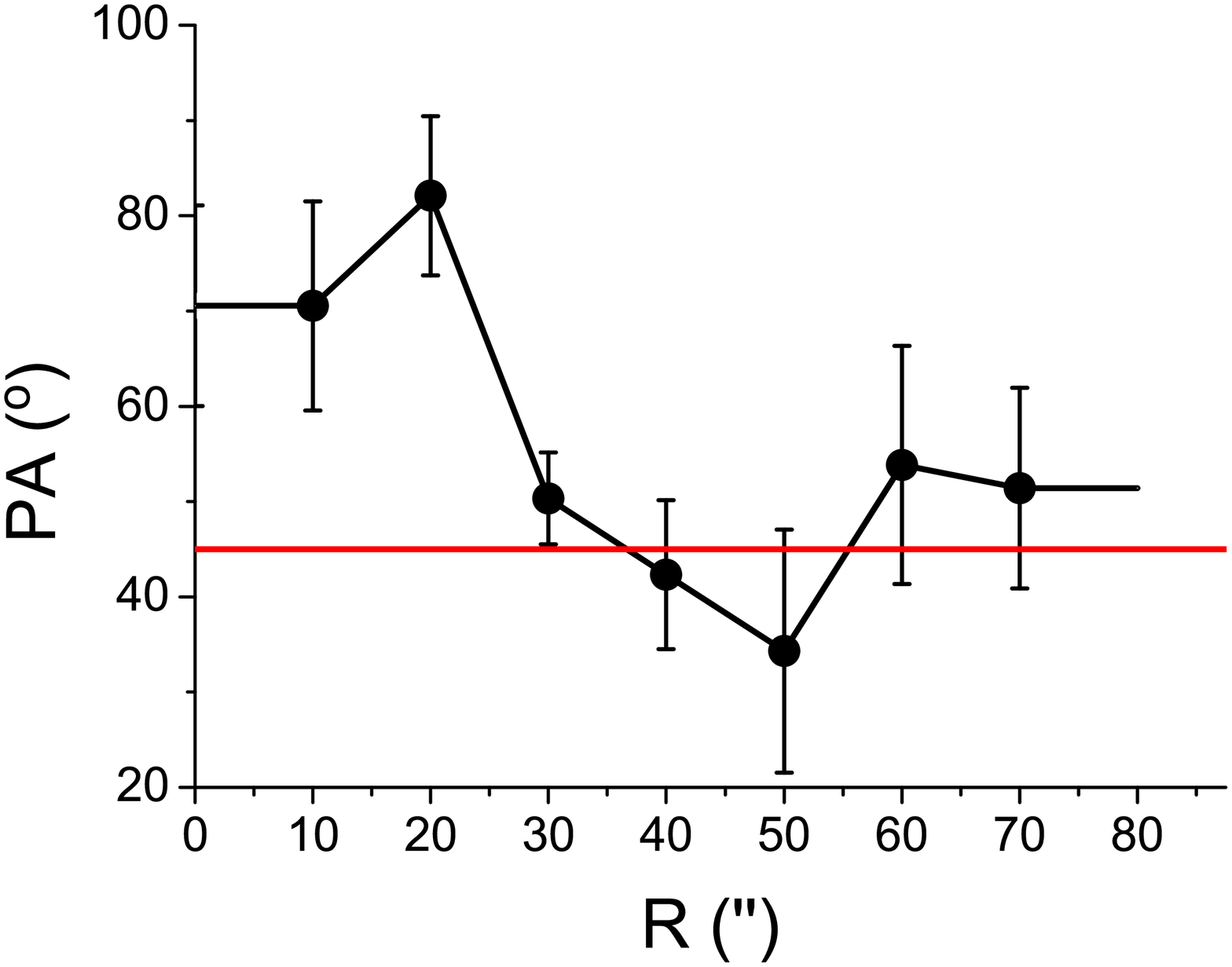} 
\includegraphics[width=7.cm,keepaspectratio]{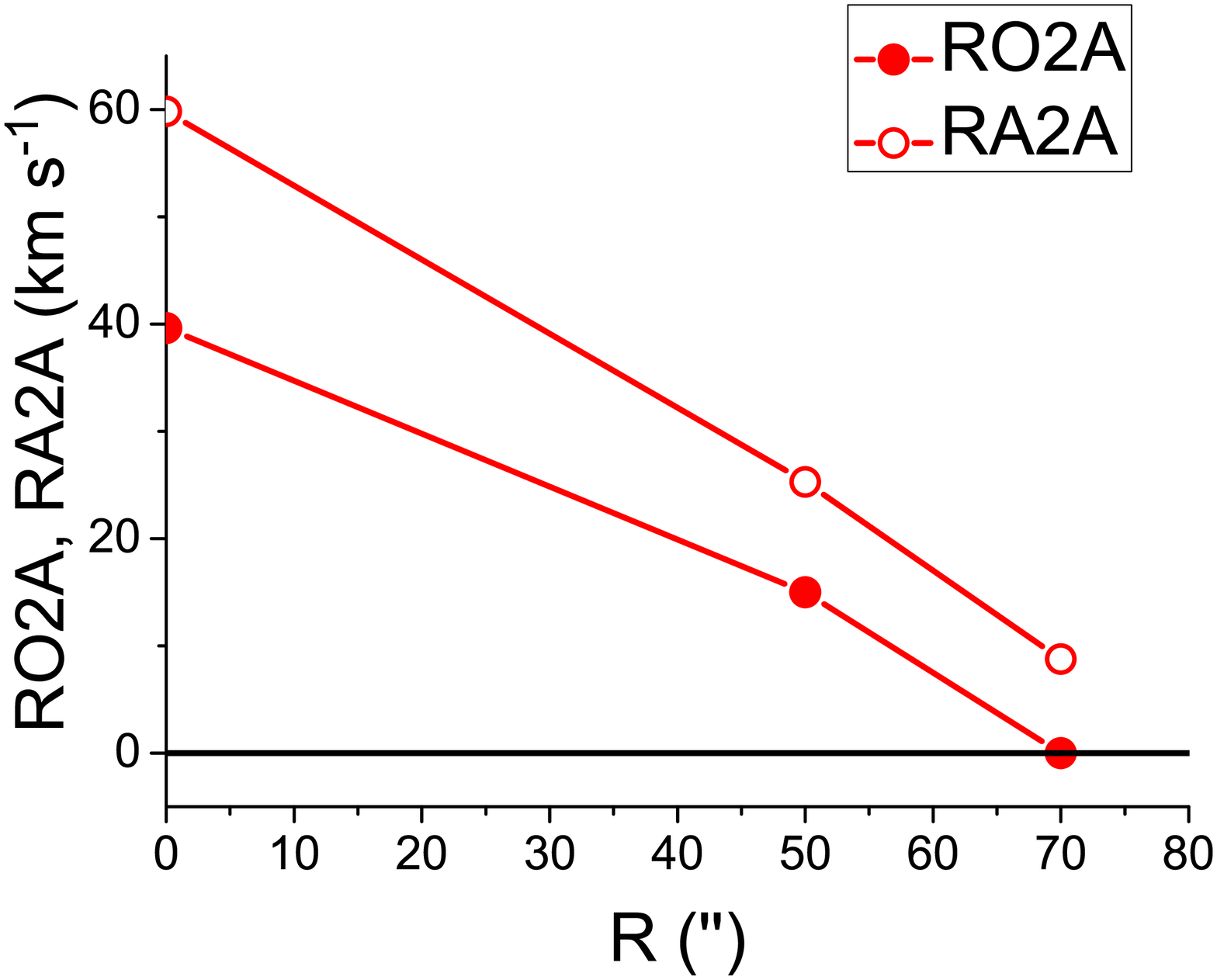}
\includegraphics[width=7cm,keepaspectratio]{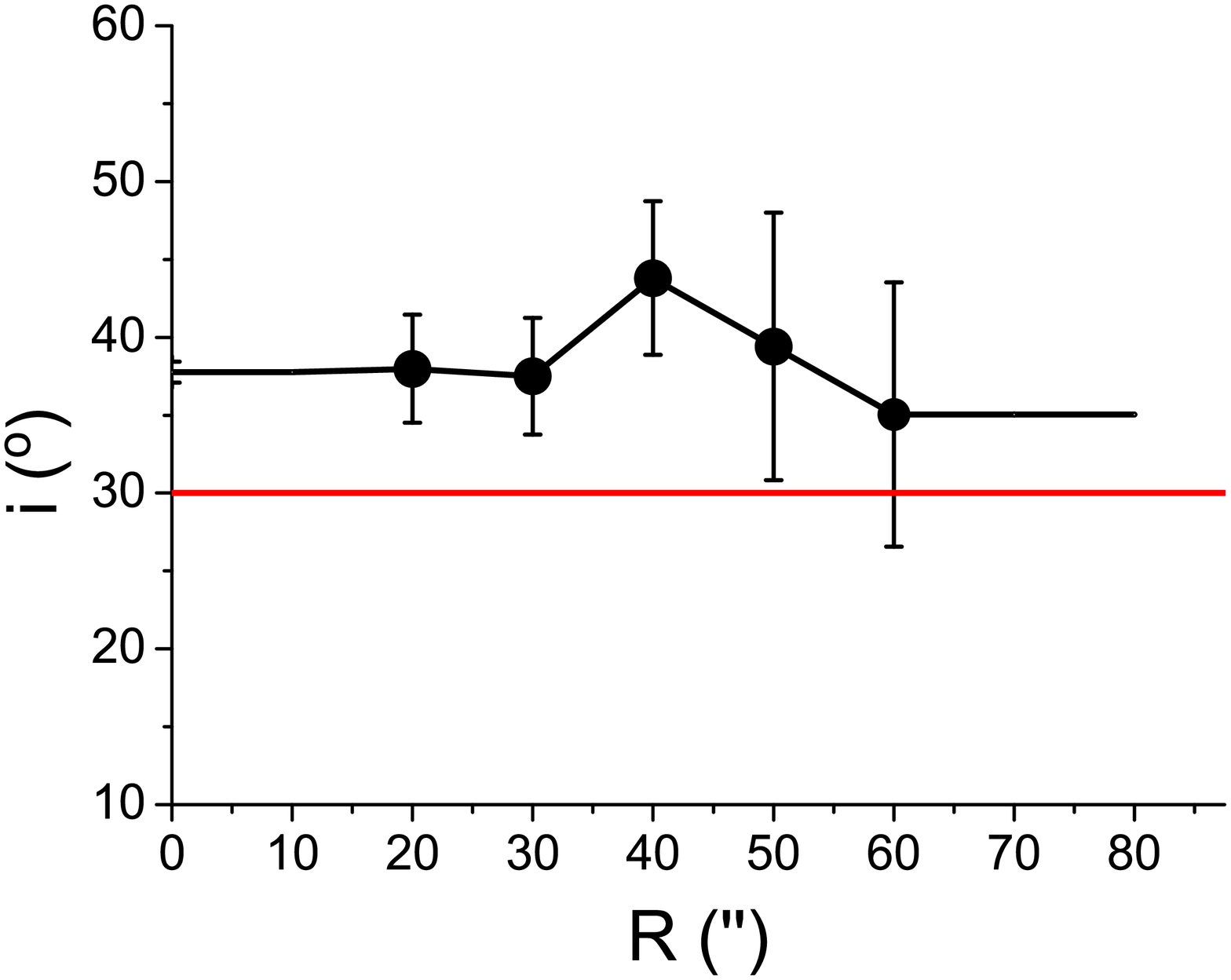}
\includegraphics[width=7.cm,keepaspectratio]{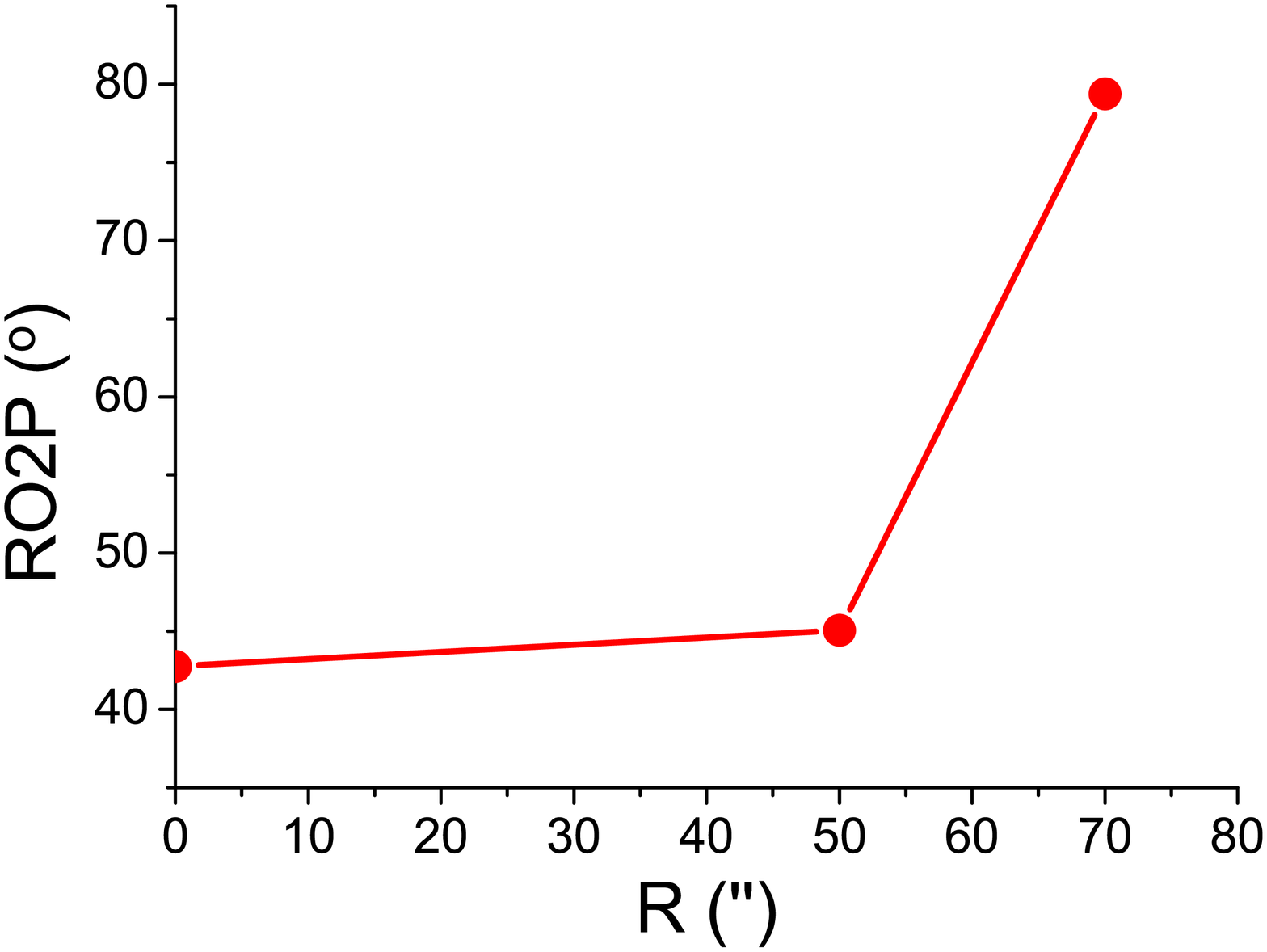} 
\includegraphics[width=7cm,keepaspectratio]{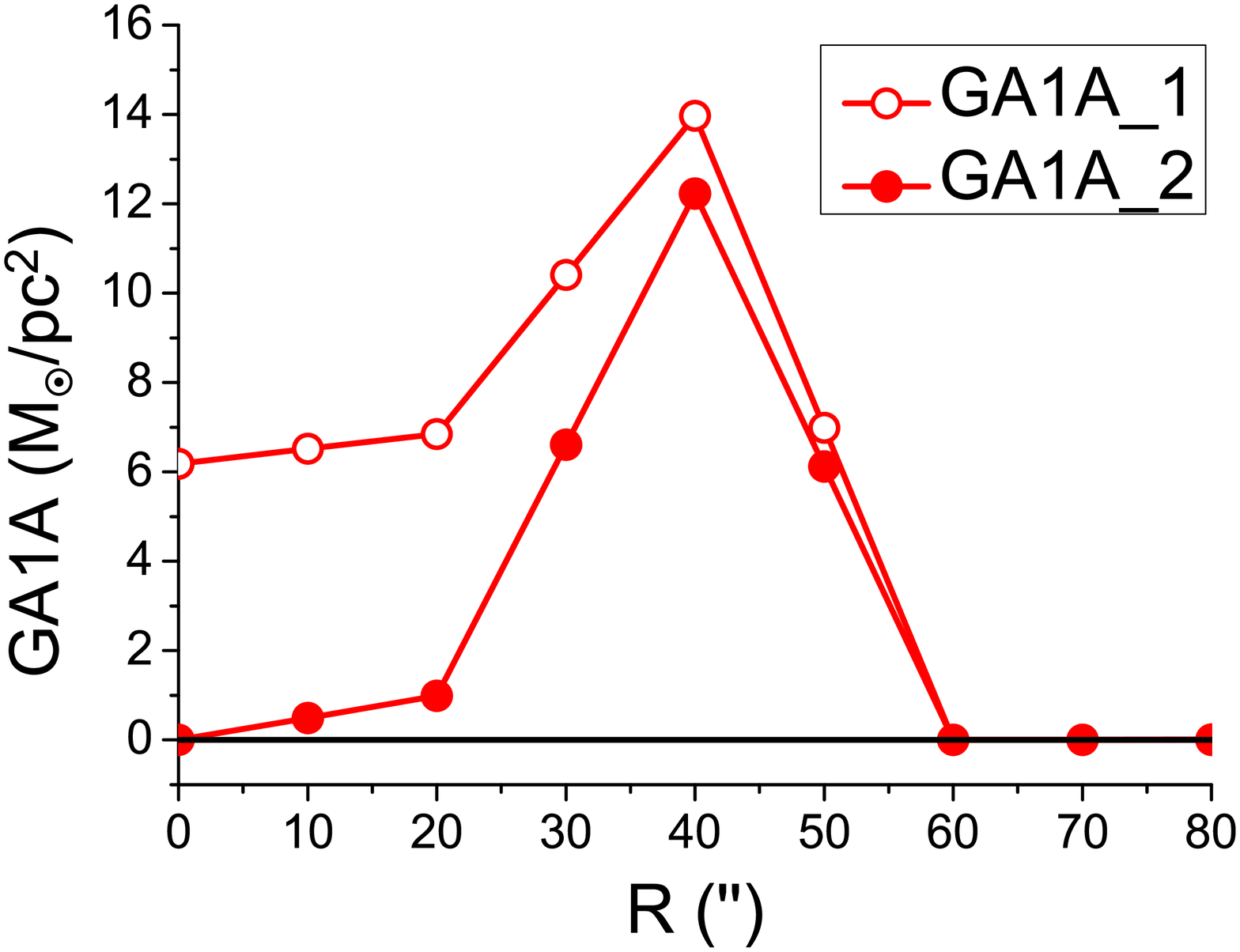}
\includegraphics[width=7cm,keepaspectratio]{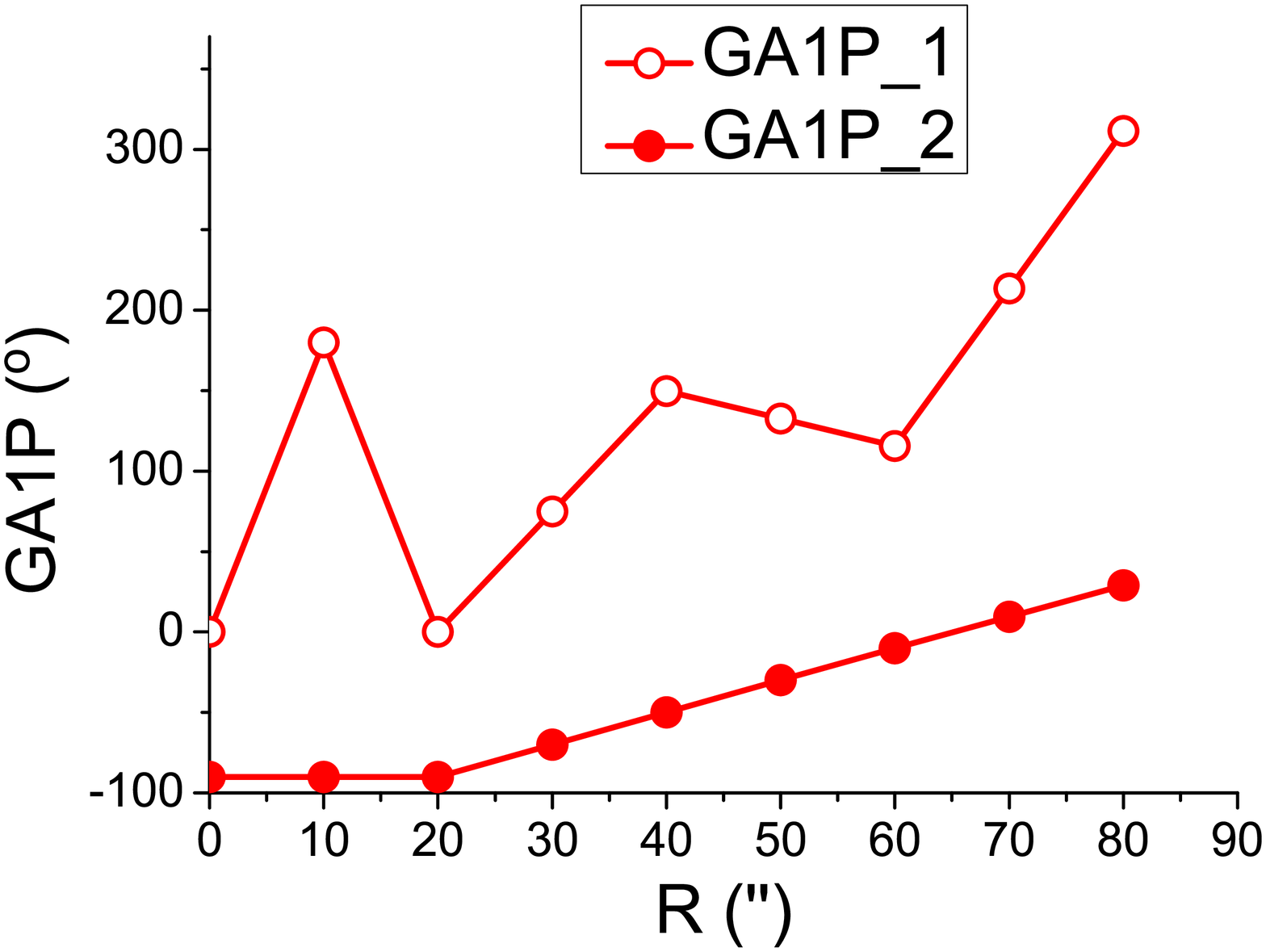}
\caption{Radial variation of parameters resulting from the
  \ion{H}{I} data cube modeling for UGC~11919 for the warped model (black) and bisymmetric model 
  (red in online version, grey in printed version). Surface gas densities and velocity curves are by
    construction corrected for inclination. RA2A and RO2A are the amplitudes of the second order harmonics of radial and tangential velocity, RO2P is the phase of RO2A. G1A is the amplitude of a Gaussian
  surface-brightness distortion representing two spiral arms 
 and G1P is the phase (azimuthal angles) of these distortion.}
\label{fig7}
\end{figure*}

\subsection{Kinematic results}

To compare 
our photometrical mass estimates with dynamical
models (from a mass-decomposition), rotation curves are obtained from our
\ion{H}{I} data, using tilted-ring modeling. For this purpose, we employ
TiRiFiC \citep{J2007}, a software
  allowing for a direct fit of modified
tilted-ring models to data cubes. Figures \ref{fig5} and \ref{fig6}  show
a comparison of data and  models for selected images of the \ion{H}{I} data cubes. Blue and pink (green) (black and grey (light grey) in printed version) contours represent the 0.75, 1.5, and 3
mJy/beam levels of the observed and model data cubes,
respectively.  As described below, for
  UGC~11919 two alternative model data cubes were created. Selected images of the data cubes demonstrating the difference between the two models are shown in 
  Fig. \ref{fig6}.

 The data cube of UGC~11919  
indicates 
strong deviations from a flat disk with circular rotation. This is 
evident from the moment maps, which quite prominently show an inner
kinematical twist (Fig. \ref{fig0}, right panel), as well as an outer ring, possibly an
extension of the spiral arms of the galaxy (Fig. \ref{fig0}, middle
panel). We consider two possibilities for this appearance. In a first attempt to explain this
structure, a classical tilted-ring
model is fitted where the inclination and the position angle vary with radius,
hence representing a symmetric S-shaped warp (see the resulting radial variation of the model parameters in Fig. \ref{fig7}). Using this approach, it is assumed that the apparent deviation
  from cylindric symmetry, best seen in the total-intensity map and the velocity
  field, is a projection effect. Comparing model and observed data
  cubes, we achieve a good fit. As expected, the model shows a strong warp in the
  center (reflected by a strong radial change in the 
inclination and position angle). However, one of the consequences
  is that the position angle of the \ion{H}{I} model deviates strongly
from our isophotal fits. The position angle found from isophotal fitting after masking the spiral structure is $PA\approx 20 \degr$, whereas from kinematical modeling we obtain $PA>40 \degr$ (see Fig. \ref{fig7}).
\begin{figure*}
\centering
\begin{minipage}[h]{1\linewidth}
\includegraphics[trim= 0 0 0 0,clip,height=5.60cm]{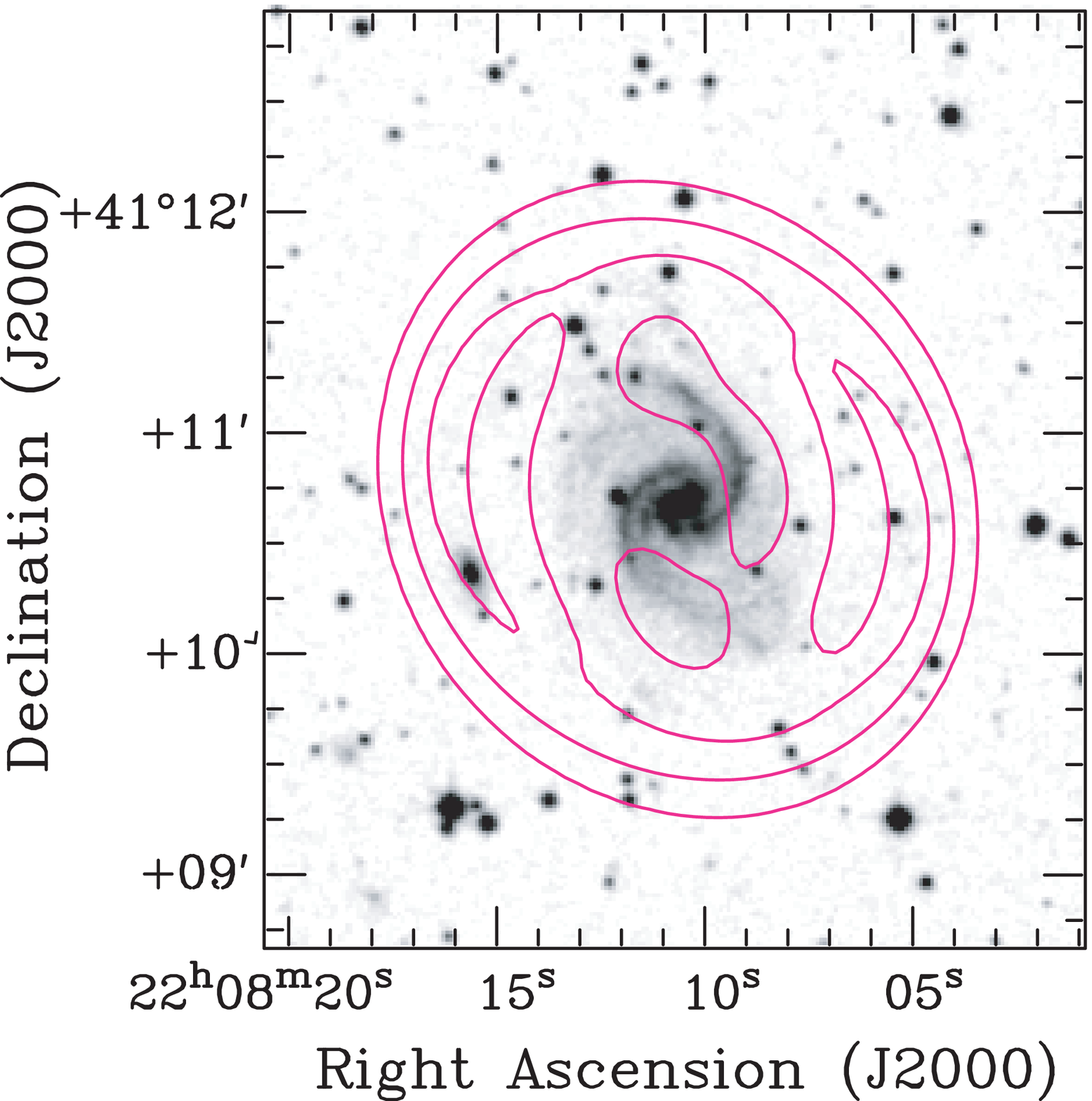}
\includegraphics[trim= 0 0 0 0,clip,height=5.60cm]{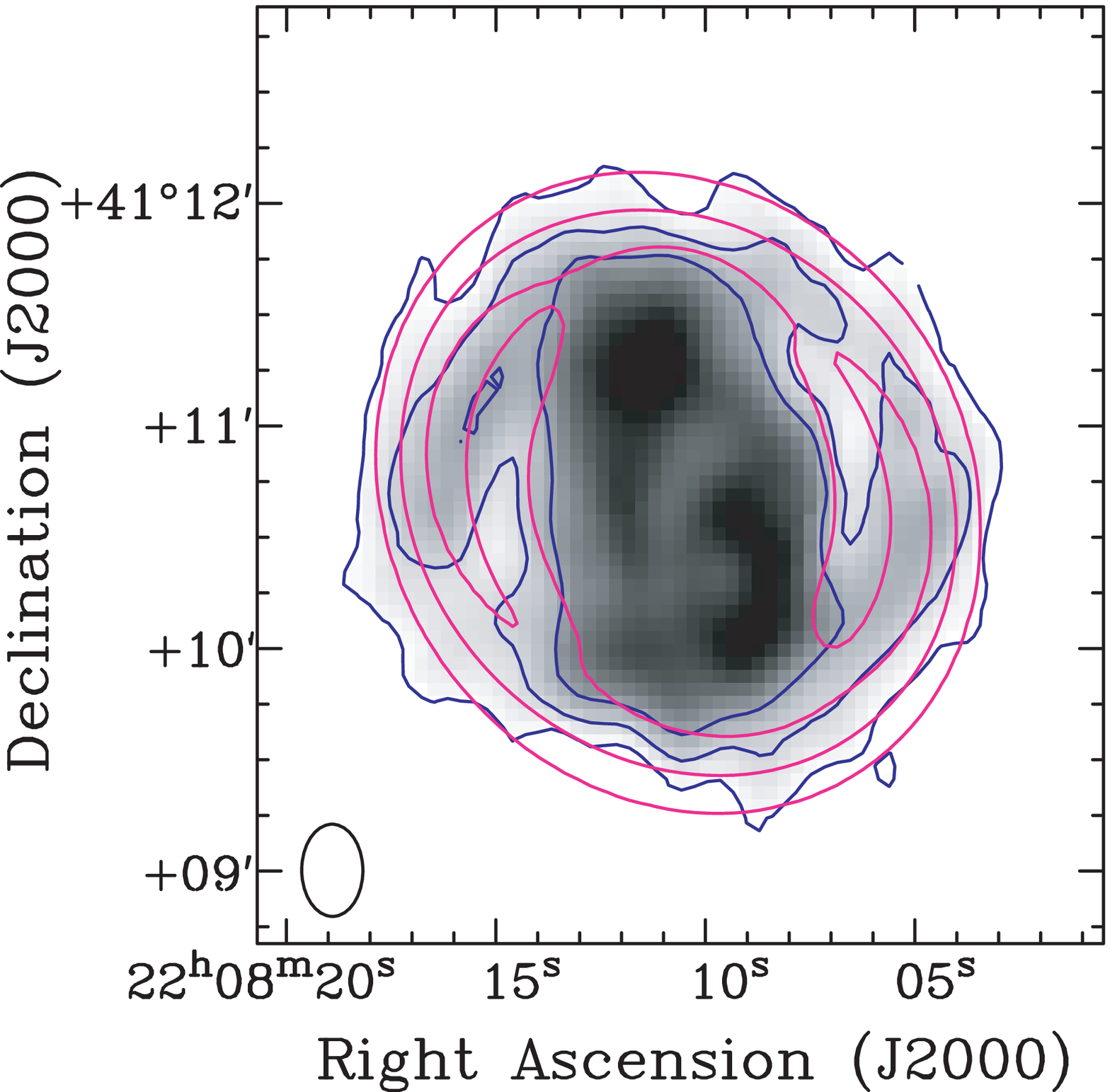}
\includegraphics[trim= 0 0 0 0,clip,height=5.60cm]{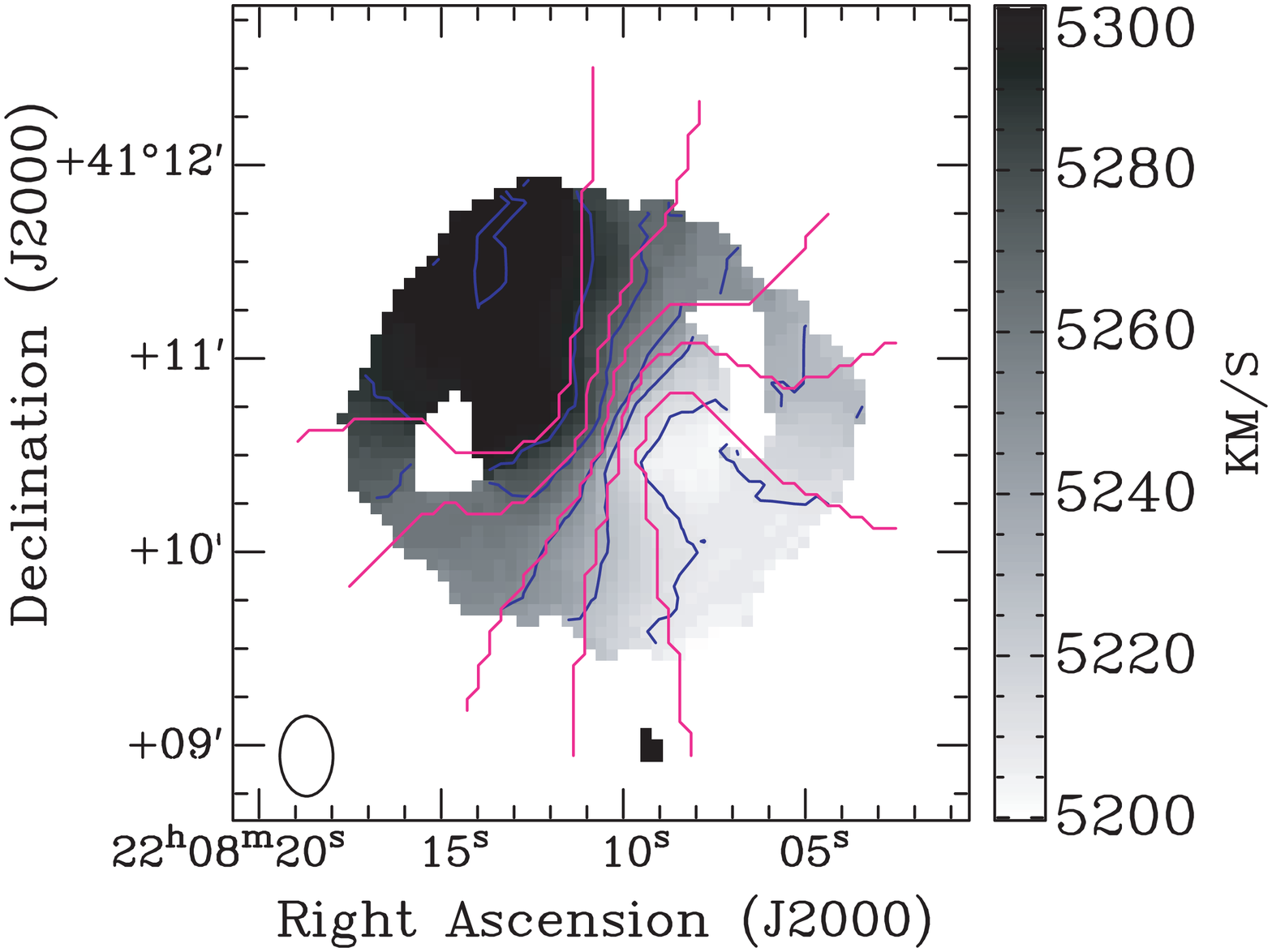}
\end{minipage}
\caption{Left panel: R-band optical image of UGC~11919 and
  \ion{H}{I} total intensity with   overplotted contours of  the  bisymmetric model (pink in online version, grey in printed version) 
\ion{H}{I} total intensity: 5.4, 20, and 35
 $\times 10^{19}$  ${\rm atoms \, cm}^{-2}$. Middle panel: the comparison of observed and bisymmetric model intensity maps with \ion{H}{I} total intensity contours: 5.4, 20, and 35 $\times 10^{19}$  ${\rm atoms \, cm}^{-2}$.  Right: first-moment map with overplotted observed (blue in online edition, black in printed version) and bisymmetric model (pink in online version, grey in printed version) contours $V_{sys}\pm$ 0, 20, 40, and $60 ~{\rm km}\,{\rm s}^{-1}$.} 
\label{momentscomparison}

\end{figure*}

 The second 
modeling approach is  to consider that the
   galaxy possesses a prominent spiral structure and a bar in the
 center as can be seen from the optical
 image (see middle panel in Fig. \ref{fig0}). In that
   picture, one can also see the apparent extension of the optical spiral structure
 in \ion{H}{I}. We tested whether the galaxy can also be
   modeled as a flat disk in the presence of i) noncircular motions or
 bar streaming and ii) morphological deviations from cylindric symmetry. Because we have a
 limited number of data points to compare our model to, we require a
 simplified approach. For the kinematics we use a
model 
representing the effects of a bisymmetric
  potential distortion on the kinematics as
introduced by \citet{Spekkens2007} (see also \citealt{Schoenmakers97}
and \citealt{Franx94} for more general implementations). In TiRiFiC
we reproduce this model by introducing second order
harmonic terms in radial and tangential velocity with amplitudes RO2A
and RA2A and phases $RA2P = RO2P + 45 \degr$, respectively. Other
  than \citet{Spekkens2007}, we allow the orientation angle of the
  bisymmetric distortion to vary with the radius. In
  addition, to introduce azimuthal surface-brightness variations we
  include a Gaussian surface density distortion with a radially
  dependent amplitude and a
radially dependent  phase (see Fig. \ref{fig7}). 
The dispersion of this distortion
  was kept constant at 8 \arcsec. We use this possibility to 
approximate the \ion{H}{I} spiral arms. The phases of the density distortion related to the
  spiral structure (describing the azimuthal angle of the maximum
    of the distortion) were found by iteration. They agree well enough
  with the orientation of the optical spirals (see Fig. \ref{momentscomparison}, left panel).  

\begin{figure*}
\centering
\includegraphics[trim= 0 100 0  90,clip,width=7cm,keepaspectratio]{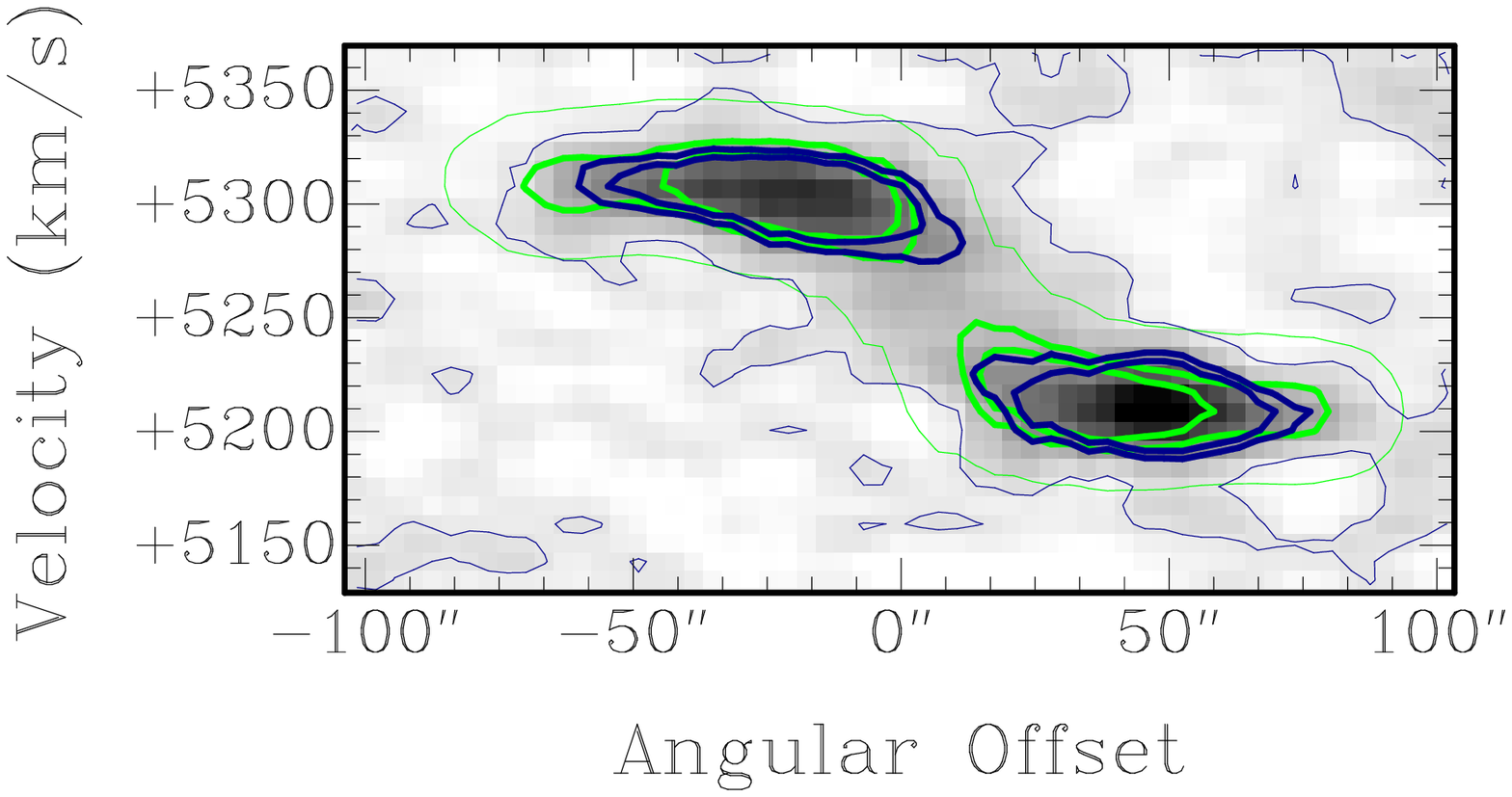}
\includegraphics[trim= 0 100 0 90,width=7cm,keepaspectratio]{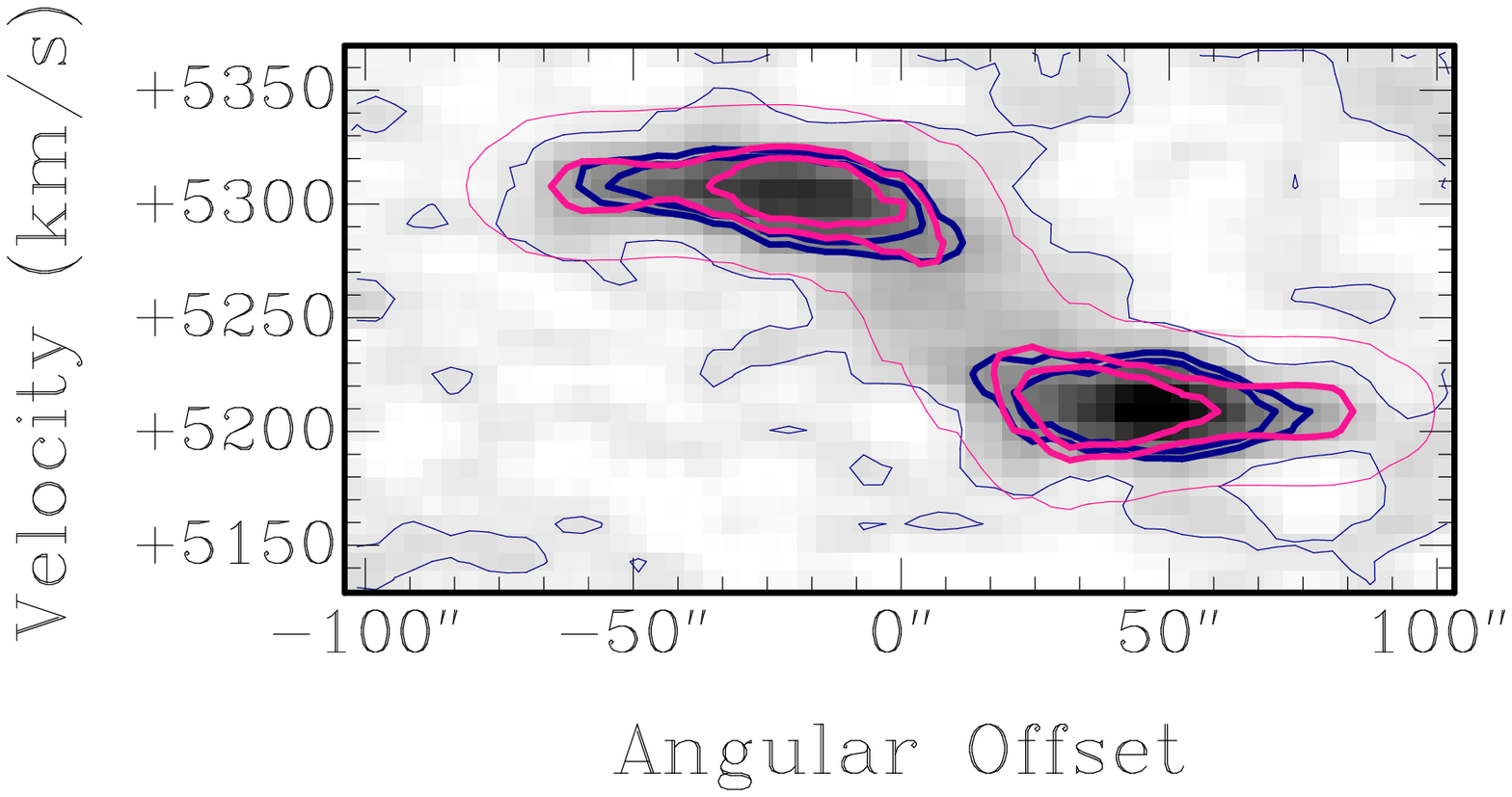}
\includegraphics[trim= 0 100 0 90,width=7cm,keepaspectratio]{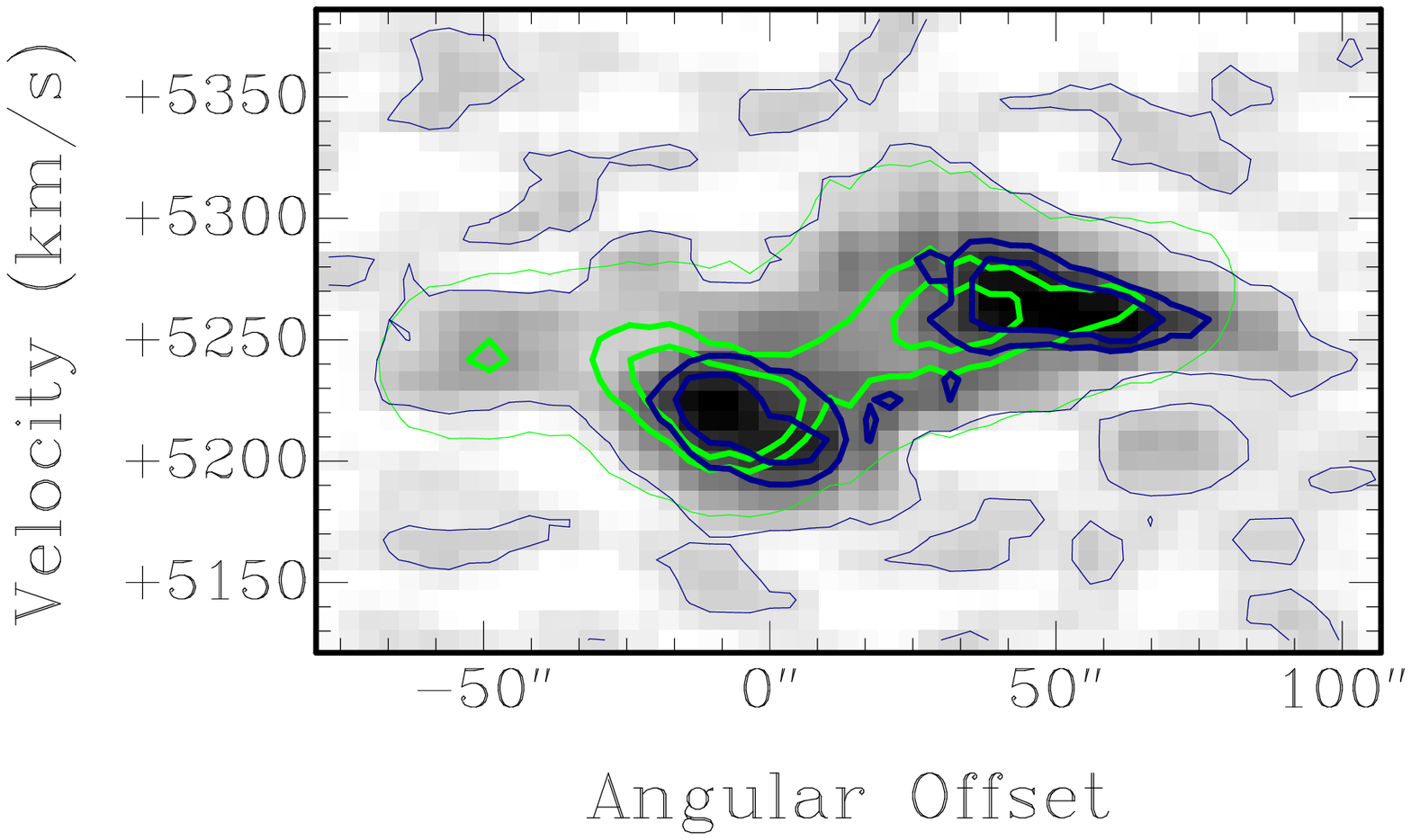}
\includegraphics[trim= 0 100 0 90,width=7cm,keepaspectratio]{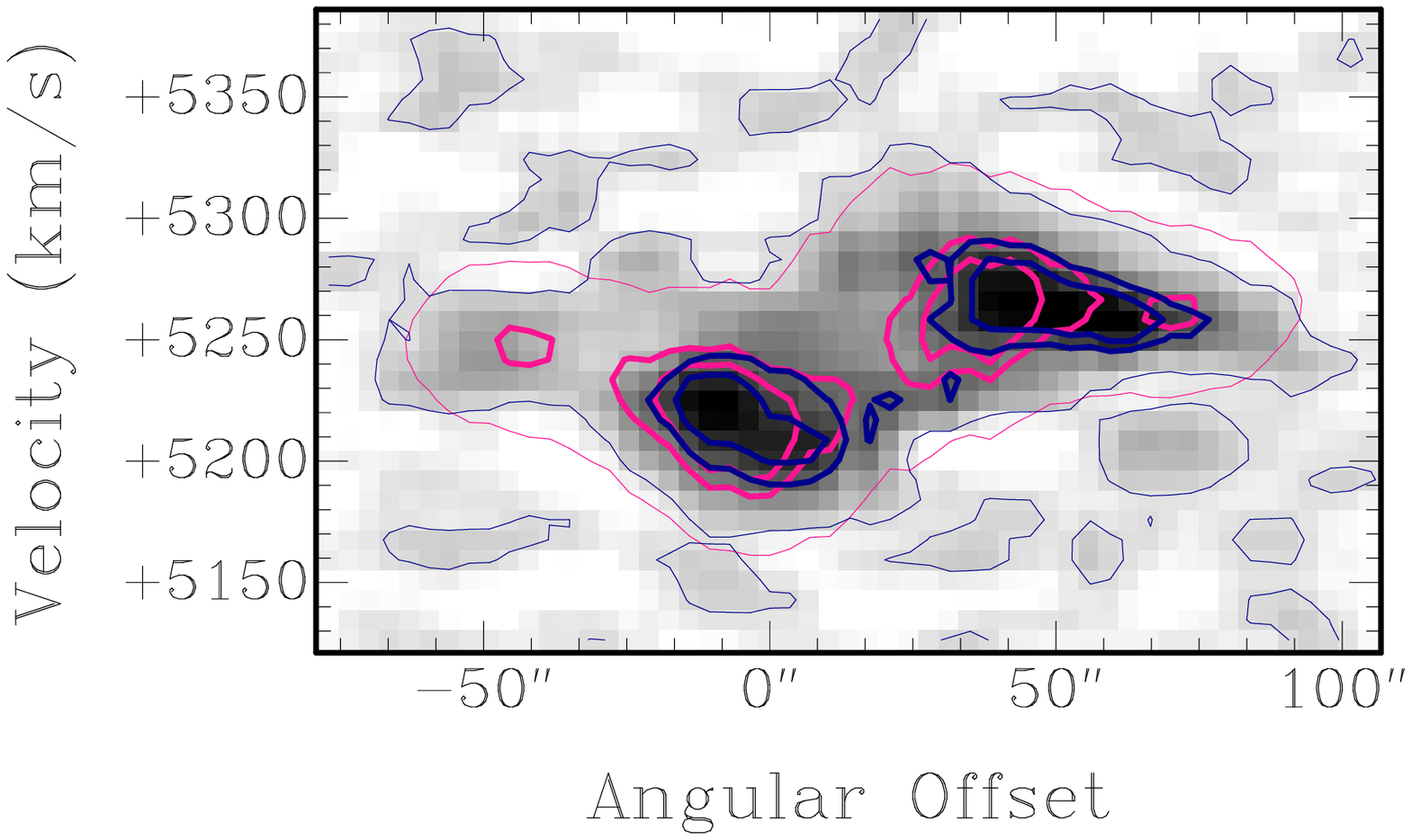}
\caption{Position-velocity diagrams along major
  (top) and minor kinematical axes (bottom) of UGC~11919. The blue,
  pink, and green (black, grey, and light grey in printed version) lines represent the 0.5 3, and 4 mJy/beam levels of the observed and
  model data cubes, respectively. Left and right panels correspond to
  the warped model and the model including non-circular motions, respectively. } 
\label{comparison}

\end{figure*}

  In Fig. \ref{comparison} we show the PV-diagrams along the major and
  minor kinematical axes with overplotted contours of \ion{H}{I}
  intensity of observed (blue in online edition, black for printed version) and model (pink and green in online version, grey and light grey for printed version) data cubes. The left and
  right panels correspond to the warped model and  the bisymmetric model, 
respectively. As can be
  seen from a direct comparison of data cube and models in Fig. \ref{fig6}, a flat disk model
  provides a better agreement with the observed data cube. The same conclusion may be
  reached from the comparison of PV-diagrams: a flat disk model gives
  a better representation of the observed data along the minor
  axis. Finally, in Fig. \ref{momentscomparison}  we show
  a comparison of the observed moment maps with the corresponding moment maps for
the bisymmetric model to show the good agreement.

 We point out that both models provide at least a reasonable fit
  to the data, such that a final decision which of these  very
  different physical interpretations to adopt is not
  straightforward. For our analysis we adopt the rotation curve
  corresponding to the model using the bisymmetric distortion, because
  of a better correspondence of the data with the model (as
  concluded by visual inspection), and
a more consistent physical picture, since streaming motions are
expected in the presence of a bar and strong spiral arms, and also surface density variations following the spiral
pattern are expected not to be the result of projection
effects only. 

A comparison of the warped and bisymmetric model parameters demonstrated in Fig. \ref{fig7}  shows that,
as expected,
some model parameters are quite different: the central inclination and
position angle of the warped disk model are higher than the uniform
parameters of the bisymmetric model (inclination $i_{\rm bi}=30\degr\pm
0.5\degr$, position angle $PA_{\rm bi}=45\degr\pm4\degr$).
Nevertheless, the resulting rotation curves, the most
important derived quantity used in our further analysis, are
comparable within the  errors (see Fig. \ref{fig7}), such
that we conclude that our following analysis is surprisingly not heavily
affected by the tilted-ring- model assumptions.

Since in this work we concentrate on the interpretation of
  the rotation curve only, and fit noncircular motions
  to disentangle their kinematical signatures rather than using the
  fitting results to further constrain the dynamics, we refrain from an
  interpretation of the corresponding coefficients.

Errors were derived by fitting all parameters for the
  receding and the approaching side independently
and calculating their deviation from the best-fit model. Final errors
were estimated by averaging the deviations of three neighboring data points.

The inclination angle of UGC~11919 found from the \ion{H}{I} data
cube modeling ($i=30\degr\pm 0.5\degr $) differs significantly from the angle obtained from
the isophotal fitting to the optical image
($i=58\degr\pm4 \degr$). The agreement becomes better when the optical
spiral structure is masked ($i\geq 35\degr$). The
position angle from the \ion{H}{I}
  modeling (bisymmetric model) is found to be $PA=45\degr\pm4
\degr$. The Uppsala General Catalog of Galaxies (\citealt{UGC}) gives
$PA=35\degr$; our isophotal fitting after masking spiral arms gives
an even lower value of $PA\approx 20 \degr$.   
From the velocity field and the optical image shown in Fig. \ref{fig0}, it can
be seen that indeed the kinematical position angle differs from
that based on the photometry (even if a judgment based on visual inspection can be
  misguided by the signatures of noncircular motion). Given that for
the optical estimates for the orientation of the galaxy a
large fraction of the data has to be masked, we trust the values
derived from the \ion{H}{I} to be the more reliable.
In the following we use
position angle $PA$ and inclination $i$ based on the 
 \ion{H}{I} analysis 
for the
quantities derived from the \ion{H}{I} data cube (\ion{H}{I} surface
density profile and rotation curve), but the optically derived
inclination and position angle for the stellar surface brightness and
color index. We
  remark that adopting
the same (kinematically derived) inclination and position angle for
all components does not change our results. 
The potential resulting
decrease in the scale length of the surface brightness in this case is not strong enough to alter
any of our conclusions. 

\begin{figure*}
\centering
\includegraphics[width=7.cm,keepaspectratio]{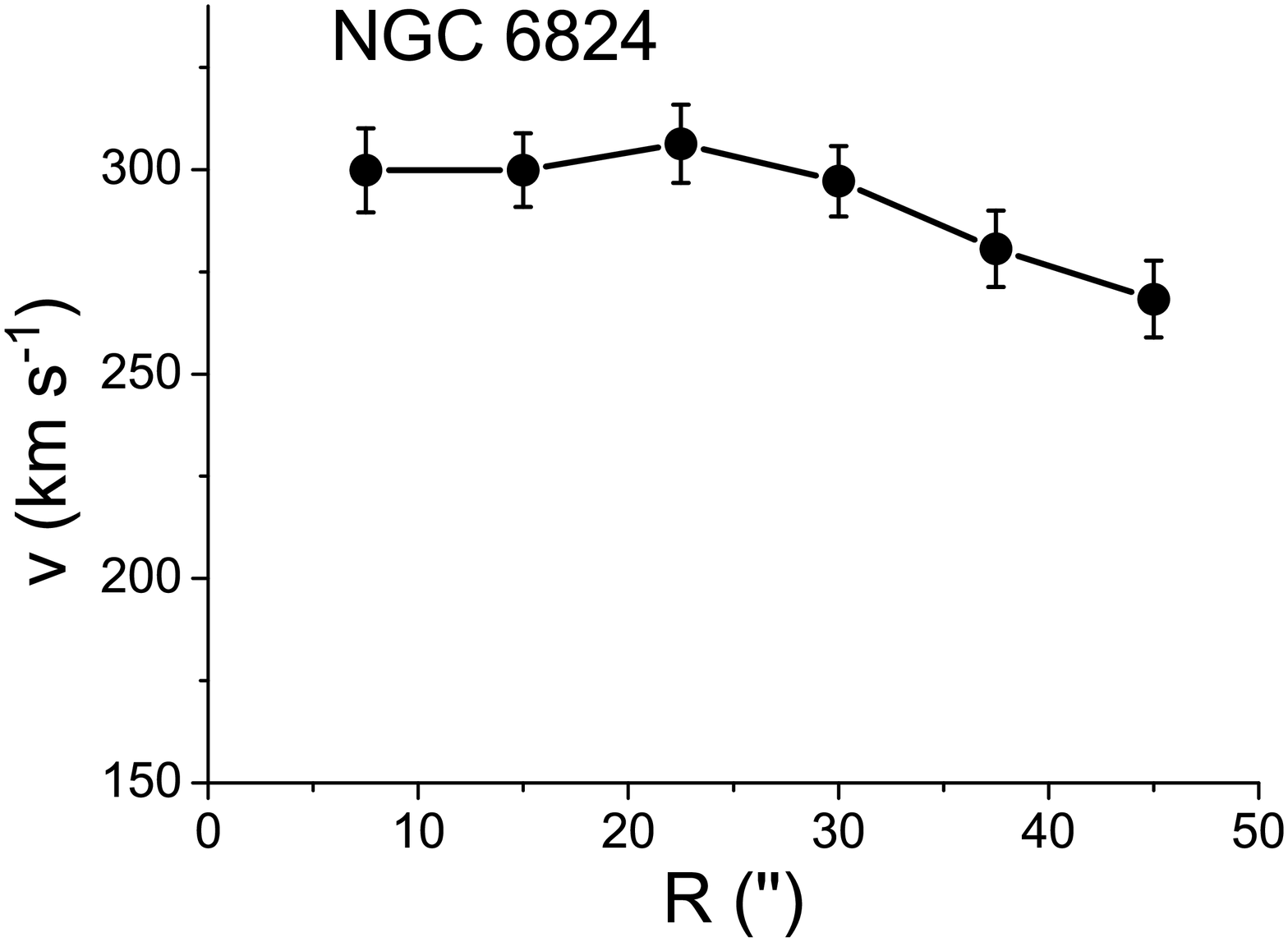}
\includegraphics[width=7.cm,keepaspectratio]{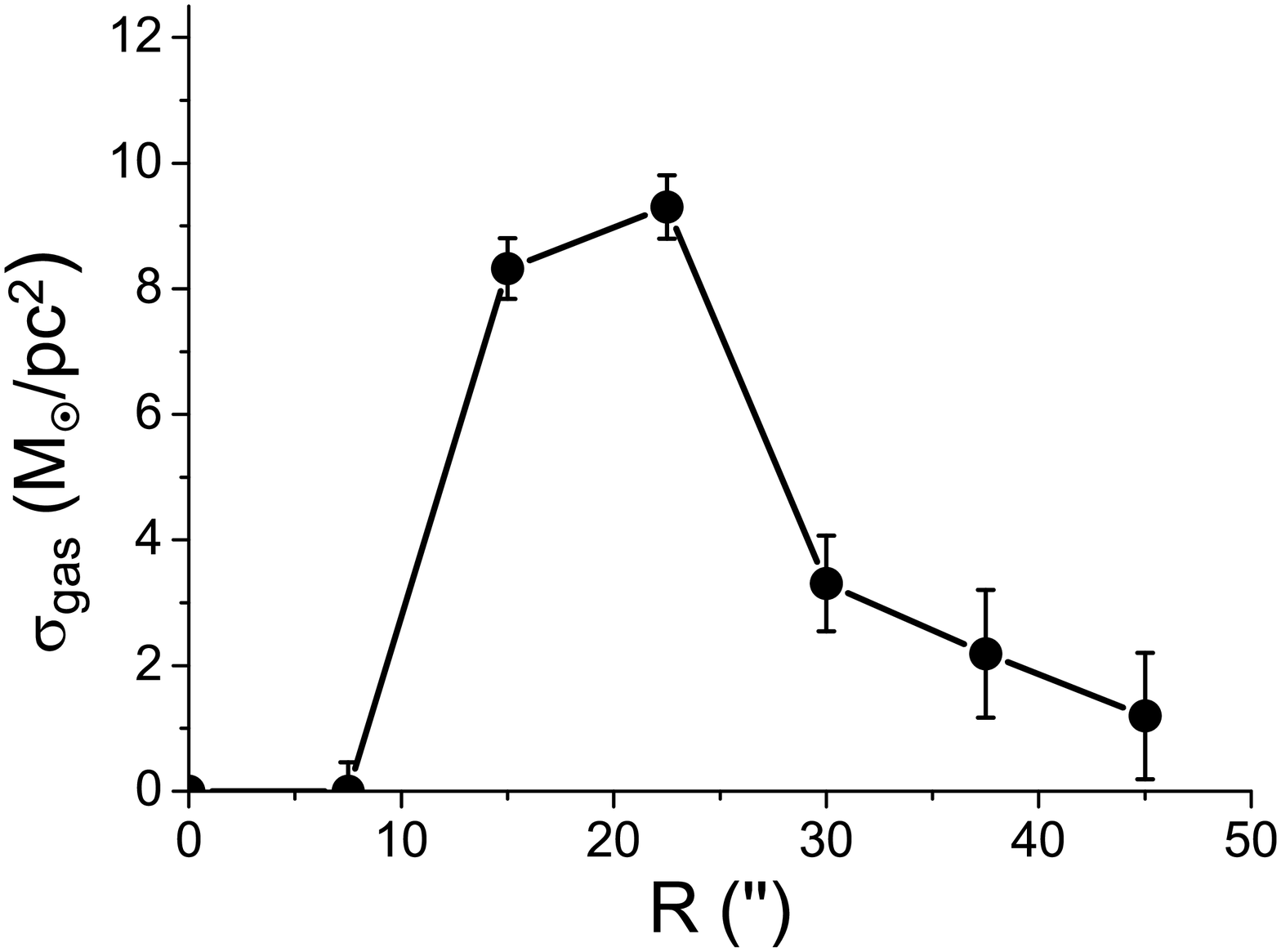}
\includegraphics[width=7.cm,keepaspectratio]{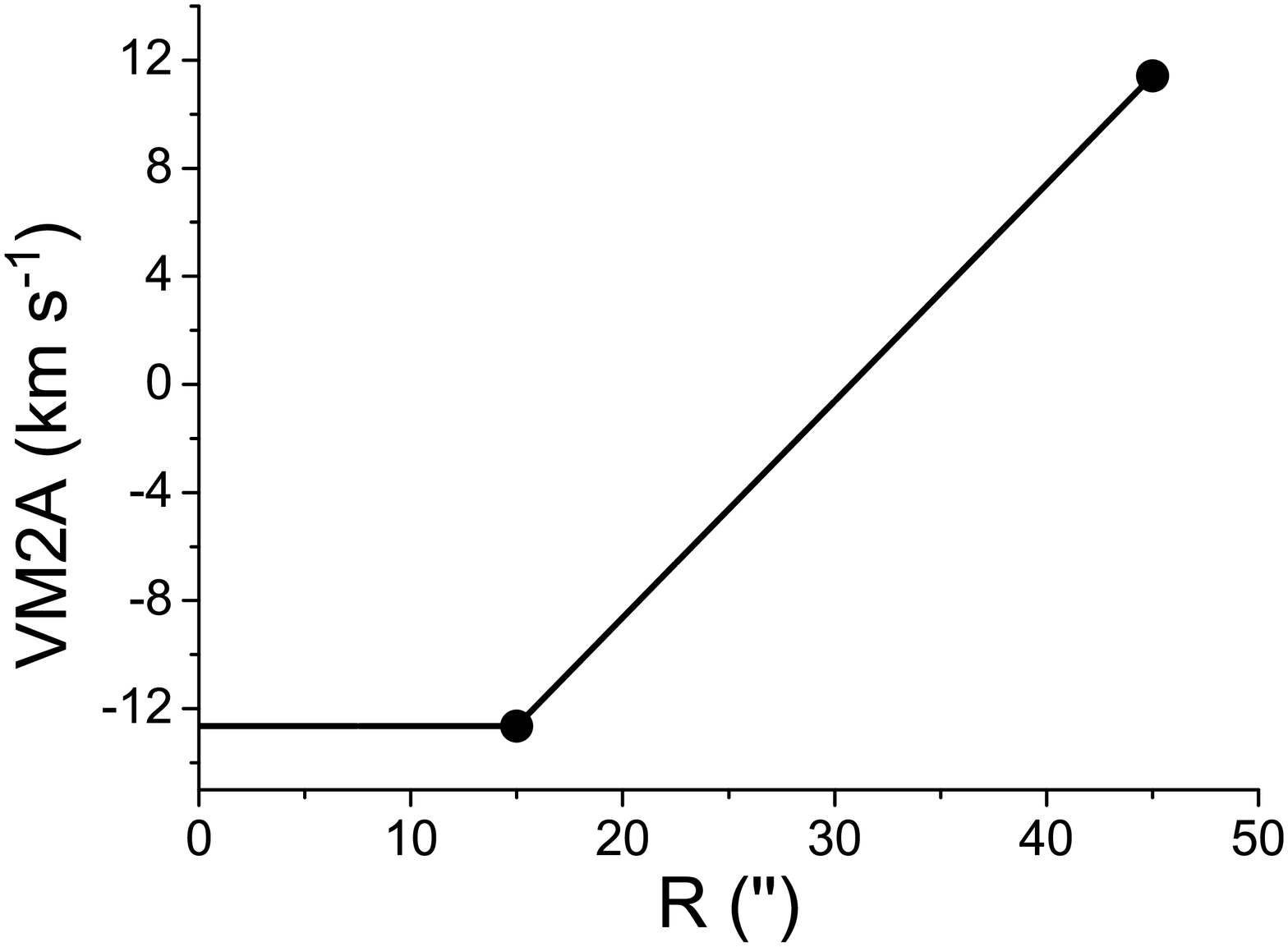}
\hspace{7cm}
\caption{Radial variation of parameters resulting from the
  \ion{H}{I} data cube modeling for 
  NGC~6824. VM2A is the amplitude of the adopted second
  order azimuthal harmonics of the line-of-sight velocity for
  NGC~6824. Surface gas density and velocity curves are corrected for inclination by
    construction.}
\label{fig7a}
\end{figure*}

For NGC~6824 we model  the \ion{H}{I} surface brightness of 
the receding and the approaching side independently. To restrict
  the number of free parameters, we assume both position angle and
  inclination to be constant with radius. The galaxy is close to edge-on, and
  does not show any obvious variation in position angle by visual
  inspection of the \ion{H}{I} data, such
  that such an assumption should not affect our derived rotation curve
by a significant factor.  
 The rotation velocity and an m=2 azimuthal distortion in the line-of-sight velocity (corresponding to the TiRiFiC parameter VM2A, see
Fig. \ref{fig7a}), which we included to achieve a good fit of the \ion{H}{I} cube, were fitted at independent radii for two sides simultaneously.  
The m=2 azimuthal distortion in the line-of-sight velocity is indicative of non-circular motions
and basically corresponds to a lopsided rotation curve. With the
introduction of a harmonic term, we attempt to disentangle the
asymmetric contribution from the rotation curve representing the
 (cylindrically symmetric part of the) potential. 

A satisfactory model corresponds to 
a constant inclination  $i=59\degr \pm 0.5 \degr$ and a constant position angle 
 $PA=45\degr \pm 2.5 \degr$, respectively. Errors in inclination, position angles, and rotation
velocity were obtained by fitting all these parameters for two sides
independently and calculating their deviation from the best-fit model. The resulting rotation curve appears to be in good agreement with
that  obtained by the GHASP project \citep{Ghasp} when corrected for the same inclination angle (see Fig. \ref{fig8}, right).

The inclination found from data cube modeling differs from that
obtained from (optical) isophotal flattening for NGC~6824 ($i=59\degr \pm 0.5 \degr$  and
$i=40\degr \pm 13 \degr$, respectively), which may result from
strong antisymmetric optical features in the galaxy.  We note that the inclination $i=59 \degr$
is in  better agreement with the Tully-Fisher relation (see Fig.
\ref{fig9}) and the photometric model of the rotation
curve. Again, we
  therefore tend to assign  a larger
  credibility to the inclination derived in our kinematical model.

\subsection{Mass decomposition}
Employing our rotation curves and photometrical data we constructed
mass models for UGC~11919 and NGC~6824. For UGC~11919, the stellar $M/L$ of bulge and disk
varies with radius, following the variation of their color indices. 
To fit the data, it is necessary to scale down the surface densities of stellar disk and bulge,
because the photometric data together with the standard $M/L$-color
relation do not agree with the observed rotation curve (see below).

The parameters describing the dark matter halo (radial scale and asymptotic velocity) of UGC~11919 were scaled to
achieve the minimal deviation of the resulting model
rotation curve from the observed curve. In this paper, we
present our results employing a pseudo-isothermal dark
halo. The situation does not change when using
as an alternative a cored profile obtained by \citealt{NFW} (NFW). Indeed, the
difference between the shapes of rotation curves for pseudo-isothermal and
NFW halos is significant only in the innermost parts of optical disks
(see, e.g., models by \citealt{deBlok}, based on the THINGS \ion{H}{I}
survey), so the total masses of these two halo types which fit
the rotation velocities far from the center, are very close. The
accuracy of our derived rotation curve 
 does not allow us to distinguish between the cases of
pseudo-isothermal and NFW halos. The density profile of the \ion{H}{I} disk was taken
from the observations and scaled by 1.3 to take into account helium in both models. 

\begin{figure*}
\centering
\includegraphics[width=8cm,keepaspectratio]{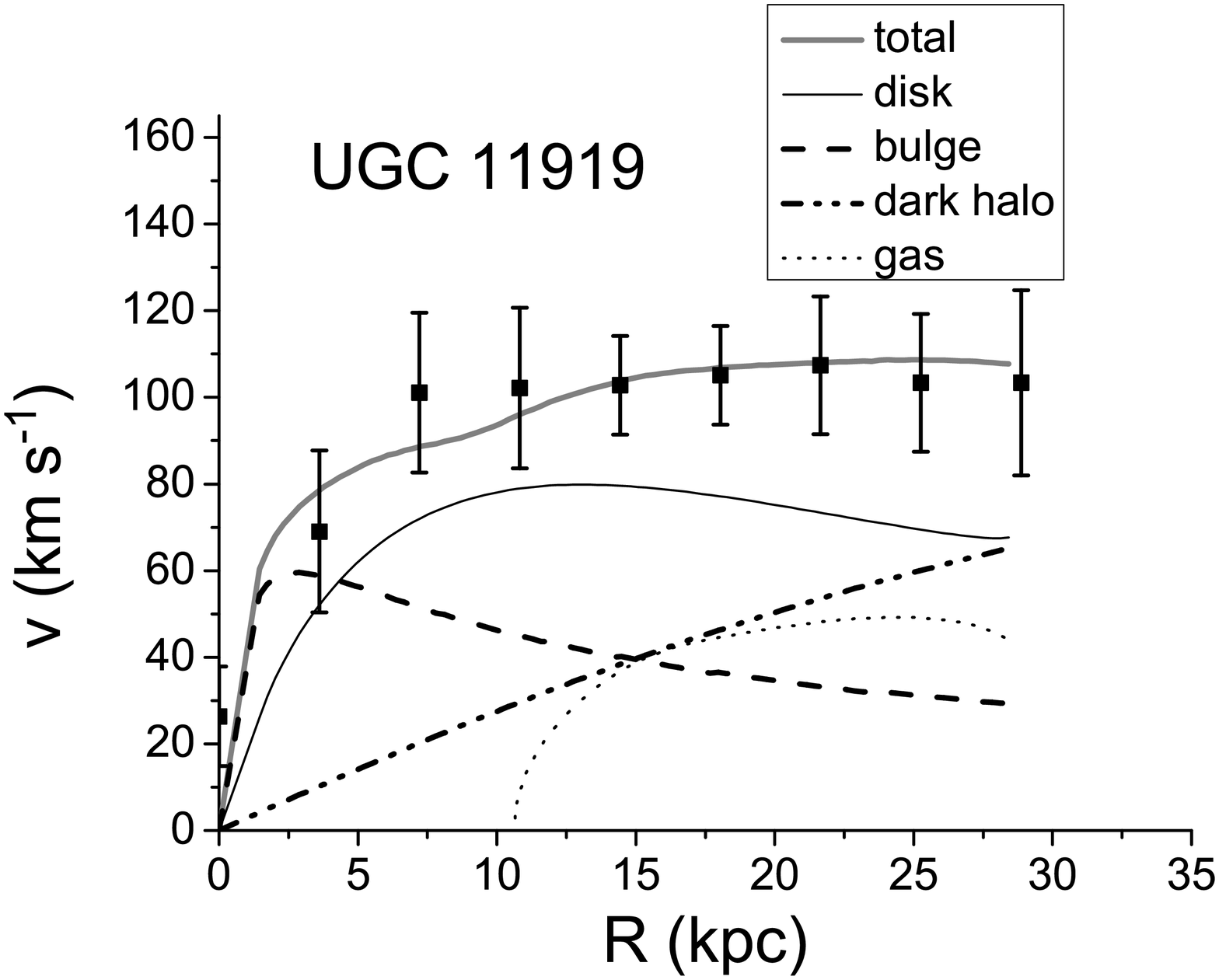}
\includegraphics[width=8cm,keepaspectratio]{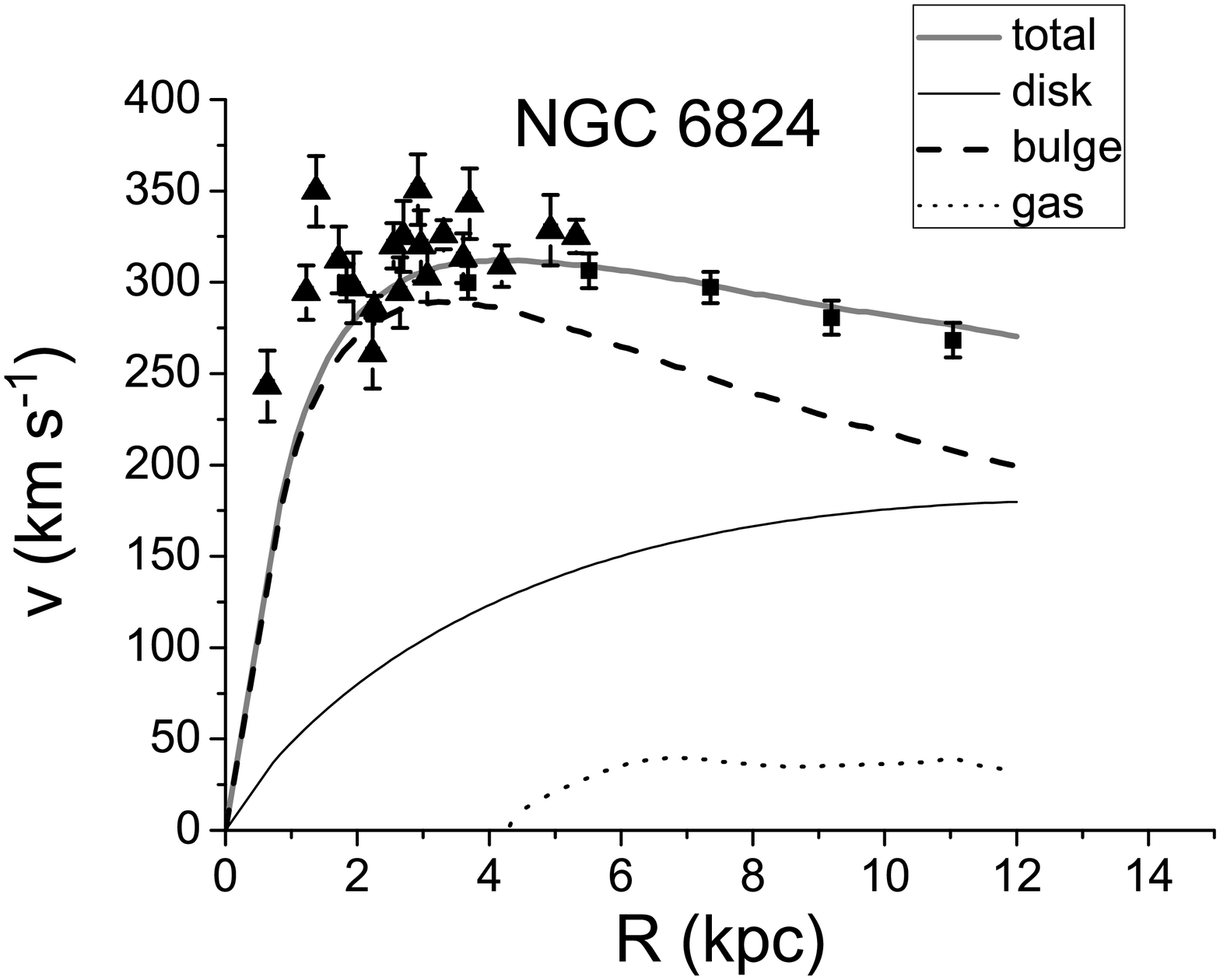}
\caption{ Top: the best-fit model of the \ion{H}{I} rotation curve
  of UGC~11919. Bottom: the model of the combined  \ion{H}{I} + $H_{\alpha}$ rotation curve (squares + triangles) of
  NGC~6824. The $H_{\alpha}$  rotation curve is taken from \citet{Ghasp}.}
\label{fig8}

\end{figure*}

For UGC~11919 no additional kinematical information is available
from the literature.  The best-fit model of our rotation curve
is consistent with a mass of
the stellar disk that is three times lower
(see Table \ref{table5}) and a bulge mass that is five times less massive 
than expected from the photometry for the given color (the mean stellar
mass-to-light ratio of the bulge is $(M/L_B)_{\rm bulge~dyn.}=0.4$ in our
model, while $(M/L_B)_{\rm
  bulge~phot.}=2$ is deduced from the photometry
). The best-fit model is shown in Fig. \ref{fig8}
(left). The dynamically (1) and photometrically (2) determined
mass-to-light ratios of the disk $(M/L_B)_{\rm disk~1}$ $(M/L_B)_{\rm
  disk~2}$ are given in Table \ref{table5}.

The maximum rotational velocity of the
disk component of the rotation curve of UGC~11919 estimated from the
photometric disk mass and the disk scalelength can be calculated via
$v_{max~disk}\approx 0.623(GM_d/R_d)^{0.5}$, where $M_d=2\pi
G\sigma_0R_d^2$ denotes the total disk mass and $\sigma_0$ is
the disk central surface density. If we assume
$\sigma_0$ and $R_d$ from our
  photometry, the expected value of the rotation velocity $ v_{max~disk}=159\pm 22$ at
$R\approx 2R_d$ is significantly higher than the maximum
rotation amplitude $v=102\pm 18$ (see
Fig. \ref{fig7}). Therefore, the result does not depend
on the used dark halo density profile, because it would persist even if no
dark matter were present in the galaxy. The
result is also robust against possible
  errors in the decomposition of the image
into bulge and disk components, because the total (disk + bulge) stellar surface
density expected for a normal stellar $M/L$ corresponds to a
higher rotation velocity than is observed. The situation also 
does not
change if we neglect the innermost two points of the rotation curve,
where the kinematics may be dominated by non-circular motions, because
the radius at which the disk component with an exponential surface
density profile reaches its maximum rotation velocity ($R\approx 10$ kpc) is
beyond this region. Furthermore, as elaborated above, we show
  that even under the assumption of 
a strong warp instead of a
  bisymmetric distortion the rotation curve stays the same
within the errors (see Fig.~\ref{fig7}). Hence, a standard photometric approach strongly
overestimates the stellar mass, and this galaxy most
likely has indeed a
light stellar component in both disk and bulge
with a non-standard IMF. The total dynamical
mass-to-light ratio is $(M/L_B)_{tot}=1.0$ for UGC~11919 inside the
optical radius $R_{25}$ (according to our estimates the apparent B-band magnitude within
$R_{25}$), corrected for Galactic extinction is $m_B=13.5$. Hyperleda
data (see Table \ref{tab1}) lead to an even lower 
$M/L_B$.

For NGC~6824 we achieve a consistent model by using
the photometrically determined mass of the stellar components (bulge and disk)
for a standard IMF. The previously
suspected large $M/L$ ratio for this galaxy was evidently a result of
an underestimation of the optically determined inclination and
the underestimation of the color index given in Hyperleda.

A decomposition of the rotation curve of NGC6824 for
photometrically estimated masses of stellar components is shown
in Fig. \ref{fig8} (right). The presented model is close to the
best-fit model, which corresponds to the minimum discrepancy between the
observed and model rotation curves.  The squares correspond to the
\ion{H}{I} rotation curve found in this work and the triangles show
the $H_{\alpha}$ rotation curve obtained in the GHASP project by
\citet{Ghasp}.  From the model it follows that the mass
  distribution of NGC~6824 is 
dominated by the visible matter inside a radius of
$R=12$ kpc. Within a
  radius of $R\approx 2 R_d$, the rotation curve of
this galaxy can be explained solely by
baryonic matter without the contribution of a DM halo (see Fig. \ref{fig8}).

The total dynamical $M/L$ for this galaxy as derived
from the mass-decomposition is $(M/L_B)_{\rm tot}=3.4$ within the
optical radius $R_{25}$. This is based on our
measurements in which we determined
the apparent B-band magnitude within
$R_{25}$ (corrected for Galactic extinction) to be $m_B=12.1$. The model
rotation curve of NGC~6824, shown in Fig. \ref{fig8}, corresponds to the
stellar mass-to-light ratios of the bulge $(M/L_R)_{\rm bulge~dyn.}=2.5$
and $(M/L_B)_{\rm bulge~dyn.}=3.3$, in good agreement with the values
obtained from the color indices $(B-R)_0$ and $(B-V)_0$ of the bulge and a
standard $M/L$ -- color model relation ($(M/L_R)_{\rm bulge~phot.}=2.6$
and $(M/L_B)_{\rm bulge~phot.}=3.6$, respectively. The same conclusion is valid for the stellar disk
 (see Table \ref{table5}).

There are two main sources of systematic errors for dynamically determined masses.
These are non-circular gas motion and
disk asymmetries, which in turn lead to
an uncertainty in the measurement of the
inclination, and with that of the rotation velocity and
  surface-brightness profiles. However,
   we explicitly include kinematical asymmetries in our models,
  such that we can assume to minimize errors of this kind.
  Another source of uncertainties could be related to the
correction for the internal extinction. However, the extinction 
affects the $M/L_B$ -- color index
diagram only slightly, because it
increases $M/L$ and the color index simultaneously,
and these shifts nearly compensate each other (see, e.g., \citealt{bdj}).

We note that the evolutionary models recently created by
\citet{Into} where the Thermally Pulsing AGB phase is
  taken into account, provide optical $M/L$ -- color
relationships which are nearly identical to those obtained earlier by \citet{bdj}.

Table \ref{table5} presents the results of the
  mass decomposition within the optical radius $R_{25}$ for the galaxies. For UGC~11919 we also give the masses of the components up to the last measured point of the rotation curve.

\begin{table*}
\small \caption{Modeling results. (1) Name;
(2) radius $R$ for the mass estimate (kpc);
(3) disk mass inside $R$ ($10^{10}\,M_\sun$);
(4) dark matter halo mass inside $R$ ($10^{10}\,M_\sun$);
(5) bulge mass inside $R$ ($10^{10}\,M_\sun$);
(6) gaseous disk mass inside $R$ ($10^{10}\,M_\sun$);
(7) total mass inside $R$ ($10^{10}M_\sun$);
(8) disk $M/L$ obtained from the observed
$(B-V)$ color of the disk, according to \citet{bdj};
(9) disk $M/L$ corresponding to the best-fit model of
the rotation curve.
\label{table5}}
  \begin{center}
    \begin{tabular}{c c c c c c c c c}
    \hline \hline
Name&$R$, & $M_{\rm disk}$    & $M_{\rm halo}$ & $M_{\rm bulge}$ & $M_{\rm gas}$ &$M_{\rm tot}$& $(M/L_B)_{\rm disk~1}$&$(M/L_B)_{\rm disk~2}$\\
(1) & (2) & (3) & (4)&(5)&(6)&(7)&(8)&(9) \\
 \hline
NGC~6824&14&8.0&0.0&11&0.2&19&2.5 $\pm 0.3$&2.7\\
\hline
UGC~11919&15&2.0&0.5&0.7&0.4&3.6&1.7$\pm 0.3$&0.5\\
\hline
UGC~11919&29&2.4&2.9&0.7&0.9&6.9&1.7$\pm 0.3$&0.5\\
\hline
 \end{tabular}
  \end{center}
\end{table*}
\section{Environment}\label{env}
For the targeted galaxies we  construct  \ion{H}{I}  total-intensity maps of the environment using the data
cubes with a robust weighting of 0.4, a tapering of 30
arcsec, and averaging four channels. 

\begin{figure*}
\includegraphics[width=8cm,keepaspectratio]{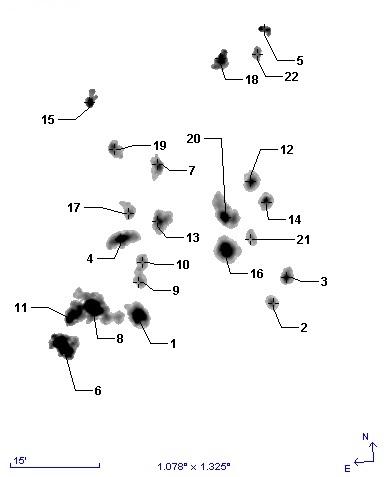}
\includegraphics[width=8cm,keepaspectratio]{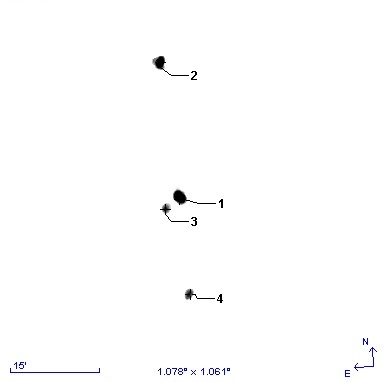}
\caption{\ion{H}{I} intensity maps of the environment of
  UGC~11919 (left) and NGC~6824 (right), obtained from the data cube with robust weighting of 0.4, tapering of 30 \arcsec, and averaged by four channels with the labeled objects.}\label{group}

\end{figure*}

From Fig. \ref{group}  (left) a rather dense environment within the bounds of our data cube becomes evident. We derive the main integral parameters, position, systemic velocity, total \ion{H}{I} flux, \ion{H}{I} mass, and \ion{H}{I} mass-to-light ratio (B-band where available) of all objects detected in this cube, shown in Table \ref{table6}.  Because our data cube is limited to a minimum velocity of $4367 {\rm km}\,{\rm s}^{-1}$, we do not cover the complete velocity range of this group. In Table \ref{table6} we mark galaxies, for which the \ion{H}{i} lies partly outside the (velocity) bounds of our data cube with an asterisk. Velocities of these objects denote only the averaged velocity of the detected \ion{H}{i} and hence do not reflect the systemic velocity. 

Table \ref{table6} and Fig. \ref{group}   (left) reveal a rich group of galaxies around object 8 (NGC 7223, see \citealt{ Crook}) to the south-east of UGC 11919. A simple comparison of velocities shows that UGC~11919 probably  does not belong to this group. From Table 6 it is evident that the systemic velocities of detected objects in the data cube cluster in two ranges (objects 1-15 and objects 16-22). The average velocity (cz) in range 1 (objects 1-15) is $4558 \pm 112 ~{\rm km}\,{\rm s}^{-1}$, while in range 2 (objects 16-22)  it is $5550 \pm130 ~{\rm km}\,{\rm s}^{-1}$.  The systemic velocity of UGC~11919 differs by $800~{\rm km}\,{\rm s}^{-1}$ from the average velocity in range 1, which is seven times larger than the standard deviation of the measured velocities in range 1 (which may be used as a proxy for the velocity dispersion in the group, despite the fact that we are not measuring the velocities of each group member correctly, see below) and two times larger than the linking velocity of $350 ~{\rm km}\,{\rm s}^{-1}$ that acts as one selection criterion to identify two galaxies as a coupled pair in \citealt{Crook}, (see also \citealt{Ramella}). Selecting by this typical linking velocity (\citealt{Crook}, \citealt{Ramella}) alone, none of the galaxies 1-15 would be identified as belonging to the same group as UGC 11919. We point out that our table does not give a full account of all objects in the group, since it is at the edges of our velocity bounds. However, this result will probably still hold if we were able to take the systemic velocities of objects 1-5 properly into account, since they are currently biased towards velocities  that are too high (closer to UGC~11919).

The galaxy UGC~11919 possesses two
companions, object 20 (CGCG~530-011) and  object 21 at the projected
distances of $r<100 \,{\rm kpc}$. The velocity differences from UGC~11919 are $355~{\rm km}\,{\rm s}^{-1}$ (20) and $252~{\rm km}\,{\rm s}^{-1}$ (21).
The galaxy CGCG~530-011 has a peculiar \ion{H}{I} structure which
might originate from an interaction. While our target galaxy is not a member of a rich group, it is also not necessarily an isolated galaxy. The \ion{H}{I} ring found for UGC 11919 could hence also be a result of gravitational interaction.

The galaxy NGC~6824 is a member of a group with mean velocity $V_{\rm lg}=4065~{\rm km}\,{\rm s}^{-1}$. The asymmetric structure and the m=2 distortion in the line-of-sight velocity of this galaxy could be due to the gravitational interaction with the members of the group (see Fig.\ref{group}, right,  and Table~\ref{table6}).  

We therefore conclude that neither UGC~11919 nor NGC~6824 is a
  particularly isolated galaxy. Both show irregular velocity
  structure, possibly as a result of a recent interaction. Whether this
  interaction might have led to the peculiar $M/L$ for UGC~11919 is
  questionable. The interaction with neighbor galaxies has already experienced a significant part of all existing galaxies possessing the standard IMF, so by itself it cannot explain the paucity of low-mass stars. It is worth mentioning that the bulk of galactic stars have formed in the past, and so the cause of the existence of abnormally light-weighted disks lays in the properties of cold star-forming gas which existed in the epoch of disk formation, and which are poorly understood now. 
\begin{table*}
\small \caption{Environmental characteristics of UGC~11919 and
    NGC~6824.
(1)  Number of detected object, as labeled in Fig.~\ref{group};
(2)  Name of object if known;
(3)  Right ascension J2000;
(4)  Declination J2000;
(5)  Optical recession velocity;
(6)  Total  \ion{H}{I} flux $F$ in ${\rm Jy}\,{\rm km}\,{\rm s}^{-1}$;
(7)  Velocity corrected to the centroid of the Local Group using the formula following \citet{Yahil};
(8)  \ion{H}{I} total mass calculated from the total flux via $ M =
2.36\cdot 10^5(V_{\rm lg}/(75 {\rm km/s}))^2F$;
(9)  \ion{H}{I} mass-to-light ratio, the B-band magnitudes were taken from Hyperleda.
Galaxies with the velocities lying at the edge of the observed data cube (so the systemic velocity is uncertain for them) are marked by asterisks.  
 \label{table6}}
  \begin{center}
\begin{tabular}{c c c c c c c c c}
\hline \hline 
\#& Object designation & RA & Dec & $cz$ & \ion{H}{I} flux
(${\rm Jy}\,{\rm km}\,{\rm s}^{-1} $) & $V_{\rm lg}$ & $M_{\rm H\,{\small
    I}}$ ($10^{9}\,M_\sun$) & $M_{\rm \ion{H}{I}}/L_{\rm B}$ \\
(1)&(2)&(3)&(4)&(5)&(6)&(7)&(8)&(9)\\
\hline
&&&& UGC~11919& &&&\\
\hline
1$^*$	&	UGC~11927	&	22:09:27.86	&	+40:59:56.7	&	4501	&	7.46	&	4800	&	7.21	&	0.43	\\

2$^*$	&		&	22:07:30.80	&	+41:01:55.4	&	4398	&	0.44	&	4698	&	0.4	&	--	\\

3$^*$	&		&	22:07:17.93	&	+41:06:00.1	&	4448	&	1.16	&	4747	&	1.09	&	--	\\

4$^*$	&	2MASX~J22094543+4112200	&	22:09:43.18	&	+41:12:38.5	&	4536	&	4.78	&	4836	&	4.69	&	--	\\

5$^*$	&		&	22:07:36.15	&	+41:47:26.8	&	4388	&	1.15	&	4688	&	1.06	&	--	\\

6	&	CGCG~530-014	&	22:10:34.14	&	+40:54:37.7	&	4525	&	18.03	&	4824	&	17.6	&	0.81	\\

7	&	UGC~11925	&	22:09:10.35	&	+41:24:41.1	&	4548	&	0.97	&	4848	&	0.96	&	0.16	\\

8	&	NGC~7223	&	22:10:08.31	&	+41:01:12.0	&	4697	&	18.4	&	4997	&	19.3	&	0.3	\\

9	&		&	22:09:26.87	&	+41:05:31.7	&	4465	&	0.36	&	4765	&	0.34	&	--	\\

10	&		&	22:09:25.16	&	+41:08:51.8	&	4568	&	0.4	&	4868	&	0.4	&	--	\\

11	&		&	22:10:26.16	&	+40:59:44.6	&	4647	&	7.62	&	4947	&	7.83	&	--	\\

12	&		&	22:07:48.81	&	+41:22:01.3	&	4620	&	1.43	&	4920	&	1.45	&	--	\\

13	&	2MASX~J22090972+4115259	&	22:09:09.99	&	+41:15:26.7	&	4772	&	1.97	&	5072	&	2.12	&	--	\\

14	&		&	22:07:35.56	&	+41:18:57.7	&	4693	&	1.04	&	4993	&	1.09	&	--	\\

15	&		&	22:10:12.18	&	+41:35:17.2	&	4703	&	1.77	&	5003	&	1.86	&	--	\\

16	&	UGC~11919	&	22:08:10.51	&	+41:10:47.4	&	5354	&	7.17	&	5654	&	9.61	&	0.3	\\

17	&	LEDA~3087958 	&	22:09:36.12	&	+41:16:52.6	&	5447	&	0.45	&	5747	&	0.63	&	0.07	\\

18	&		&	22:08:15.09	&	+41:42:28.9	&	5499	&	3.23	&	5799	&	4.56	&	--	\\

19	&		&	22:09:51.82	&	+41:27:48.0	&	5499	&	0.88	&	5799	&	1.24	&	--	\\

20	&	CGCG~530-011    	&	22:08:12.18	&	+41:16:19.6	&	5709	&	4.63	&	6009	&	7.01	&	0.35	\\

21	&		&	22:07:48.03	&	+41:12:45.0	&	5606	&	0.23	&	5906	&	0.34	&	--	\\

22	&		&	22:07:43.38	&	+41:43:25.6	&	5735	&	0.51	&	6035	&	0.78	&	--	\\

\hline
&&&& NGC~6824& &&&\\
\hline
1	&	NGC~6824	&	19:43:41.27	&	+56:06:34.0	&	3563	&	1.88	&	3835	&	1.16	&	0.02	\\
2	&		&	19:44:04.82	&	+56:29:03.9	&	3493	&	0.74	&	3766	&	0.44	&	--	\\
3	&	LEDA~214725	&	19:43:57.39	&	+56:04:33.9	&	3949	&	0.16	&	4221	&	0.12	&	0.16	\\
4	&		&	19:43:29.69	&	+55:50:22.7	&	4164	&	0.18	&	4436	&	0.15	&	--	\\
\hline
\end{tabular}
  \end{center}
\end{table*}

\section{Conclusions}
We carried out  \ion{H}{I} observations together with
multicolor surface photometry for two galaxies, UGC~11919 and NGC~6824,
for which  in a catalog search we found earlier 
indications of a peculiar $M/L$ for a given color index
$(B-V)_0$. We model the \ion{H}{I} data cubes in
  the scope of an extended tilted-ring model. Using this model we constructed the rotation curves of the galaxies with the aim of deriving their kinematically-based mass distributions. 

Both galaxies have some disk asymmetries and rather
complex gas kinematics. The main problem in 
estimating the rotational velocity of NGC 6824 is the ambiguity
of its disk inclination angle $i$. 
An optical isophotal fit gives $ i=40\degr \pm 13 \degr$,
whereas the kinematical fit to the radio data favors  $i=59\degr \pm 0.5
\degr$.  We prefer the kinematical fit, because this galaxy is far from being optically axisymmetric having two open spiral arms and
a dust lane. However, if
one assumes the lower value for 
inclination,  the rotation
velocity would become, under standard assumptions, too high
for its luminosity, which in turn could give evidence of a massive
dark matter component  in the
galaxy, provided that the $M/L$ ratios of the stellar
components correspond to their observed color. 
However, our trusted value for the inclination,
$i=59\degr$, agrees with the $M/L$ expected from the
photometry. And there is no firm evidence of any peculiarity
of the stellar IMF in this galaxy. 

Evidently NGC~6824 is dynamically dominated by visible matter. It possesses
an extended pseudobulge and quite normal mass-to-light ratios of 
disk and bulge. The photometrically determined contributions of
stellar disk and bulge are in good agreement with the observed
rotation curve.  To get a good fit of the \ion{H}{I} data cube for
NGC~6824 we include an m=2 distortion in the line-of-sight velocity.

For the second galaxy, UGC~11919, the situation is different. 
The analysis of our
photometric and kinematic data indicates that the disk of this
galaxy,  under standard assumptions,
is unusually light for its luminosity, 
independent of the mass of
the dark matter halo. Determining the mass of its
disk using its \ion{H}{I} kinematics, corrected for non-circular
motions, yields a disk
mass-to-light ratio of $(M/L_B)_{\rm disk}=0.5$ in the best-fit model. In this model the rotation curve corresponds to 
a
disk which is three times lighter, and a bulge lighter by
a factor of five than is
expected from the photometric data. This galaxy may therefore
possess a bottom-light stellar mass function. If we assume
an inclination found from isophotal flattening ($i=58\degr$)
instead of the kinematically determined one of $i =30 \degr$, we 
obtain an even lower $M/L$ for the galaxy and its
components. The employment of an alternative kinematical model reproducing
  the observations by means of a regularly rotating but heavily
  warped \ion{H}{I} disk also results in a similar rotation curve (with a slightly
  decreased rotation velocity) and does not, therefore, alter our main conclusion. We also found that this galaxy has an outer \ion{H}{I} ring. 
It
is possible that this structure originated from the gravitational
interaction with two companions.

 Since we determine a luminosity and rotation velocity of our target systems, we are also able to highlight our findings
in the context of the classical Tully-Fisher relation (TFR). The term
  was introduced by 
\citet{Tully77} and follow-ups. The TFR appears
  to hold for a large variety of galactic systems. Any galaxy
  with an unusual IMF should also be an outlier on this
  relation. 
\begin{figure}[h!]
\includegraphics[width=8cm,keepaspectratio]{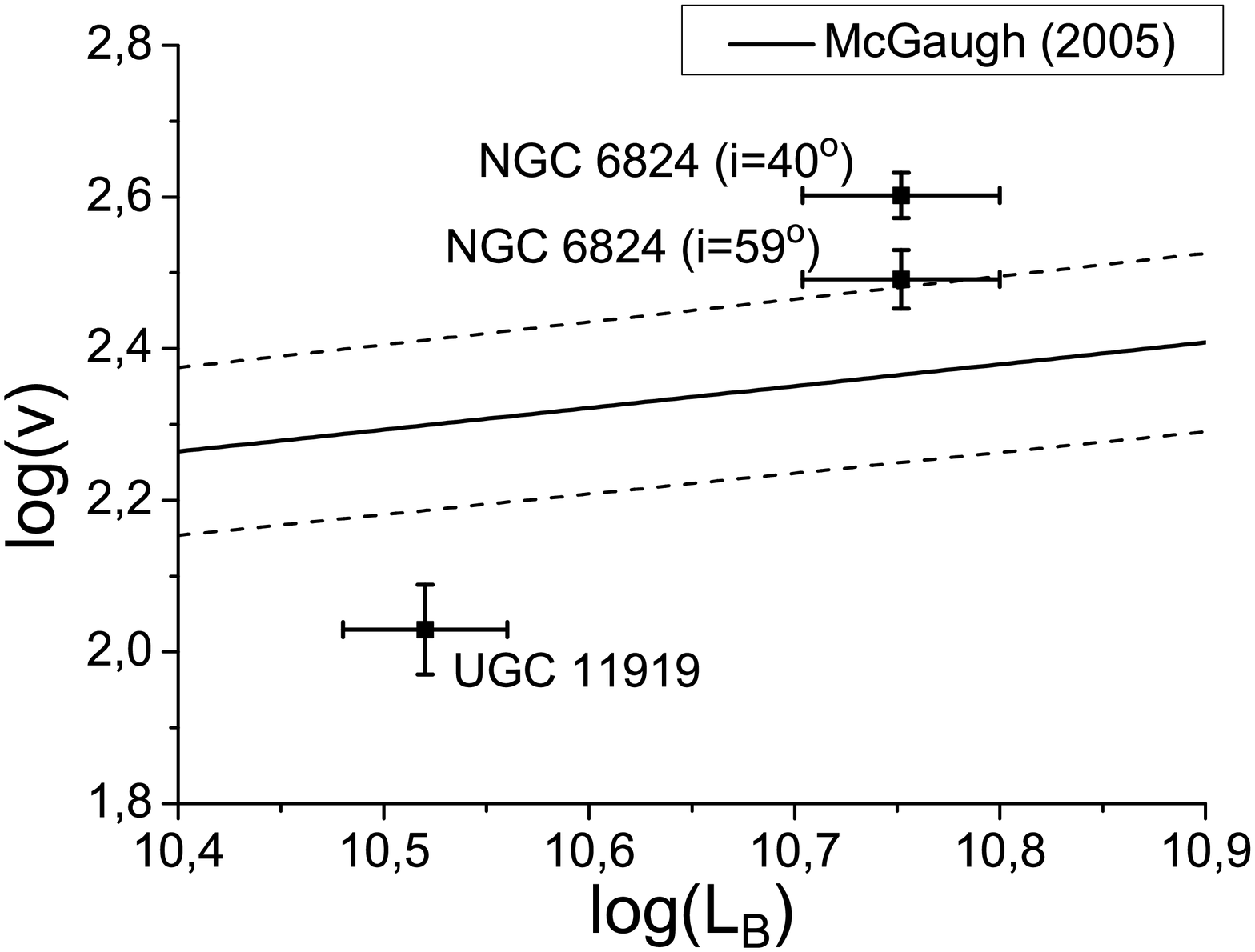}
\caption{Position of NGC~6824 (for two values of inclination) and UGC~11919 on the Tully-Fisher diagram (\citealt{McGaugh2005}). The dashed lines correspond to the formal uncertainties.}
\label{fig9}
\end{figure}
Figure \ref{fig9}  shows that while NGC~6824, for which we found the photometric and
dynamically determined masses to be consistent, lies on the upper
edges of the relation, UGC~11919 is a clear outlier.

  In accordance with our earlier conclusions 
 ( \citealt{Saburova2011A}), we claim that spiral galaxies of high brightness with a strongly bottom
heavy or bottom light stellar IMF should occur rarely.
In the present  paper we confirm the
  existence of a galaxy which has a very low $M/L$ for its old (red)
stellar population apparently being depleted of low mass stars. These
systems deserve a more detailed investigation.

\begin{acknowledgements}

We thank the referee, Albert Bosma for his important remarks, which allowed us to improve the paper.\\
 We thank Alexei Moiseev for his useful comments.\\
We acknowledge the possibility of using the HyperLeda database (http:\\leda.univ-lyon1.fr/).\\
The photometric data is based on observations obtained with the Apache Point Observatory
0.5-meter telescope, which is owned and operated by the Astrophysical
Research Consortium.\\
Anna Saburova thanks ASTRON (Netherlands) for the possibility of visiting it and for
help with the \ion{H}{I} data reduction.\\

The Westerbork Synthesis Radio Telescope is operated by the ASTRON
(Netherlands Foundation for Research in Astronomy) with support from the
Netherlands Foundation for Scientific Research (NWO).\\
This work was partly supported by Russian Foundation for Basic Research, grants 11-02-12247 and 12-02-31452, and the Dynasty fund.\\
TiRiFiC is publicly available from http:\\www.astron.nl/~jozsa/tirific/.
\end{acknowledgements} 

\bibliographystyle{aa}   
\bibliography{ref}

\end{document}